\documentstyle[graphicx, amssymb, psfig, 11pt]{article}

\begin{document}

\title{Universal Lossless Compression with Unknown Alphabets - The
Average Case
 \footnote{This work was partially supported by the University of Utah,
 ECE Department, startup fund and NSF Grant CCF-0347969.
 Parts of the material in this paper were presented at
 the 40th Annual Allerton Conference
 on Communication, Control, and Computing, Monticello, IL, October
 2-4, 2002, the IEEE International
 Symposium on Information Theory, Chicago, IL, June 27 - July 2, 2004, and
 the Data Compression Conference, Snowbird, Utah, U.S.A., March 23-25, 2004.
 }}
\author{Gil I. Shamir \\
 Department of Electrical and Computer Engineering \\
 University of Utah \\
 Salt Lake City, UT 84112, U.S.A \\
 e-mail: gshamir@ece.utah.edu.}
\date{}
\maketitle

\begin{abstract}
Universal compression of \emph{patterns\/} of sequences generated by
independently identically distributed
(i.i.d.) sources with unknown, possibly large, alphabets
is investigated.
A pattern is a sequence of indices that
contains all consecutive indices in increasing
order of first occurrence.
If the alphabet of a source that generated a sequence is unknown,
the inevitable cost of coding the unknown alphabet symbols
can be exploited to create the pattern of the sequence.
This pattern can in turn be compressed by itself.
It is shown that if the alphabet size $k$ is essentially
small, then the \emph{average\/} minimax and maximin redundancies
as well as the
redundancy of every code for almost
every source, when compressing a pattern,
consist of at least $0.5 \log \left (
n / k^3 \right )$ bits
per each unknown probability parameter, and if all alphabet letters are likely
to occur, there
exist codes whose redundancy is at most
$0.5 \log \left ( n / k^2 \right )$ bits
per each unknown probability parameter, where
$n$ is the length of the data sequences.
Otherwise, if the alphabet is large,
these redundancies are essentially at least $O \left ( n^{-2/3} \right
)$ bits per symbol, and there exist codes that achieve
redundancy of essentially $O \left ( n^{-1/2} \right )$
bits per symbol.
Two sub-optimal low-complexity sequential algorithms for compression
of patterns are presented and their description lengths analyzed, also
pointing out that the pattern average universal description length can decrease
below the underlying i.i.d.\ entropy for large enough alphabets.
%Their analysis demonstrates the relationship between i.i.d.\ compression
%and that of patterns for various alphabet sizes.
%Finally, the entropy of patterns is compared to that of i.i.d.\ sequences,
%and upper and lower bounds on the pattern description length are obtained
%as functions of the i.i.d.\ entropy.


\end{abstract}
%\clearpage

% Internal clause new commands
% ----------------------------

\newcommand{\be}{\begin{equation}}
\newcommand{\ee}{\end{equation}}

\newcommand{\bea}{\begin{eqnarray}}
\newcommand{\eea}{\end{eqnarray}}

\newcommand{\bi}{\begin{itemize}}
\newcommand{\ei}{\end{itemize}}

\newcommand{\ben}{\begin{enumerate}}
\newcommand{\een}{\end{enumerate}}

\newcommand{\bef}{\begin{figure}[htbp]}
\newcommand{\enf}{\end{figure}}

\newcommand{\bt}{\begin{tabular}{lcllcl}}
\newcommand{\et}{\end{tabular}}

\newcommand{\bd}{\begin{description}}
\newcommand{\ed}{\end{description}}

% Basic New Commands
% ------------------

\newcommand{\ul}{\underline}

\newcommand{\eref}[1]{(\ref{#1})}       % Equation reference.

% Ordinary Mathematical New Commands
% ----------------------------------

\newcommand{\dfn}{\stackrel{\triangle}{=}}  % Equal by definition.
\newcommand{\eqex}{\stackrel{\cdot}{=}}     % Exponentially equal.
\newcommand{\half}{\frac{1}{2}}                 % half.

\newcommand{\comb}[2]{\left (
 \raisebox{-4pt}{$\stackrel{\mbox{\large $#1$}}{#2}$} \right ) }

% Special new commands.
% ---------------------

\newcommand{\vvec} {{\mathbf v}}
\newcommand{\pvec} {\mbox{\boldmath $\theta$}}
\newcommand{\svec}{\mbox{\boldmath $\gamma$}}
\newcommand{\uvec} {{\mathbf u}}
\newcommand{\tvec} {{\mathbf t}}
\newcommand{\tvecm}{{\mathbf t}^-}
\newcommand{\tvecp}{{\mathbf t}^+}
\newcommand{\rvec} {\mbox{\boldmath $\pi$}}
\newcommand{\dvec} {{\mathbf d}}
\newcommand{\ivec} {{\mathbf i}}
\newcommand{\avec}{\mbox{\boldmath $\psi$}}
\newcommand{\sigvec}{\mbox{\boldmath $\sigma$}}
\newcommand{\phivec}{\mbox{\boldmath $\phi$}}
\newcommand{\vphivec}{\mbox{\boldmath $\varphi$}}
\newcommand{\bvec} {{\mathbf b}}
\newcommand{\aavec} {{\mathbf a}}

\newcommand{\Pvec} {{\mathbf \Theta}}
\newcommand{\Svec} {{\mathbf \Gamma}}
\newcommand{\Uvec} {{\mathbf U}}
\newcommand{\Tvec} {{\mathbf T}}
\newcommand{\Rvec} {{\mathbf \Pi}}
\newcommand{\Dvec} {{\mathbf D}}
\newcommand{\Avec} {\Psi}
\newcommand{\Yvec} {{\mathbf Y}}

\newcommand{\pvece}{\hat{\pvec}}
\newcommand{\svece}{\hat{\svec}}
\newcommand{\uvece}{\hat{\uvec}}
\newcommand{\tvece}{\hat{\tvec}}
\newcommand{\tvecme}{\hat{\mathbf t}^-}
\newcommand{\tvecpe}{\hat{\mathbf t}^+}
\newcommand{\dvece}{\hat{\dvec}}
\newcommand{\avece}{\hat{\avec}}

\newcommand{\Pvece}{\hat{\Pvec}}
\newcommand{\Svece}{\hat{\Svec}}
\newcommand{\Uvece}{\hat{\Uvec}}
\newcommand{\Tvece}{\hat{\Tvec}}
\newcommand{\Dvece}{\hat{\Dvec}}
\newcommand{\Avece}{\hat{\Avec}}

\newcommand{\uvect}{\tilde{\uvec}}
\newcommand{\uvecp}{\breve{\uvec}}
\newcommand{\svect}{\tilde{\svec}}
\newcommand{\svecb}{\bar{\svec}}
\newcommand{\ivecb}{\bar{\ivec}}
\newcommand{\rvect}{\tilde{\rvec}}
\newcommand{\Uvect}{\tilde{\Uvec}}

\newcommand{\pvecspace}{\Theta \hspace{-6.8pt} \mbox{{\small 0}}
 \hspace{2pt}}

\newcommand{\avecspace}{\Lambda}
\newcommand{\avv}{\avecspace'}
\newcommand{\avn}{\avecspace_{n}}
\newcommand{\avdd}{\avecspace_{\dvec}}
\newcommand{\avd}{\avecspace_{d}}
\newcommand{\avq}{\avecspace_{q}}
\newcommand{\avk}{\avecspace_{k}}
\newcommand{\avqd}{\avecspace_{d,q}}
\newcommand{\avqdd}{\avecspace_{\dvec,q}}
\newcommand{\avqxi}{\avecspace_{\xi}}
\newcommand{\aveps}{\avecspace_{\varepsilon}}
\newcommand{\avxinot}{\bar{\avecspace}_{\xi}}
\newcommand{\avqt}{\avecspace'_{q}}
\newcommand{\avqtt}{\avecspace''_{q}}
\newcommand{\avqnot}{\bar{\avecspace}_{q}}
\newcommand{\avg}{\avecspace_{\xi,\gamma}}
\newcommand{\avp}{\avecspace_{\xi,\theta}}
\newcommand{\avt}{\avecspace_{\xi,t}}
\newcommand{\avm}{\avecspace_{\xi,m}}
\newcommand{\fxig}{F_{\xi, \gamma}}
\newcommand{\fxip}{F_{\xi, \theta}}
\newcommand{\avgn}{\bar{\avecspace}_{\xi,\gamma}}
\newcommand{\avgna}{\bar{\avecspace}_{\xi,\gamma,1}}
\newcommand{\avgnb}{\bar{\avecspace}_{\xi,\gamma,2}}
\newcommand{\avpn}{\bar{\avecspace}_{\xi,\theta}}
\newcommand{\avtn}{\bar{\avecspace}_{\xi,t}}
\newcommand{\avmn}{\bar{\avecspace}_{\xi,m}}
\newcommand{\fxign}{\bar{F}_{\xi, \gamma}}
\newcommand{\fxipn}{\bar{F}_{\xi, \theta}}

\newcommand{\Tgi}{{\cal T}}
\newcommand{\Tgrid}{\mbox{\boldmath $\Tgi$}}
\newcommand{\Agrid}{\mbox{\boldmath $\Omega$}}
\newcommand{\tgrid}{\mbox{\boldmath $\tau$}}
\newcommand{\Ggrid} {{\mathbf G}}
\newcommand{\ggrid} {{\mathbf g}}

\newcommand{\phiset}{\mbox{\boldmath $\varphi$}}
\newcommand{\Phiset}{{\mathbf \Phi}}
\newcommand{\phis}{\varphi}
\newcommand{\Phis}{{\mathbf \Phi}}

\newcommand{\eventA}{{\cal A}}
\newcommand{\eventF}{{\cal F}}
\newcommand{\eventT}{{\cal T}}
\newcommand{\Xab}{{\cal X}}

% Theorems Lemmas and Examples counters
% -------------------------------------
\newtheorem{theorem}{Theorem}
\newtheorem{lemma}{Lemma}[section]
\newtheorem{corollary}{Corollary}
\newtheorem{proposition}{Proposition}
\newtheorem{condition}{Condition}

\renewcommand{\thecondition}{\Alph{condition}}

\section{Introduction}
\label{sec:introduction}

Classical universal compression \cite{davison} usually considers
coding sequences that were generated by a source with a known
alphabet but with some unknown statistics. In this paper, we
consider the universal coding problem, where an independently
identically distributed (i.i.d.)\ source generates data from an
alphabet that is totally unknown to both encoder and decoder, and
whose size $k$ can grow with $n$. In this case, the cost of coding
the alphabet letters is inevitable and depends strictly on the
alphabet letters themselves.  However, after coding of the
alphabet letters, the data sequence can be uniquely represented by
its \emph{pattern\/}. The pattern of a sequence is a sequence of
pointers that point to the actual alphabet letters, where the
alphabet letters are assigned \emph{indices\/} in order of first
occurrence. For example, the pattern of the sequence ``lossless''
is ``12331433''. A pattern sequence thus contains all positive
integers from $1$ up to a maximum value $k$ in increasing order of
first occurrence, and is also independent of the alphabet of the
actual data. One can separate the coding of the alphabet symbols
from that of the pattern, and use universal coding techniques to
encode patterns. The universal coding cost of a totally unknown
alphabet is inevitable regardless of the code used, and depends
strictly on the actual alphabet letters. Therefore, the more
interesting universal coding problem becomes that of efficiently
encoding the alphabet independent patterns.

To the best of our knowledge, the idea of separating the
description of the alphabet symbols from the representation of
the pattern of a sequence for universal coding first appeared in the literature
in \cite{aberg}.  This procedure was motivated in \cite{aberg} by
the \emph{multi-alphabet\/} coding problem
\cite{shtarkov1}, i.e., the problem in which
a sequence is generated by a known alphabet, but contains only a small
subset of the alphabet letters.   A separate description was used to inform
the decoder which symbols from the alphabet have occurred in a sequence,
and then their pattern was coded separately.
However, no theoretical
evidence was
provided to show that such a technique has
advantage over other multi-alphabet coding techniques,
as those proposed in \cite{shtarkov1}.

Stronger motivation for coding patterns of sequences was first given
by Jevti\'c, Orlitsky, and Santhanam \cite{orlitsky} (see also
\cite{orlitsky1}-\cite{orlitsky4o}),
who motivated this problem by the problem of universal coding of
sequences generated by sources over alphabets that are initially unknown
to both the encoder and the decoder.
The encoder then has
to send the decoder complete information about the alphabet
letters, and can utilize this inevitable cost to improve the
coding performance by representing the actual data sequence by its
pattern.  This problem can be strongly motivated by many practical applications
that compress sequences generated by either a small or a large alphabet.  For example,
consider transmission of text in a language that was never seen before.  The graphical
structure of the letters must first be transmitted.  If it is transmitted in the order
of first occurrence, the pattern of the text can then be compressed.  This application
further motivates the problem of pattern compression over large alphabets because
in text the natural alphabet unit can be a word instead of a letter.
Another example is compression of sequences of species first seen on another planet.
Since there is no prior knowledge of their forms, they can be designated by their pattern,
i.e., the first specie encountered is number $1$, the second number $2$, and so on.

The i.i.d.\ case is the simplest one to consider.  However,
coding of patterns whose underlying process is i.i.d.\ is
different from coding of i.i.d.\ sequences because the constraints
that are imposed by the definition of a pattern result in a
non-i.i.d.\ probability mass function over the patterns that is
different from the i.i.d.\ one of the original sequence.  This
allows shorter representations for patterns than those that would
be used for the underlying i.i.d.\ sequences. Of course, this
improvement is not free, and it only comes because of the
inevitable price of representing the alphabet itself. However,
while it was shown by Kieffer in \cite{kieffer} that if the
alphabet size is very large (goes to infinity), no universal code
exists, i.e., no code can achieve vanishing redundancy for i.i.d.\
sequences, this is not the case for the resulting patterns, as was first
shown by Orlitsky et.\ al.\ in \cite{orlitsky}, \cite{orlitsky1}-\cite{orlitsky4},
because only at most $n$ letters of the actual alphabet appear in
the pattern sequence. Furthermore, better universal compression performance
is also possible in the case where the alphabet size $k$ is
sub-linear in $n$ or even fixed.  Moreover, even better non-universal
compression is sometimes possible because every pattern represents
a collection of many sequences, thus reducing the overall pattern
entropy (see, e.g., \cite{gil11}, \cite{gil111}, \cite{shamir_dcc04},
\cite{shamir_allerton04}, \cite{shamir_itw05}).

The classical setting of the universal lossless compression
problem \cite{davison} assumes that a sequence $x^n$ of length $n$
that was generated by a source $\pvec$ is to be compressed without
knowledge of the particular $\pvec$ that generated $x^n$ but with
knowledge of the \emph{class\/} $\avecspace$ of all possible sources
$\pvec$. The average performance of any given code, that assigns a
length function $L(\cdot)$, is judged on the basis of the
\emph{redundancy\/} function $R_n \left ( L, \pvec \right )$,
which is defined as the difference between the expected code
length of $L \left ( \cdot \right )$ with respect to (w.r.t.)\ the
given source probability mass function $P_{\theta}$ and the
$n$th-order entropy of $P_{\theta}$ normalized by the length $n$
of the un-coded sequence.

Naturally,
the lack of knowledge of the source parameters
in universal coding results
in some redundancy when coding data emitted by any or almost
any unknown source from a known class.
To measure the universality of such a class, some notion of
this redundancy is used to represent the best possible performance
for some worst case, i.e., the redundancy expected from the best
code for the worst case.
This notion of redundancy thus serves as a lower bound on the
worst case redundancy of any code for this class of sources.
Two such notions are the
\emph{maximin\/} redundancy and the \emph{minimax\/} redundancy,
defined in Davisson \cite{davison}.
In the maximin Bayesian approach, the parameter $\pvec$
is considered random, and the
maximin redundancy is obtained by the worst distribution
that maximizes the minimum \emph{expected\/} redundancy, i.e.,
the worst distribution for the best code.
The minimax approach considers the parameter to be
deterministic, and defines the minimax redundancy
as the redundancy of the best code for the worst choice of $\pvec$.
A third stronger notion of redundancy for ``most'' sources in
a class was later established by Rissanen \cite{rissa}.
This notion describes the
performance of the best possible code for almost every source in
the class except a subset of the class whose probability under the
uniform \emph{prior\/} (i.e., distribution in
$\avecspace$) is negligible, and for which smaller redundancy can
be obtained.
A different approach to the study of universal codes is that
of individual sequences.
The minimax redundancy for individual sequences \cite{shtarkov}
is the redundancy of the best code for the worst possible sequence
$x^n$ that can be generated by any source in the class.  In this paper,
however, we focus on average redundancies.

Several publications
\cite{davison}, \cite{davison2}, \cite{davison1},
\cite{gala1}, \cite{rissa}
investigated the average redundancy performance in standard compression
of classes of parametric sources and in particular i.i.d.\ sources
over alphabets of size $k$, which are governed by $k-1$ parameters.
It was shown that for a \emph{finite\/} size alphabet,
each unknown probability parameter costs
at least $0.5 \log n$ extra redundancy bits.  This lower bound
applies in all average senses: minimax and maximin (which were
demonstrated to be identical), and for almost all sources
in the class.  It also applies in the minimax individual sense.
Furthermore, it was shown to be achievable, and in particular by
using a linear complexity (fixed per symbol) sequential coding
scheme that combines the \emph{universal mixture\/} based
\emph{Krichevsky-Trofimov\/} (KT) probability estimators \cite{krichevsky} with
arithmetic coding \cite{arit1}.
Recently, \cite{gil_dimacs}, \cite{gil9}, \cite{gil91}, we extended the average results
and showed that if the alphabet size is allowed to grow sub-linearly with
$n$, each probability parameter costs $0.5 \log (n/k)$ bits in all
average senses, and also this redundancy is achievable even sequentially with
the KT estimators.
At the same time, related
results have been independently obtained for the
individual case by Orlitsky, Santhanam, and Zhang \cite{orlitsky3}, \cite{orlitsky4o}.

While standard universal compression, in particular that of i.i.d.\ sources, has been
extensively researched, the problem of compression of patterns has only been
addressed recently, with focus, until now, only on the individual sequence case.
The initial work on this problem was presented in \cite{aberg}.
However,
Jevti\'c, Orlitsky, Santhanam, and Zhang \cite{orlitsky},
\cite{orlitsky1}-\cite{orlitsky4o} have recently achieved significant
progress in understanding this problem.  In particular, they considered the performance
of the best universal code for the worst sequence
over all possible patterns generated by underlying i.i.d.\ sequences
of length $n$.
Using combinatoric techniques, they have
shown that the minimax individual redundancy is lower bounded
by $O \left ( n^{-2/3} \right )$ bits per symbol and upper bounded
by $O \left ( n^{-1/2} \right )$ bits per symbol.
They have also
derived a high complexity sequential algorithm that achieves the order of the upper
bound and a sub-optimal computationally heavy low complexity
sequential algorithm that achieves
redundancy of $O \left ( n^{-1/3} \right )$ bits per symbol.

In this paper, we focus, unlike previous work,
on the average redundancy performance of
universal codes for coding patterns.
We also consider the different
behavior for different alphabet sizes $k$, and
investigate the actual \emph{description length\/} required for patterns.
First, lower bounds on the average minimax/maximin redundancies are obtained
as a function of the alphabet size $k$.  (These bounds
naturally apply also to the worst case individual redundancies.)
Then, we derive lower bounds on the redundancy for most sources.
Next, we obtain upper bounds on the redundancy focusing on the case in which
all actual alphabet symbols are likely to be observed in the coded sequence.
Although we use techniques that are much different from those
used in \cite{aberg}, \cite{orlitsky}, \cite{orlitsky1}-\cite{orlitsky4o} for the
derivation of the minimax lower bound and the upper bound, the
average case results we obtain in this paper demonstrate similar
behavior of the redundancy in the average cases
to that of the individual worst case.
This is very important, because it demonstrates
that the expected behavior for the worst setting
is not much better than the worst sequence behavior.  Hence,
when coding patterns, like when coding standard sequences, one cannot expect to perform
significantly better for the worst source than the performance for the worst sequence.
Next, two sub-optimal low-complexity sequential algorithms are presented.  The actual
description length of these algorithms is studied (where the displacement
relative to the i.i.d.\ source entropy, defined as the \emph{modified redundancy\/} for
patterns is considered).  The description length for these algorithms demonstrates an
interesting result, where the pattern entropy for large enough alphabets must decrease
compared to the i.i.d.\ one.  Subsequently to the work presented here
(see also \cite{shamir_dcc04}), pattern entropy
and entropy rate have been extensively studied, first in \cite{gil111}, and later
in \cite{gemelos04}-\cite{gemelos05isit}, \cite{orlitsky04itw}-\cite{orlitsky05isit},
\cite{gil11}, \cite{shamir_allerton04}-\cite{shamir_itw05}.

To derive the lower bounds, we use the
relations between redundancy and capacity that are
presented in Section~\ref{sec:background} based on
\cite{davison}, \cite{feder}, \cite{merhav1}.
The minimax/maximin bound we obtain for larger $k$'s is larger than
that obtained for most sources.  This is because we must use
different techniques
to derive the two bounds, where the more demanding conditions
to obtain the bound for most sources result in a smaller bound.
This hints to the fact that it may be possible that in the case
of patterns, it may cost more redundancy beyond the entropy
to code the worst source than it
costs to code most other sources in the class.
The upper bounds are obtained by a constructive approach.  For small
$k$'s it combines Rissanen's approach \cite{rissa} with our
recent approach from \cite{gil9}, \cite{gil91} and with the
more demanding conditions in coding patterns.

For readability and convenience, each of the sections that contain heavy
analysis is structured such that the results and their properties are described
first.  Then, a short description of the structure of the proof is given.
Finally, each such section is concluded with the technical proofs, where
steps that require much technical detail are relegated to appendices.

The outline of the paper is as follows.
In Section~\ref{sec:note_def}, we define the notation.
Section~\ref{sec:background} reviews the individual sequence results of
coding patterns, and the techniques we use to derive the new
results.
Section~\ref{sec:main_results} summarizes the main results in the paper.
Sections~\ref{sec:maximin} and~\ref{sec:lb_most} contain the derivations
of the minimax/maximin lower bounds and the bounds for most sources,
respectively.  In Section~\ref{sec:upper_bounds}, we derive upper bounds
on the redundancy with focus on the class of sources for which all symbols are
likely to be observed.  In Section~\ref{sec:seq},
we present the sequential algorithms and study their description lengths and their
displacements from the i.i.d.\ entropy.
%Section~\ref{sec:entropy} contains a study of
%the entropy of patterns.
Then, in Section~\ref{sec:discuss}, a discussion about the
results is presented.  Finally, some concluding remarks are brought in
Section~\ref{sec:summary}.

\section{Notation and Definitions}
\label{sec:note_def}

\subsection{Universal Coding}

Let $x^n \dfn \left ( x_1, x_2, \ldots, x_n \right )$ denote a sequence
of $n$ symbols over an unknown alphabet $\Sigma$ of size $k$.
The class
of all i.i.d.\ sources that can generate any sequence $x^n$ over $\Sigma$
will be denoted by $\avecspace$.  The subclass of i.i.d.\ sources that
generate up to $k$ alphabet symbols will be denoted by $\avk$.
The subclass of sources that generate $k$ symbols that are likely to be
observed with probability greater than $1 - o(k/n)$
will be denoted by $\tilde{\avecspace}_k$.
A parameter
$\pvec \in \avk$ is a vector of $k-1$ probability parameters
$\pvec \dfn \left ( \theta_1, \theta_2, \ldots, \theta_{k-1} \right )$.
For convenience, we will sometimes use the constrained component $\theta_k$
of $\pvec$.  All $k$ components of $\pvec$ are non-negative and sum up to $1$.
In general, boldface letters will denote vectors, whose components will
be denoted by their indices in the vector.  We will use \emph{hat\/} to denote
the \emph{Maximum Likelihood\/} (ML) estimator of a parameter obtained from
the data sequence $x^n$, e.g. $\pvece$ will denote the ML estimator of $\pvec$.
Capital letters will denote random variables.

Let $\pvec \in \avk$ be a parameter vector that determines the statistical
parameters of some source in the class $\avk$.
Let $x^n$ be a sequence of $n$ symbols generated by
the source $\pvec$.
The average $n$th-order redundancy obtained
by a code
that assigns length function $L (\cdot )$ for source $\pvec$ is defined as
\be
 \label{eq:redundancy_def}
 R_n \left (L, \pvec \right ) \dfn
  \frac{1}{n}
  E_{\theta} L \left [ X^n \right ] -
  H_{\theta} \left [  X \right ],
\ee
where $E_{\theta}$ denotes expectation w.r.t.\ the parameter $\pvec$, and
$H_{\theta} \left [ X \right ]$ is the (per-symbol) entropy of the source.
(We will also use $H_{\theta} \left [ X^n \right ]$ as the $n$th-order
sequence entropy of $\pvec$, where in the i.i.d.\ case,
$H_{\theta} \left [ X^n \right ] = n H_{\theta} \left [ X \right ]$.)
It has been established in the literature
(see, e.g., \cite{krichevsky}, \cite{merhav1}, \cite{rissa})
that assigning a universal
probability $Q \left ( x^n \right )$ is identical to designing a universal
code for coding $x^n$, because entropy
coding techniques can be used to code the sequence
using a number of bits that
equals, up to integer length constraints,
to the negative logarithm to the base of $2$
of the assigned probability.  In particular,
one can use \emph{arithmetic coding\/} \cite{arit1} to allow sequential
coding with sequential probability assignment schemes.  We will thus
ignore integer length constraints, and in places consider the redundancy
as a function of the probability assignment scheme $Q \left ( \cdot \right )$
instead of the code $L \left ( \cdot \right )$.

We can also define the \emph{individual\/} sequence
redundancy (see, e.g., \cite{shtarkov})
of a code with length function $L \left ( \cdot \right )$ per sequence $x^n$ as
\be
 \label{eq:individual_red_def}
 R_n \left (L, x^n \right ) \dfn
  \frac{1}{n} \left \{ L \left ( x^n \right ) +
  \log P_{ML} \left ( x^n \right )
  \right \},
\ee
where the logarithm function is taken to the base of $2$, here and elsewhere,
and $P_{ML} \left ( x^n \right ) \dfn P_{\hat{\theta}} \left ( x^n \right )$
is the probability of $x^n$ given
by the ML estimator $\pvece$ of the governing parameters.
The negative logarithm
of this probability is the smallest possible
code length for a particular sequence under a given statistical model (in our
case the i.i.d.\ one).

The average \emph{minimax\/} redundancy of the class $\avk$ is defined as
\be
 \label{eq:minimax_red}
 R_n^+ \left ( \avk \right ) \dfn
  \min_L \sup_{\pvec \in \avk} R_n \left ( L, \pvec \right ).
\ee
Similarly, we can define the \emph{individual minimax\/} redundancy as that
of the best code $L \left ( \cdot \right )$ for the worst sequence $x^n$, i.e.,
\be
 \label{eq:minimax_individual}
 \hat{R}_n^+ \left ( \avk \right ) \dfn
  \min_L \sup_{\pvec \in \avk} \max_{x^n}
  \frac{1}{n} \left \{ L \left ( x^n \right ) +
  \log P_{\theta} \left ( x^n \right ) \right \}.
\ee

To define the \emph{maximin\/} redundancy of $\avk$,
let us assign a probability
measure (prior) $w \left ( \cdot \right )$ on $\avk$
and let us define the mixture source
\be
 \label{eq:mixture_source}
 P_{w} \left ( x^n \right ) \dfn \int_{\avk}
  w \left ( d\pvec \right )P_{\theta} \left ( x^n \right ).
\ee
The average redundancy associated with a length function
$L \left ( \cdot \right )$ is defined as
\be
 \label{eq:mixture_redundancy}
 R_n \left (L, w \right ) \dfn
  \int_{\avk}
  w \left ( d\pvec \right ) R_n \left ( L, \pvec \right ).
\ee
The minimum expected redundancy for a given prior $w$
(which is attained by the ideal code length w.r.t.\ the mixture,
$L \left (x^n \right ) = - \log P_{w} \left ( x^n \right )$) is
defined as
\be
 R_n \left ( w \right ) \dfn
  \min_L R_n \left ( L,w \right ).
\ee
Finally, the maximin redundancy of the class
$\avk$ is the worst case minimum expected
redundancy among all priors $w$, i.e.,
\be
 \label{eq:maximin_redundancy}
 R_n^- \left ( \avk \right ) \dfn
  \sup_{w} R_n \left ( w \right ).
\ee

\subsection{Patterns}

The \emph{pattern\/} of a sequence $x^n$ will be denoted by
$\Psi \left ( x^n \right )$.  Many different sequences over
the same alphabet (and over different alphabets) have the same pattern.
For example, for
the sequences $x^n =$``lossless'', $x^n =$``sellsoll'',
$x^n =$``12331433'', and $x^n =$``76887288'', the pattern is
$\Psi \left ( x^n \right ) =$``12331433''.  Therefore, for given
$\Sigma$ and $\pvec$, the probability of a pattern \emph{induced\/}
by an i.i.d.\ underlying probability is given by
\be
\label{eq:pattern_probability}
 P_{\theta} \left [ \Psi \left (x^n \right) \right ] =
  \sum_{y^n: \Psi (y^n) = \Psi (x^n)} P_{\theta} \left ( y^n \right ).
\ee
We note that the probability in \eref{eq:pattern_probability} is dominated
by some of the sequences, where others only contribute negligibly.  This
fact is used to derive an upper bound in Section~\ref{sec:upper_bounds}.
The per sequence (block)
\emph{pattern entropy\/} of order $n$ of a source $\pvec$ is thus defined
as
\be
\label{eq:pattern_entropy}
 H_{\theta} \left [ \Psi \left (X^n \right ) \right ] \dfn
 -\sum_{\Psi \left (x^n \right )} P_{\theta} \left [ \Psi \left (x^n \right) \right ]
 \log P_{\theta} \left [ \Psi \left (x^n \right) \right ].
\ee

In order to define the redundancy function of patterns for a given code and
a given source $\pvec$, we need to realize that a vector $\pvec'$ that is
a permutation of another vector $\pvec$ produces similar typical patterns, and
is, in fact, the same source in the pattern domain.  Therefore, we can define
the notation $\avec \left ( \pvec \right )$ as the permutation of
$\pvec$ which is ordered in non-decreasing order of components,
i.e., $\psi_1 \left ( \pvec \right ) \leq \psi_2 \left ( \pvec \right ) \leq
\ldots \leq \psi_k \left ( \pvec \right )$.
For example, if $\pvec = \left ( 0.7, 0.1, 0.2 \right )$,
then $\avec \left ( \pvec \right ) = \left ( 0.1, 0.2, 0.7 \right )$.
We can also, alternately, view a
vector $\sigvec$ as a permutation vector of indices, and use
$\theta \left ( \sigma_i \right )$ to denote the $i$th component of the
permuted vector $\pvec$, permuted according to $\sigvec$.
For the example above, if we define $\sigvec = \left (3, 1, 2 \right )$,
then $\pvec \left ( \sigvec \right ) = \left ( 0.2, 0.7, 0.1 \right )$
and $\theta \left ( \sigma_2 \right ) = \theta_1 = 0.7$.
In most sections, we will consider the original vector $\pvec$ to
be already ordered non-decreasingly, and therefore the identity
permutation $\sigvec = \left ( 1,2, \ldots, k \right )$ will
give $\avec \left ( \pvec \right ) = \pvec = \pvec \left ( \sigvec \right )$.
All vectors $\avec \left ( \pvec \right )$ for all $\pvec \in \avk$ will
constitute the pattern space $\Psi \left ( \avk \right )$, and similarly,
we can define $\Psi \left ( \avecspace \right )$ as the pattern
space induced by (or projected from) the class $\avecspace$.

The \emph{average pattern redundancy\/} for coding patterns generated
by a source $\pvec$ using a code that assigns a representation of length
$L \left [ \Psi \left ( x^n \right ) \right ]$ to the pattern of
sequence $x^n$ is defined as
\be
 \label{eq:pattern_redundancy_def}
 R_n \left [L,  \avec \left ( \pvec \right ) \right ] \dfn
  \frac{1}{n}
  E_{\theta} L \left [\Psi \left( X^n \right )\right ] -
  \frac{1}{n}
  H_{\theta} \left [ \Psi \left ( X^n\right )\right ].
\ee
Similarly to \eref{eq:individual_red_def}, we can define the
\emph{individual pattern redundancy\/} for a given code as
\be
 \label{eq:pattern_individual_red_def}
 R_n \left [L, \Psi \left (x^n \right ) \right ]\dfn
  \frac{1}{n} \left \{ L \left [ \Psi \left ( x^n \right )\right ] +
  \max_{\pvec}
  \left \{ \log P_{\theta} \left [ \Psi \left ( x^n \right ) \right ]
  \right \} \right \}.
\ee
Note that the ML probability is now different from that for the simple
i.i.d.\ case, because the ML is taken over the pattern probability and not
over the i.i.d.\ one.

Even in the simplest i.i.d.\ underlying case, it becomes very difficult
to derive closed form expressions beyond \eref{eq:pattern_probability}
on the probability of a pattern, (except for very specific patterns).  It
will therefore be useful to define quantities that relate a
code length to the i.i.d.\ entropy in the average case and to the i.i.d.\
ML probability in the individual case.  We will refer to these quantities
as the \emph{modified\/} redundancies.   The modified redundancy will be
studied in Section~\ref{sec:seq}, as part of the study of the description
length of the proposed sequential schemes. The average \emph{modified redundancy\/}
for a code $L \left ( \cdot \right )$ that codes patterns of a source
$\pvec$ is defined as
\be
 \label{eq:pattern_modified_redundancy_def}
 \tilde{R}_n \left [L,  \avec \left ( \pvec \right ) \right ] \dfn
  \frac{1}{n}
  E_{\theta} L \left [\Psi \left( X^n \right )\right ] -
  H_{\theta} \left [X\right ].
\ee
The individual pattern modified redundancy is defined as
\be
 \label{eq:pattern_individual_modified_red_def}
 \tilde{R}_n \left [L, \Psi \left (x^n \right ) \right ]\dfn
  \frac{1}{n} \left \{ L \left [ \Psi \left ( x^n \right )\right ] +
  \max_{\pvec}
  \left \{ \log P_{\theta} \left [ x^n \right ]
  \right \} \right \}.
\ee
We should note that unlike the regular redundancy, the modified redundancy
does not actually satisfy conditions that must be satisfied by redundancy functions.
In particular, it can be negative also in the average case.  If this happens,
it simply means that
one can universally
describe patterns using shorter descriptions than the entropy of
the underlying i.i.d.\ source.  We will see this phenomenon in
Section~\ref{sec:seq} and in \cite{gil11}.
The modified redundancy thus becomes handy for bounding the description length
a code can assign to a pattern, i.e,
\be
\label{eq:pattern_description_length}
 E_{\theta} L \left [\Psi \left( X^n \right )\right ] =
 H_{\theta} \left [ \Psi \left ( X^n\right )\right ] +
 n R_n \left [L,  \avec \left ( \pvec \right ) \right ] =
 H_{\theta} \left [X^n\right ] +
 n \tilde{R}_n \left [L,  \avec \left ( \pvec \right ) \right ],
\ee
and we can use either equalities to bound this description length.

Using the definition of the average pattern redundancy in
\eref{eq:pattern_redundancy_def},
we can replace $R_n \left (L, \pvec \right )$ by
$R_n \left [ L, \avec \left ( \pvec \right ) \right ]$ in
\eref{eq:minimax_red} to define the average minimax
pattern redundancy $R_n^+ \left [ \Avec \left ( \avk \right )\right ]$.
Similarly, we can define the average maximin pattern redundancy
$R_n^- \left [ \Avec \left ( \avk \right )\right ]$ by the same
substitution in \eref{eq:mixture_redundancy}.
Taking the maximum
of \eref{eq:pattern_individual_red_def}
on $x^n$ and the minimum on $L \left ( \cdot \right )$,
similarly to \eref{eq:minimax_individual}, we obtain the
individual minimax pattern redundancy
$\hat{R}_n^+ \left [ \Avec \left ( \avk \right )\right ]$.
Note that all these redundancies can also be obtained w.r.t.\
the class of all i.i.d.\ sources $\avecspace$ regardless of the alphabet
size.  Naturally,
$R_n^+ \left [ \Avec \left ( \avecspace \right )\right ]$,
$R_n^- \left [ \Avec \left ( \avecspace \right )\right ]$,
and $\hat{R}_n^+ \left [ \Avec \left ( \avecspace \right )\right ]$
will take the maximal redundancy value over all alphabet sizes $k$.

\section{Technical Background}
\label{sec:background}

\subsection{Individual Pattern Redundancy}

To the best of our knowledge, universal compression of patterns
was first introduced by {\AA}berg, Shtarkov and Smeets \cite{aberg}.
{\AA}berg et.\ al.\ addressed the compression problem of
individual pattern sequences.  In particular, they used the
individual sequence minimax approach
developed by Shtarkov \cite{shtarkov} to design the best code
in the individual minimax sense.  For standard sequence compression,
this approach assigns to an $n$-symbols sequence $x^n$ probability
$Q \left ( x^n \right )$ that
equals its ML probability
normalized by the sum of the
ML probabilities over all possible sequences, i.e.,
\be
\label{eq:individual_ml}
 Q  \left ( x^n \right ) \dfn
 \frac{P_{ML} \left ( x^n \right )}
 {\sum_{ y^n}
 P_{ML} \left ( y^n \right )},
\ee
where $P_{ML} \left ( x^n \right )$ is the ML probability of $x^n$, and
the summation is over all possible sequences $y^n$ of
length $n$.  This approach guarantees (under negligible integer length
constraints) individual redundancy of
\be
\label{eq:minimax_ind_red}
 R_n \left (Q, x^n \right ) = \frac{1}{n} \log
 \frac{P_{ML} \left ( x^n \right ) }{Q \left ( x^n \right )} =
 \frac{1}{n} \log \left \{ \sum_{y^n} P_{ML} \left ( y^n \right ) \right \}
\ee
for every sequence $x^n$.  Equation \eref{eq:minimax_ind_red} is true
in particular for the worst sequence $x^n$ for which this redundancy
is the minimal attainable.  Therefore, this approach achieves the minimax
redundancy.

The approach above was modified for patterns by modifying
\eref{eq:individual_ml} to
\be
\label{eq:individual_pattern_ml}
 Q  \left [ \Psi \left ( x^n \right ) \right ] \dfn
 \frac{P_{\widehat{\psi\left ( \theta \right )}}
 \left [ \Psi \left ( x^n \right ) \right ]}
 {\sum_{ \Psi \left ( y^n \right ): \theta \in \Psi \left ( \avk \right )}
 P_{\widehat{\psi\left ( \theta \right )}}
 \left [ \Psi \left ( y^n \right ) \right ]},
\ee
where $P_{\widehat{\psi\left( \theta \right )}}
\left [ \Psi \left ( x^n \right ) \right ]$ is the
ML pattern probability for the pattern of the sequence $x^n$, and the normalization
factor is the sum of all ML probabilities for all possible patterns of
sequences generated by sources $\pvec \in \avk$.  Restricting the derivation
to $\avk$ (and not the wider i.i.d.\ class $\avecspace$), it was shown
in \cite{aberg} that the normalizing sum is approximately lower bounded by
\be
 \label{eq:aberg_bound}
 \sum_{ \Psi \left ( y^n \right ): \theta \in \Psi \left ( \avk \right )}
 P_{\widehat{\psi\left ( \theta \right )}} \left [ \Psi \left ( y^n \right ) \right ]
 \gtrsim
 \frac{1}{k!} \cdot
 \frac{\sqrt{\pi}}{\Gamma \left (k/2 \right )} \cdot
 \left (n / 2 \right )^{(k-1)/2},
\ee
where $\Gamma \left ( \cdot \right )$ is the Gamma function.
If further analysis steps are performed beyond
those in \cite{aberg}, this yields
a lower bound on the individual minimax pattern redundancy for patterns
with at most $k$ different alphabet symbols of
\be
 \label{eq:aberg_bound1}
 \hat{R}_n^+ \left [ \Psi \left ( \avk \right )\right ] \gtrsim
 \left ( 1 - \varepsilon \right )
 \frac{(k-1)}{2n} \log \frac{n}{k^3},
\ee
where $\varepsilon > 0$ can be made arbitrarily small.
This bound is, of course, useful only for $k = o\left (n^{1/3} \right )$ and becomes
negative for larger alphabet sizes.
Based on this result and prior results in \cite{shtarkov1}, {\AA}berg
et.\ al.\ also proposed a sequential scheme for coding patterns, for which
they provided empirical results.  The computational requirements of this scheme
appear to be rather demanding.

Major progress in the research of individual pattern compression
has been recently
obtained by Jevti\'c, Orlitsky, Santhanam, and Zhang \cite{orlitsky},
\cite{orlitsky1}-\cite{orlitsky4}.
The approach used in those papers was similar to that in \cite{aberg} based on Shtarkov's
minimax results and on combinatoric techniques.  These papers considered the compression
of patterns generated by any source from the whole class $\avecspace$, independently
of the alphabet size $k$, i.e., the maximum number of different indices
in the pattern.
First, it was shown \cite{orlitsky} that probability
assignment of
\be
\label{eq:individual_modified_pattern_ML}
 \tilde{Q}  \left [ \Psi \left ( x^n \right ) \right ] \dfn
 \frac{P_{ML} \left ( x^n \right )}
 {\sum_{ \Psi \left (y^n \right ) : \theta \in \Psi \left ( \avecspace \right )}
 P_{ML} \left ( y^n \right )},
\ee
where $P_{ML} \left ( x^n \right )$ is
the i.i.d.\ ML probability (not the pattern ML probability) but the
summation is only on all possible patterns,
results in modified individual redundancy of
\be
\label{eq:minimax_modified_red}
 \tilde{R}_n \left [ \tilde{Q} , \Psi \left ( x^n \right ) \right ] =
  \frac{1.5 \log e}{n^{2/3}} +
 o \left ( \frac{1}{n^{2/3}}\right ).
\ee
This redundancy is obtained for every pattern of length $n$
independently of the number of indices in the pattern, and is
also the minimax modified individual pattern redundancy.
Then, Orlitsky et.\ al.\ \cite{orlitsky1}-\cite{orlitsky4} demonstrated that this modified
redundancy is, in fact, a lower bound on the actual pattern redundancy.
(Note that if the analysis in \cite{aberg} is modified
to the whole class $\avecspace$, one can obtain the same bound.)
Using integer partitioning of
a sequence of length $n$, it was also shown in
\cite{orlitsky1}-\cite{orlitsky4} that
there exist codes that achieve individual
minimax pattern redundancy of at most $O \left ( n^{-0.5} \right )$.  Summarizing
all these results, it was shown that there exist codes for which
\be
 \label{eq:orlitsky_bounds}
 \frac{1.5 \log e}{n^{2/3}} +
 o \left ( \frac{1}{n^{2/3}}\right ) \leq
 \hat{R}_n^+ \left [ \Psi \left ( \avecspace \right ) \right ]
 \leq
 \frac{\pi \sqrt{2/3} \log e}{\sqrt{n}}.
\ee
Finally, a computationally demanding high complexity sequential scheme
was shown
in \cite{orlitsky2}-\cite{orlitsky4} to achieve the order of the upper
bound
in \eref{eq:orlitsky_bounds},
as well as a low-complexity
sequential scheme that achieves minimax individual redundancy
of $O \left ( n^{-1/3} \right )$.

\subsection{Average Case - Background}

Unlike the prior results on compression of patterns, we focus on the average
case problem in compression of patterns induced by sequences generated by
i.i.d.\ sources.  To derive lower and upper bounds, we will use
techniques that are based on Davisson's \cite{davison}
and Rissanen's \cite{rissa} approaches, and their extension
\cite{feder}, \cite{merhav1}.  In particular, the well established
connection between
universal coding redundancy and channel capacity will be used to obtain
lower bounds on the average pattern redundancy.
In \cite{davison}, it was established that the maximin redundancy of a class
$\avk$ is bounded from below by (and asymptotically equals to)
the normalized capacity of the ``channel'' defined by the conditional probability
$P_{\theta} \left ( x^n \right)$, i.e., the channel
whose input is the parameter
$\pvec$ and whose output is the data sequence
$x^n$.  It was further established that the average minimax redundancy is
lower bounded by the maximin redundancy.  Using Gallager's later result
\cite{gala1} that shows that the minimax and maximin
redundancies are essentially equivalent, this leads to the bound
on both minimax and maximin redundancies of
\be
 \label{eq:maximin_red_cap}
 R_n^+ \left ( \avk \right ) = R_n^- \left ( \avk \right ) \geq
  \sup_{w} \frac{1}{n} I_{w} \left ( \Pvec ;~X^n \right ),
\ee
where $I_{w} \left ( \Pvec ;~X^n \right )$ is the mutual
information induced by the joint measure
$w \left ( \pvec \right ) \cdot P_{\theta} \left (x^n \right )$.
Using \eref{eq:maximin_red_cap}, any lower bound on the capacity
of the channel defined by $P_{\theta} \left ( x^n \right)$ can be
used to bound the minimax and maximin redundancies.  In particular,
one can pick a set $\Agrid$ of $M$ points
$\pvec \in \avk$.  If these points can
be shown to be
\emph{distinguishable\/} by the sequence $X^n$, then
$(\log M)/n$ can serve as a lower bound on the
normalized capacity of the respective channel,
and thus on the minimax and maximin redundancies.
This lower bound is specifically implied by Fano's Inequality using
the fact that the error probability goes to $0$ (see, e.g., \cite{merhav1}).
Distinguishability in a set of points $\Agrid$ is defined (in a stronger sense than
needed to the result above) as follows.  Let $\pvec \in \Agrid$ be a point that generates
the random sequence $X^n$.  Let $\pvece = f \left (X^n \right )$ be an estimator of
$\pvec$ from $X^n$, and let $\pvece_{\Omega} = g \left (\pvece \right )$
be a point in $\Agrid$ that is used
to estimate $\pvec$ from the estimator $\pvece$, where $\pvece$ is not necessarily a point in
$\Agrid$.  Then, there exist functions $f(\cdot)$ and $g(\cdot)$, such that
$P_{\theta} \left (\pvece_{\Omega} \neq \pvec \right ) \rightarrow 0$ as
$n\rightarrow \infty$, for every $\pvec \in \Agrid$.  In words,
there exists an estimator of $\pvec$
out of the points in $\Agrid$, such that
the probability
that a sequence that was generated by one point in the set would appear to have
been generated by a different point in the set vanishes with $n$.

The approach described above
will be adopted to patterns in order
to derive the bound in Section~\ref{sec:maximin}.
In the patterns case, we will consider the set of sources
$\pvec \in \Psi \left ( \Agrid \right )$, and the pattern source
estimator $\avec \left (\pvece \right )$ will be
defined as a function of the pattern, i.e.,
$\avec \left (\pvece\right ) = f \left [ \Psi \left ( X^n \right ) \right ]$,
since the sequence itself is not
observed.  Then, the estimator
$\pvece^{\psi}_{\Omega} = g \left [ \avec \left ( \pvece \right ) \right ]$
must be in the pattern source space
$\Psi \left ( \Agrid \right )$.
Since the minimax
and maximin average redundancies are essentially the same, we will
consider only the minimax one, and the results will apply to both.

Merhav and Feder \cite{merhav1}
extended the concept of the
redundancy-capacity and derived a strong version of the
redundancy-capacity theorem.  They showed that
if it is possible to partition the class $\avk$ into disjoint sets
of sources $\pvec$, each of at least $M$ points that are distinguishable
by $X^n$, then the redundancy is lower bounded by
\be
 \label{eq:red_cap}
 R_n \left ( L, \pvec \right ) \geq \left ( 1 - \varepsilon \right )
    \frac{\log M}{n},
\ee
for every code $L \left ( \cdot \right )$,
and almost every $\pvec \in \avk$, where
$\varepsilon > 0$ is arbitrarily small.  In order to be able to use this
result, one needs to make sure that the points in each set are
uniformly distributed within the set, and every point in $\avk$
is included in one set (see also \cite{gil5}-\cite{gil_dimacs}).
Sometimes such an assumption cannot
be made unless a non-uniform prior is assumed within the class.  In such
cases the result in \eref{eq:red_cap} does not apply to most sources in
the class, but to all sources in the class except a subset whose probability
under the prior assumed vanishes.  The technique that will be presented in
Section~\ref{sec:maximin} for patterns will suffer from this problem, and thus cannot be used
to obtain a lower bound on the redundancy of most sources.  Therefore, a different
technique that uses Merhav and Feder's theorem will be applied in
Section~\ref{sec:lb_most} to derive a lower bound for most sources.
As in Section~\ref{sec:maximin}, the ideas described in this paragraph for
standard compression will be applied to patterns in a similar manner to that described
in the preceding paragraph.

Both versions of the redundancy-capacity theorem presented above can be
used by taking grids of points from the class $\avk$, and showing that
the points in each grid are distinguishable.  Then, the normalized
logarithm of the number of grid points gives a lower bound on the required
redundancy.  For the minimax redundancy, one such grid is sufficient using
the weak version of the theorem.  For the redundancy for most sources, we need
to show how we shift the grid to cover the whole class without violating the
conditions of the strong version of the redundancy-capacity theorem, where the
points in each shift of the grid remain distinguishable.  For standard compression
with fixed alphabet size $k$,
a uniform grid with spacing of $n^{-0.5 ( 1 -\varepsilon ) }$ for an arbitrarily
small $\varepsilon > 0$ is sufficient for distinguishability.  This yields the well
known bound, for which the cost of each unknown probability parameter is
$0.5 \log n$ bits.  Recently, we showed \cite{gil9}, \cite{gil91}
that in the case of large alphabets,
the simple grid used to achieve the fixed $k$ bound is not sufficient.  In
the minimax case, a non-uniform grid with increasing spacing
in each dimension was created, and resulted
in a cost of $0.5 \log (n/k)$ bits for each unknown probability parameter.
The same cost with smaller second order term resulted for most
sources using sphere packing
\cite{conway} considerations to create a grid (or lattice) of distinguishable
points.  (Note that this idea is in line with Rissanen's proof for a parametric source
with a finite number of parameters \cite{rissa}.)
The ideas that led to these bounds will be modified in
Sections~\ref{sec:maximin} and~\ref{sec:lb_most} for lower bounding the
minimax and most sources redundancies of patterns.

In \cite{feder}, Feder and Merhav showed that there exist classes that
consist of different subclasses, each with different redundancy within itself.
For example, a union of subclasses $\avk$ constitute the class $\avecspace$.
If all the subclasses are coded as one class, the redundancy adapts to the
worst one among the subclasses even if the actual source
is from a subclass within which smaller redundancy can be obtained.
However, in most simple cases, the cost of
distinguishing between subclasses is negligible w.r.t.\ the universal cost within
each subclass.  Hierarchical coding first distinguishes between the
different subclasses and then between sources within each subclass.  For example,
if the class $\avecspace$ is considered, the encoder will first code the
alphabet size $k$ and then perform universal coding within the subclass $\avk$.
Such an approach yields lower costs for coding sources in many subclasses than
the cost of coding the whole class.  Therefore, unlike the results
in \cite{orlitsky}, \cite{orlitsky1}-\cite{orlitsky4}, we will consider
the subclass $\Psi \left ( \avk \right )$ and analyze the pattern redundancy
for each $k$.  If $k$ is initially unknown, $\left ( 1 + \varepsilon \right )\log k$
bits can be used to relay to the decoder the number of
indices in the pattern using Elias's \cite{elias} coding of the integers.

One technique that will be used in Section~\ref{sec:upper_bounds} to design
a code for coding patterns will use ideas as in Rissanen's quantization two-part code
method \cite{rissa}.  This technique estimates the ML parameters from the sequence $X^n$
and then
quantizes them onto a grid of points.  Then, only the quantized version of the
ML parameters is
relayed to the decoder, and entropy coding is used w.r.t.\ this version as
if the quantized parameters are the true source parameters.  The redundancy
of this code consists of the cost of relaying the quantized ML estimators and
the cost caused by the quantization of the ML parameters.  The latter results
from the deviation of the quantized parameters from the actual parameters.
Usually, the quantization cost can be made negligible by tuning the grid
spacing properly.  Unlike Rissanen's approach, we will need to use
a non-uniform grid for the quantization, as in \cite{gil9}, \cite{gil91}, although,
unlike these references,
we will be concerned with index probabilities for patterns
and not the actual letter probabilities.

\section{The Main Results}
\label{sec:main_results}

The paper contains the following main results:
\bi
 \item a lower bound on the maximin and minimax redundancy for universal coding of patterns,
 \item a lower bound on the redundancy for most sources when coding patterns,
 \item an upper bound on the redundancy of coding patterns, specifically for not
 very large alphabets where all alphabet letters are likely to occur in a sequence,
 \item two sub-optimal sequential low-complexity methods for coding patterns with
 upper bounds on the displacements of their description lengths from the i.i.d.\
 ML description length and also with implications to the pattern entropy.
\ei
Each of the above results is studied in a separate subsequent section.

In particular, we show that
the $n$th-order maximin and minimax average universal coding redundancies for
patterns induced by i.i.d.\ sources
with alphabet size $k$ are lower bounded by
\be
  R^+_n \left [ \Psi \left ( \avk \right ) \right ]
%  = R^-_n \left [ \Psi \left ( \avk \right ) \right ]
  \geq
  \left \{ \begin{array}{ll}
    \frac{k-1}{2n}
    \log \frac{n^{1-\varepsilon}}{k^3} + \frac{k-1}{2n}
    \log \frac{\pi e^3}{2} -
    O \left ( \frac{\log k }{n} \right ), & \mbox{for } k \leq
    \left ( \frac{\pi n^{1-\varepsilon}}{2} \right )^{1/3} \\
    \left ( \frac{\pi}{2} \right )^{1/3} \cdot
    (1.5 \log e) \cdot n^{-(2+\varepsilon)/3} -
    O \left ( \frac{\log n}{n} \right ), &
    \mbox{for } k >
    \left ( \frac{\pi n^{1-\varepsilon}}{2} \right )^{1/3}
   \end{array} \right ..
\ee
The $n$th-order average universal coding redundancy is lower
bounded by
 \be
  R_n \left [ L, \avec \left ( \pvec \right ) \right ]\geq
   \left \{ \begin{array}{ll}
    \frac{k-1}{2n} \log \frac{n^{1-\varepsilon}}{k^3} -
    \frac{k-1}{2n} \log \frac{8 \pi}{e^3} -
    O \left ( \frac{\log k}{n} \right ), & \mbox{for } k \leq
    \frac{1}{2} \cdot \left ( \frac{n^{1-\varepsilon}}{\pi} \right )^{1/3} \\
    \frac{1.5 \log e}{2 \pi^{1/3}} \cdot n^{-(2+\varepsilon)/3} -
    O \left ( \frac{\log n}{n} \right ), &
    \mbox{for } k > \frac{1}{2} \cdot
    \left ( \frac{n^{1-\varepsilon}}{\pi} \right )^{1/3}
   \end{array} \right .
 \ee
for every code $L(\cdot)$ and
almost every i.i.d.\ source $\pvec \in \avk$.  Both lower bounds
demonstrate that for small $k$, each parameter costs at least
$0.5 \log \left ( n /k^3 \right )$ bits.  For larger alphabets, the cost
is at least $O \left (n^{(1-\varepsilon)/3}\right )$ bits overall.

Next, it is shown that there exist codes with length function
$L^* \left ( \cdot \right )$ that achieve redundancy
 \be
  R_n \left [ L^*, \avec \left ( \pvec \right ) \right ] \leq
  \left \{
   \begin{array}{ll}
    \left ( 1 + \varepsilon \right ) \frac{k-1}{2n} \log \frac{n^{1+\varepsilon}}{k^2}, &
    \mbox{for } k \leq \sqrt{n}^{1 - \varepsilon} ~\mbox{and}~
    \pvec \in \tilde{\avecspace}_k \\
    \frac{\pi \sqrt{2/3} \log e}{\sqrt{n}} + O\left (\frac{1}{n}\right ), &
%    \left ( 1 + \varepsilon \right ) \frac{2}{\sqrt{\ln 2}}
%    \frac{\sqrt{\log n}}{\sqrt{n}}, &
    \mbox{for } k \geq \sqrt{n}^{1 - \varepsilon}
    ~\mbox{or}~
    \pvec \not \in \tilde{\avecspace}_k
   \end{array}
  \right .
 \ee
for patterns induced by any i.i.d.\ source $\pvec \in \avk$.  Namely,
for small $k$, each parameter costs at most $0.5 \log \left ( n / k^2 \right )$ bits,
and for large $k$, $O \left (\sqrt{n} \right )$ overall.

Next, a linear (per sequence) complexity
sequential method (with prior knowledge of $k$)
is shown with modified individual redundancy that satisfies
\be
  \tilde{R}_n \left [ Q_k, \Psi \left ( x^n \right ) \right ] \leq
  \frac{k}{2n} \log \frac{n}{k^3} +
  \left ( \frac{19}{12} \log e \right ) \frac{k}{n} -
  \frac{1}{2n} \log n + O \left ( \frac{k^2}{n^2} \right ).
\ee
for every pattern
$\Psi \left (x^n \right )$ of a sequence $x^n$ with $k$ distinct
indices and for every $k \leq n$.  With increased complexity, identical performance
is also achieved without prior knowledge of $k$.
However, a second linear complexity scheme achieves similar asymptotic performance
in $k$, with only second order penalty without prior knowledge of $k$.
Finally, the implications of these bounds on the pattern entropy are noticed, in particular,
indicating that the pattern entropy must decrease from the i.i.d.\ one if $k$ is
larger than $c n^{1/3}$, for some constant $c$.

\section{A Maximin and Minimax Lower Bound}
\label{sec:maximin}

In \cite{gil9}, \cite{gil101}-\cite{gil91}, it was established that
for a large known alphabet of size $k$, choosing a set $\Agrid$ of $M$
sources $\pvec$ whose $k-1$ free components are placed only at
points on a non-uniform grid of increased spacing in each dimension
yields a set of distinguishable sources if the grid spacing is
properly chosen. The $k-1$ components of grid points take values
only from the \emph{grid\/} vector $\tgrid \dfn \left ( \tau_1,
\tau_2, \ldots, \tau_b, \ldots, \tau_B \right )$. The components of
$\tgrid$ satisfy $\tau_1 < \tau_2 < \cdots < \tau_b < \cdots <
\tau_B$, and the spacing between every two consecutive components
increases with $b$. The advantage of such a grid is that it yields a
tighter bound on the redundancy, as we can include more points in
regions of $\avk$ in which closer points are distinguishable, i.e.,
for small probability parameters. For coding patterns, we can build
a similar grid of sources.  However, we need to verify
distinguishability in the pattern domain, as explained in
Section~\ref{sec:background}. A valid grid point $\pvec \in \Agrid$
and a non-identity permutation $\pvec' = \pvec \left ( \sigvec
\right ) \neq \pvec$, $\pvec' \in \Agrid$,  of $\pvec$ will not be distinguishable in the
pattern domain, as they are likely to generate similar patterns.
Hence, in order to build a grid of sources which are distinguishable
in the pattern domain, we can take the grid $\Agrid$ for i.i.d.\ sources, but
keep only one point for each
set of permutations of the same source vector $\pvec \dfn \avec
\left ( \pvec \right )$ (which is ordered in non-decreasing order of
components).  We then consider a new grid $\Psi \left ( \Agrid \right )$
that contains only points $\pvec \dfn \avec \left ( \pvec \right )$ in which
the components are ordered in nondecreasing order.
In this section, we will show how such a grid can be
obtained, and then will use
the weak-version of the redundancy-capacity theorem to derive a
lower bound on the minimax redundancy using this grid. We start by stating the main
result that lower bounds the minimax pattern redundancy, and then
present its proof \footnote{The initial derivation of a related bound
to that of \eref{eq:minimax_pattern_bound} appears in \cite{gil101}, and was
done subsequently to the derivation of the
individual sequence
minimax lower bound in \cite{orlitsky4} (see, e.g., \cite{orlitsky1}).
The bound in \cite{gil101}
was later improved.  A problem with the second region
of both bounds (in \cite{gil101} and the improved one) was pointed
out by Ortlitsky and Santhanam in October 2003.  Consequently, the
improved bound and its proof were corrected resulting in the
second region of \eref{eq:minimax_pattern_bound}.}.

\begin{theorem}
\label{theorem_maximin_patterns}
 Fix an arbitrarily small $\varepsilon > 0$, and let $n \rightarrow \infty$.  Then,
 the $n$th-order maximin and minimax average universal coding redundancies for
 patterns induced by i.i.d.\ sources
 with alphabet size $k$ are lower bounded by
\be
  \label{eq:minimax_pattern_bound}
  R^+_n \left [ \Psi \left ( \avk \right ) \right ]
%  = R^-_n \left [ \Psi \left ( \avk \right ) \right ]
  \geq
  \left \{ \begin{array}{ll}
    \frac{k-1}{2n}
    \log \frac{n^{1-\varepsilon}}{k^3} + \frac{k-1}{2n}
    \log \frac{\pi e^3}{2} -
    O \left ( \frac{\log k }{n} \right ), & \mbox{for } k \leq
    \left ( \frac{\pi n^{1-\varepsilon}}{2} \right )^{1/3} \\
    \left ( \frac{\pi}{2} \right )^{1/3} \cdot
    (1.5 \log e) \cdot n^{-(2+\varepsilon)/3} -
    O \left ( \frac{\log n}{n} \right ), &
    \mbox{for } k >
    \left ( \frac{\pi n^{1-\varepsilon}}{2} \right )^{1/3}
   \end{array} \right ..
\ee
\end{theorem}

Theorem~\ref{theorem_maximin_patterns} shows that as long as $k$ is small
(of $o \left (n^{1/3} \right )$), each index probability parameter
costs at least $0.5 \log \left ( n / k^3 \right )$ extra code bits.  However, if
the alphabet size is larger, a threshold phenomenon occurs, and the redundancy
is of $O \left ( n^{-2/3} \right )$ overall.  Note that this result applies
even if $k > n$, because regardless of the actual alphabet size, the number of indices
that will occur in a pattern is upper bounded by $n$.
The bound in the first region coincides with the \emph{individual\/} minimax bound
obtained from \cite{aberg}, described in \eref{eq:aberg_bound1}.  The
second region points to the same behavior as
the worst case bound in \eref{eq:orlitsky_bounds}.
The average lower bound naturally applies to the individual minimax worst case
redundancy, but not
the other way around.  Theorem~\ref{theorem_maximin_patterns} shows that
we are unlikely to
gain much in the average case over the worst sequence at least for the minimax
redundancy.  The $n^{-\varepsilon/3}$ gap may
indicate a true small gap between the individual worst case and the average worst
case, but may also be due to sub-optimal bounding.

The proof of Theorem~\ref{theorem_maximin_patterns} builds a non-uniform grid $\Agrid$
of points as in the i.i.d.\ minimax case.  Then, the grid size is reduced by a factor
of $k!$ eliminating all permutations of any grid point $\pvec$ except the
ordered permutation $\avec \left ( \pvec \right )$, resulting in a new grid
$\Psi \left ( \Agrid \right )$ in the induced patterns space.  This elimination is a worst
case one, since sources for which there are identical components
$\theta_i = \theta_j$ for $j \neq i$ have less than $k!$ permutations in the
original i.i.d.\ grid.  The elimination of more grid points than necessary
becomes significant for $k = O \left ( n^{1/3} \right )$ or larger.  For
alphabets of these sizes, most distinguishable grid points in the i.i.d.\
standard compression grid contain identical components.  Therefore, we reduce
the bound on the grid size by a factor that is too large.  This results in
a useless bound that is smaller than $1$ on the number of grid points $M$
in the pattern grid, and requires
adaptation of the largest bound on $M$ as
a function of $k$ to all large $k$'s.

A second issue that needs to be addressed in order to use the weak version of
the redundancy-capacity theorem is that of distinguishability of the grid points in the
pattern domain, as described in Section~\ref{sec:background}.
Although the grid we will use is a subset of the distinguishable i.i.d.\ grid,
we need to have distinguishability in the pattern domain, i.e.,
if point $\pvec$ generated the sequence $X^n$, the pattern
$\Psi \left ( X^n \right )$ needs to appear as if it were generated by
$\avec \left ( \pvec \right )$.  If we observe the sequence $X^n$
and obtain the ML estimator $\pvece$ of $\pvec$ in the i.i.d.\ domain,
we may have sequences for which
$\hat{\theta}_i > \hat{\theta}_j$ for $j > i$.  For such sequences, $\pvece_{\Omega}$
may still be equal $\pvec$
in the i.i.d.\ domain.  In the pattern domain, however,
if this happens, by observing $\Psi \left (X^n \right )$,
$\hat{\theta}_i$ will appear to be the estimate of $\theta_j$ and
$\hat{\theta}_j$ of $\theta_i$.
We thus need to show, that despite that, distinguishability is
still maintained, and thus by the restriction that
$\pvece^{\psi}_{\Omega} \in \Psi \left ( \Agrid \right )$, we will still
have $\pvece^{\psi}_{\Omega} = \avec \left ( \pvec \right )$ for all
cases in which $\pvece_{\Omega} = \pvec$.
This will be done as the last step of the proof of the theorem.
The proof of Theorem~\ref{theorem_maximin_patterns}
follows and concludes this section.

\noindent
{\bf Proof of Theorem~\ref{theorem_maximin_patterns}:}
Let $\avk$ be the class of i.i.d.\ sources with an alphabet of size $k$,
and let $\Psi \left ( \avk \right )$ denote its induced pattern
class. First, let us consider a non-uniform grid $\Agrid$ of points
in $\avk$. Also, at this point, let us assume that $k \leq n^{1 - 2\varepsilon}$.
This assumption will be justified later on, and then it will be shown how we can
still obtain a bound for the redundancy over $\Psi \left ( \avk \right )$ for larger
values of $k$.
Let $\tgrid$ be a vector of grid components, such that the first
$k-1$ components $\theta_i,~i =1,\ldots, k-1$,
of $\pvec \in \Agrid$ must satisfy $\theta_i \in \tgrid$.
Let $\tau_b$ be the $b$th point in $\tgrid$, and define it as
\be
 \label{eq:grid_point}
 \tau_b \dfn
 \sum_{j=1}^{b} \frac{2 (j - \frac{1}{2})}{n^{1-\varepsilon}} =
 \frac{b^2}{n^{1-\varepsilon}}.
\ee
Then, for the $b$th point in $\tgrid$,
\be
 \label{eq:grid_point2}
 b = \sqrt{\tau_b} \cdot \sqrt{n}^{1-\varepsilon},
\ee
and also, the spacing $\Delta \left ( \tau_b \right )$ between points $\tau_b$ and
$\tau_{b-1}$ satisfies
\be
 \label{eq:grid_spacing}
 \Delta \left ( \tau_b  \right ) \dfn \tau_b - \tau_{b-1}
  = \frac{2\left ( b - \frac{1}{2} \right )}{n^{1-\varepsilon}}
  = \frac{2 \left ( \sqrt{\tau_b} \sqrt{n}^{1-\varepsilon} - 0.5 \right )}
  {n^{1-\varepsilon}} \geq
  \frac{\sqrt{\tau_b}}{\sqrt{n}^{1-\varepsilon}},
\ee
where the last inequality is obtained because $\tau_b \geq n^{-(1 -\varepsilon)}$.
From \eref{eq:grid_spacing}, we see that for large $b$ and $\tau_b$
the spacing between grid points is the same spacing used to obtain the well known
bounds for compression of i.i.d.\ fixed size alphabet sources.  However,
for small probability parameters, we obtain a denser grid.
Figure~\ref{fig:large_k_grid} demonstrates this non-uniform grid.
\bef
  \centerline{\includegraphics[bbllx=0pt,bblly=365pt,bburx=500pt,
    bbury=780pt,height=6cm,
    clip=]{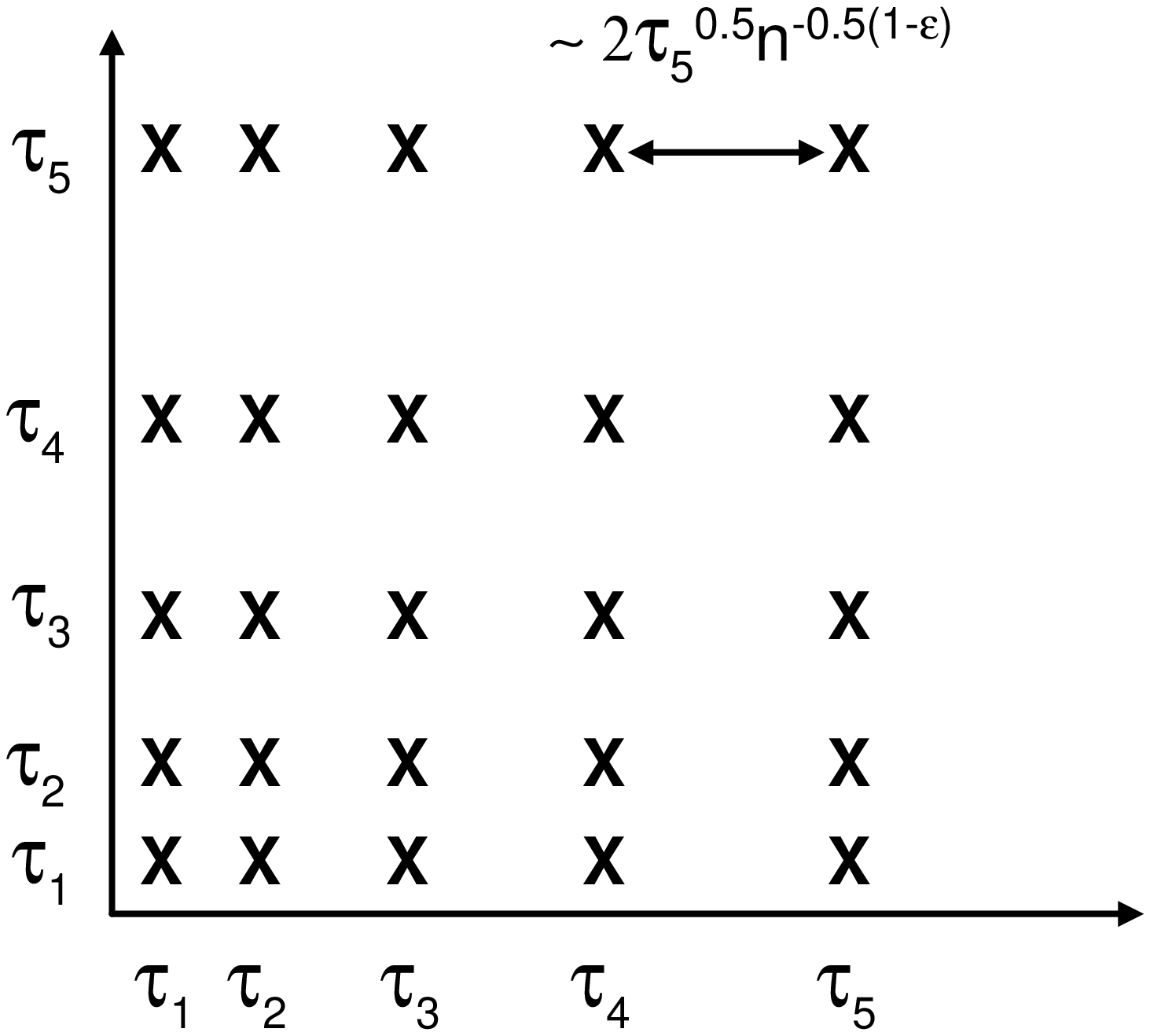}}
 \caption{Non-uniform grid for a large alphabet}
 \label{fig:large_k_grid}
\enf

Let us first lower bound the number of points in the standard i.i.d.\ grid.
Let $\pvec = \left ( \theta_1,
\theta_2, \ldots, \theta_{k-1} \right )$ be a point on the grid $\Agrid$.
Let $b_i$ be the index of $\theta_i$ in $\tgrid$, i.e.,
$\theta_i = \tau_{b_i}$.
Then, from \eref{eq:grid_point}-\eref{eq:grid_point2},
\be
 \label{eq:ball_condition1}
 \sum_{i=1}^{k-1} \theta_i =
 \sum_{i=1}^{k-1} \tau_{b_i} =
 \sum_{i=1}^{k-1} \frac{b_i^2}{n^{1-\varepsilon}}.
\ee
Hence, there is a one-to-one mapping between
a grid point $\pvec$ and the index vector
$\bvec \dfn \left (b_1, b_2, \ldots, b_{k-1} \right )$ of positive
integers.
Since the components of $\pvec$ are probabilities, we must have
\be
 \label{eq:prob_cond}
 \sum_{i=1}^{k-1} \theta_i \leq 1.
\ee
From \eref{eq:ball_condition1} and \eref{eq:prob_cond}, it follows that if
\be
 \label{eq:ball_condition}
  \sum_{i=1}^{k-1} b_i^2 \leq n^{1 - \varepsilon},
\ee
$\pvec$ must be a valid grid point.  Hence, the total number of grid points
is the number of nonnegative integer components
vectors $\bvec$ satisfying \eref{eq:ball_condition}.
As shown in the next lemma, this number is lower bounded by the volume of
a $k-1$ dimensional sphere with radius $\sqrt{n}^{1-\varepsilon'}$,
$V_{k-1} \left ( \sqrt{n}^{1-\varepsilon'} \right )$ (see \cite{conway} for
this volume), where $\varepsilon' > \varepsilon$ and $\varepsilon'-\varepsilon$ is fixed,
divided by $2^{k-1}$ for obtaining only positive components.
Note that due to the integer length constraints on
the components of $\bvec$ we must use the greater $\varepsilon'$, and
we obtain a lower bound (i.e., we consider the volume of a smaller sphere
in order not to include integer vectors that are not in the sphere).
\begin{lemma}
\label{lemma_integer_vectors}
For the standard i.i.d.\ case with $k \leq n^{1-2\varepsilon}$,
the number of grid points satisfying
\eref{eq:prob_cond} is lower bounded by
\be
 \label{eq:M_ball_bound}
  M_{\mbox{i.i.d.}}
  \geq \frac{V_{k-1} \left ( \sqrt{n}^{1-\varepsilon'} \right )}{2^{k-1}}
  =
  \frac{1}{2^{k-1}} \cdot
  \left \{
  \begin{array}{ll}
   \frac{\pi^{(k-1)/2} \cdot n^{(1-\varepsilon')(k-1)/2}}{[(k-1)/2]!}; &
   k ~\mbox{odd}, \\
   \frac{[(k-2)/2]! \cdot \pi^{(k-2)/2} \cdot 2^{k-1} \cdot
   n^{(1-\varepsilon')(k-1)/2}}{(k-1)!}; & k ~\mbox{even}.
  \end{array}
  \right .
\ee
\end{lemma}
The proof of Lemma~\ref{lemma_integer_vectors} is in
\ref{ap:lemma_integer_vectors_proof}.
Taking the logarithm of the bound in \eref{eq:M_ball_bound},
and approximating factorials by Stirling's approximation
\be
 \label{eq:stirling}
 \sqrt{2 \pi m} \cdot \left ( \frac{m}{e} \right )^m \leq m! \leq
 \sqrt{2 \pi m} \cdot \left ( \frac{m}{e} \right )^m \cdot
 \exp \left \{ \frac{1}{12m}\right \},
\ee
we obtain
\be
 \label{eq:log_M_iid}
 \log M_{\mbox{i.i.d.}} \geq
   \frac{k-1}{2} \log \frac{n^{1-\varepsilon'}}{k} +
   \frac{k-1}{2} \log \frac{\pi e}{2} -
   \frac{1}{2} \log k - O(1).
\ee

Now, let us consider only a portion $\Psi \left ( \Agrid \right )$
of the grid $\Agrid$ for the grid of distinguishable patterns.  The
grid $\Psi \left ( \Agrid \right )$ includes all points $\pvec \in \Agrid$
for which $\avec \left ( \pvec \right ) = \pvec$, i.e., only the permutation
of any point $\pvec' \in \Agrid$ for which the components are in non-decreasing
order is included in $\Psi \left ( \Agrid \right )$.  Note that this condition
applies to the $k$ dimensions of $\pvec$ including the additional $k$th parameter
$\theta_k$.  (For this matter, if $\theta_k$ does not take a point from
$\tgrid$, the nearest neighboring points from $\tgrid$ will be considered
as its grid point value.)
The transformation from the space $\avk$ to the space
$\Psi \left ( \avk \right )$, that contains all the points
in $\Psi \left ( \Agrid \right )$, is shown in
Figure~\ref{fig:pattern_grid} for $k = 2$ and $k=3$.  In the second
case, only a projection of two components on a two dimensional space
is shown.
\bef
 \centerline{\includegraphics[bbllx=40pt,bblly=505pt,bburx=560pt,
  bbury=770pt,height=5cm,
    clip=]{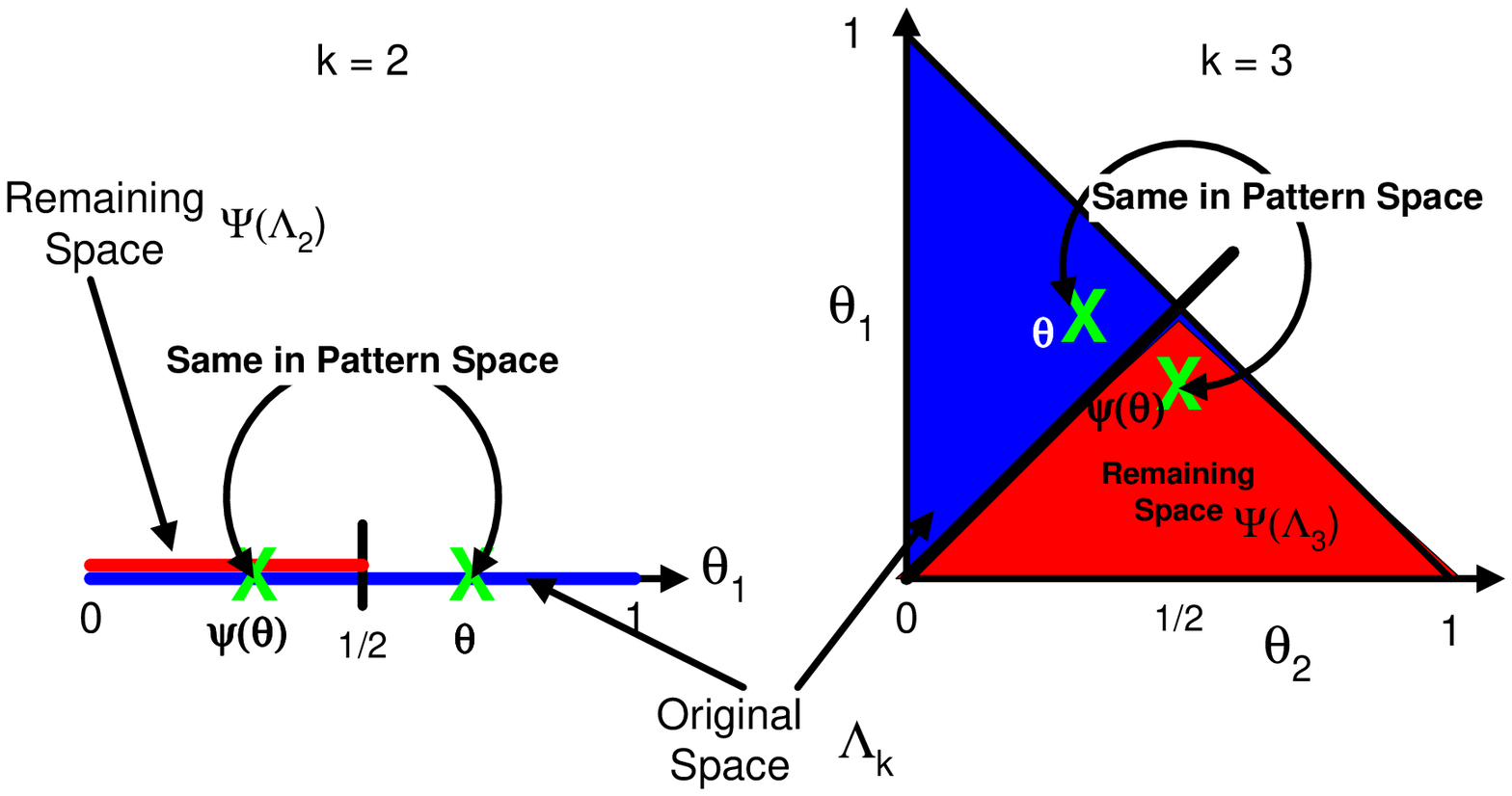}}
 \caption{Transformation from i.i.d.\ space $\avk$ to pattern space
 $\Psi \left (\avk \right )$ for $k = 2$ and $k = 3$}
 \label{fig:pattern_grid}
\enf

In order to lower bound the size $M$ of the grid $\Psi \left ( \Agrid \right )$,
we need to take out from $\Agrid$
any point $\pvec \in \Agrid$ that is a non-identity permutation of
$\avec \left ( \pvec \right )$.  For each point in $\Psi \left ( \Agrid \right )$,
there are at most $k!$ such permutations (although there may be less).  Therefore,
we can lower bound the logarithm of $M$, using Stirling's approximation
and \eref{eq:log_M_iid}, by
\bea
 \nonumber
 \log M &\geq&
 \log M_{\mbox{i.i.d.}} - \log \left ( k! \right ) \\
 \label{eq:pattern_lb_log_M}
 &\geq&
 \frac{k-1}{2} \log \frac{n^{1-\varepsilon'}}{k^3} +
 \frac{k-1}{2} \log \frac{\pi e^3}{2} - 2 \log k  - O(1).
\eea
From \eref{eq:pattern_lb_log_M}, we note that there exists a constant $c$
such that if $k > c n^{(1-\varepsilon')/3}$ the bound above
becomes negative.  The reason is that we eliminated many points from the grid
more than once.  For example, the grid point
$\pvec = \left [\tau_1, \tau_1, \cdots, \tau_1 \right ]$
only appears once in the grid $\Agrid$ but was reduced by a factor of $k!$ times
to obtain the bound in \eref{eq:pattern_lb_log_M}.  This problem is negligible
for small $k$'s, because such grid points make a negligible fraction of
$\Agrid$.  However, for large $k$'s, almost all or all (for very large $k$'s) grid
points contain many components that are identical.

To achieve a more useful bound on the logarithm of the number of grid points for large
alphabets, we can find the value of $k$ for which the maximal lower bound
is obtained from \eref{eq:pattern_lb_log_M}.  Denote it by $k_m$. Then,
for $k > k_m$ (including $k>n$), we can fix the first
$k - k_m$ components of $\pvec$ at a value of
$o \left [ 1 / (n^{1+\varepsilon}(k - k_m)) \right ]$ for all points in
$\pvec \in \Psi \left ( \Agrid \right )$, where the $i$th component,
$i \leq k - k_m$,
of all $\pvec \in \Psi \left ( \Agrid \right )$ takes the same value,
and any of these letters will appear in $x^n$ with probability going to $0$.
The other $k_m$ components
will take the values from a pattern grid for alphabet of size
$k_m$.  Note that now we can justify the assumption that $k_m \leq n^{1-2\varepsilon}$,
assumed earlier for computing the number of grid points.  In fact, $k_m$ is much
smaller as indicated earlier.  However, by fixing all other
components of $\pvec$ as described above, the bound for $k_m$ applies even
for alphabets with $k > n^{1-2\varepsilon}$.
If we now show that all points are distinguishable in the
grid for $k = k_m$, they will also be distinguishable in the grid
defined above for larger $k$.  Therefore, the bound for $k_m$ will
hold for every larger $k$ as well.

The bound in \eref{eq:pattern_lb_log_M}
attains a maximum value for $k_m = \left ( \pi / 2 \right )^{1/3}\cdot n^{(1-\varepsilon')/3}
\approx 1.16 n^{(1-\varepsilon')/3}$.  Substituting $k_m$ in \eref{eq:pattern_lb_log_M},
normalizing by $n$, and replacing $\varepsilon'$ by $\varepsilon$,
we obtain the second region of the bound
in \eref{eq:minimax_pattern_bound}.  The first region of the bound is
obtained by normalizing the bound in \eref{eq:pattern_lb_log_M} by $n$ and substituting
$\varepsilon'$ by $\varepsilon$.
To conclude the proof
of Theorem~\ref{theorem_maximin_patterns}, we only need to
prove distinguishability
in the non-uniform pattern grid.  By the weak version of the redundancy-capacity
theorem, if distinguishability is proved, then the bounds we have obtained
lower bound the minimax redundancy.

We will now show that distinguishability in the grid $\Psi \left ( \Agrid \right )$
is a direct result of the distinguishability in the grid
$\Agrid$.
Let the sequence $X^n$ be generated by
the point $\pvec = \avec \left ( \pvec \right ) \in \Psi \left ( \Agrid \right )$.
Let $\Psi \left ( X^n \right )$ be the pattern of $X^n$.  Consider the estimator
$\avec \left (\pvece \right )$ of $\pvec$ obtained as a function of $\Psi \left (X^n \right )$,
and let $\pvece^{\psi}_{\Omega}$ be the nearest point to $\avec \left (\pvece \right )$
on the grid $\Psi \left ( \Agrid \right )$.  We will show that there is an estimator
$\avec \left (\pvece \right )$ for which
$P_{\theta} \left ( \pvece^{\psi}_{\Omega} \neq \pvec \right ) \rightarrow 0$ as
$n \rightarrow \infty$.
By definition of $\Psi \left ( \Agrid \right )$,
the components $\theta_i; 1 \leq i \leq k,$ of $\pvec$ are in non-decreasing order.
Let $\pvece$ be the ML estimator of
$\pvec$ from $X^n$, and $\pvece_{\Omega}$ the closest point in $\Agrid$
to $\pvece$ that is used to estimate $\pvec$ in the standard i.i.d.\ case.
Let $\avec \left ( \pvece \right )$ be the ordered
permutation of $\pvece$, that can be obtained directly from the pattern
$\Psi \left ( X^n \right )$.  For every $\theta_i$, $i=1,\ldots, k$, let $\tau_{b_i}$
be the nearest point in $\tgrid$ to $\theta_i$ that is smaller than or
equal to $\theta_i$.  (Note that for $i < k$, $\tau_{b_i} = \theta_i$, and only
for $\theta_k$ it may be smaller than $\theta_k$.)  Define the event $A_i$
as
\be
\label{eq:Aidef}
 A_i ~:~ \left | \hat{\theta}_i - \theta_i \right | \geq
 \frac{\Delta \left ( \tau_{b_i} \right )}{2},
\ee
i.e., the event in which the ML estimate of
component $\theta_i$ is outside
an interval of length $\Delta \left ( \tau_{b_i} \right )$ centered at
$\theta_i$.  (If an error occurs in estimating $\theta_i$ by $\hat{\theta}_{\Omega i}$,
this must be true because $\Delta \left ( \tau_{b_i} \right )/2$ is at most
half the distance between $\theta_i$ and its nearest neighbors.)
Let event $\Psi \left ( A_i \right )$ be defined as
\be
\label{eq:PsiAidef}
 \Psi \left ( A_i \right ) ~:~ \left | \avec_i \left ( \pvece \right ) -
 \theta_i \right | \geq
 \frac{\Delta \left ( \tau_{b_i} \right )}{2},
\ee
where $\avec_i \left (\pvece  \right )$ denotes the $i$th ordered component of
the ML estimate of $\pvec$, where the components are ordered in non-decreasing order.
Define event $A \dfn \bigcup_i A_i$ as the union of all events $A_i$ and event
$\Psi \left ( A \right ) \dfn \bigcup_i \Psi \left (A_i \right )$
as the union of all events $\Psi \left ( A_i \right )$.
The probability that event $A$ occurs when $X^n$ is generated by $\pvec$ will
be denoted by $P_{\theta} \left ( A \right )$.  In a similar manner,
$P_{\theta} \left [ \Psi \left ( A \right ) \right ]$ will denote the probability
that $\Psi \left (A \right )$ occurs given $X^n$ is generated by $\pvec$.
By definition of $\Psi \left ( A \right )$ and \eref{eq:PsiAidef}, event
$\Psi \left ( A \right )$ implies that the ordered version $\avec \left ( \pvece \right )$
of the ML estimator $\pvece$ is outside the portion in $\Psi \left ( \avk \right )$
of the box with
edges $\Delta \left ( \tau_{b_i} \right )$, for every $i$, centered at
$\pvec$.  The following lemma, which is proved in
\ref{ap:lemma_event_A_prob_proof},
bounds $P_{\theta} \left ( A \right )$:
\begin{lemma}
\label{lemma_event_A_prob}
\be
 \label{eq:event_A_prob}
 P_{\theta} \left ( A \right ) \leq
 2^{(\log k) + (\log n) - cn^{\varepsilon/2}}
 \rightarrow 0,
\ee
where $c$ is a constant.
\end{lemma}
Note that $P_{\theta} \left ( A \right ) \geq P_{\theta} \left (
\pvece_{\Omega} \neq \pvec \right )$, but we require a bound on the larger
probability in order to apply it to the pattern space.
In the standard i.i.d.\ case, there is thus a
vanishing probability even to estimating $\pvec$ outside the defined box.
The following lemma relates between the probability of event $\Psi
\left ( A \right )$ and that of event $A$.

\begin{lemma}
\label{lemma_patterns_iid}
\be
 \label{eq:error_prob_patterns}
 P_{\theta} \left [ \Psi \left ( A \right ) \right ] \leq
 P_{\theta} \left ( A \right ).
\ee
\end{lemma}
{\bf Proof:}
We show that
$\bar{A} \rightarrow \left ( \overline{\Psi \left ( A \right )} \cap B \right )$, where
$\bar{A}$ is the complement to
$A$, and event $B$ is defined below.
Therefore,
$\left [\Psi \left ( A \right ) \cup \bar{B} \right ] =
\left \{ \left [ \Psi \left ( A \right ) \cap B \right ] \cup \bar{B} \right \} \rightarrow A$,
and thus also $\Psi \left ( A \right ) \rightarrow A$ and
$P_{\theta} \left \{ \Psi \left ( A \right ) \right \} \leq
P_{\theta} \left ( A \right )$.
The proof consists of the following steps:  First, let
$\vphivec \subseteq \left ( \theta_1, \theta_2, \ldots, \theta_{k-1}, \tau_{b_k} \right )$
be a subset of $\pvec$ with $\theta_k$ replaced by the nearest smaller grid point.
Let $\vphivec$ consist only of distinct (unequal) elements.
Let the components of $\vphivec$ be ordered in increasing order.
We show that $\bar{A}$ implies that $\hat{\vphivec}$ is also in increasing order
for any choice of $\vphivec$ as described above.  The latter event is denoted by $B$.
Hence, the respective ordered
components of $\avec \left (\hat{\vphivec} \right )$
will not be permuted from those of $\hat{\vphivec}$.
This means that if $\avec \left (\hat{\vphivec} \right ) \neq \hat{\vphivec}$ (i.e.,
$\avec \left (\hat{\vphivec} \right )$ is a non-identity permutation of $\hat{\vphivec}$ and
event $\bar{B}$ occurs),
$A$ must occur.
Then, we show that for equal components of $\pvec$,
although the components of $\pvece$ may not be ordered, if $\bar{A}$ is satisfied,
then each of the ordered components of $\avec \left (\pvece \right )$ must
satisfy $\overline{\Psi \left (A_i \right )}$.  Together with the first step, this
means that given $\bar{A}$, at least $k-2$ components of $\avec \left ( \pvece \right )$
must satisfy event $\overline{\Psi \left (A_i \right )}$.
The only remaining components of $\avec \left ( \pvece \right )$ consist of at most
$\avec_k \left ( \pvece \right )$ and one more component $\avec_l \left (\pvece \right)$
which takes the value of $\hat{\theta}_k$ if $\hat{\theta}_k$ is not the maximal ML
component of $\pvece$.  (Otherwise, the proof is complete.)
For these two components, we show that
$\left \{ \left [ \Psi \left ( A_l \right ) \cup \Psi \left (A_k \right ) \right ]
\cap B \right \} \rightarrow A$, concluding the proof.

First, assume that $\bar{A}$ occurs.  Then, for all $i; 1 \leq i \leq k$,
\be
\label{eq:Aibardef}
 \left | \hat{\theta}_i - \theta_i \right | <
 \frac{\Delta \left (\tau_{b_i} \right )}{2} ~\Rightarrow~
 -\frac{\Delta \left (\tau_{b_i} \right )}{2} <
 \hat{\theta}_i - \theta_i <
 \frac{\Delta \left (\tau_{b_i} \right )}{2}.
\ee
Let $\tau_{b_j} > \tau_{b_i}$.
Note that by definition of $\pvec$
as an ordered vector and of $\tau_{b}$,
$\tau_{b_j} > \tau_{b_i}$ implies that $\theta_j > \theta_i$ (and also that $j>i$).
(The other direction is true for $j < k$.)
Given $\bar{A}$, we thus have,
\be
 \hat{\theta}_j - \hat{\theta}_i =
 \left (\hat{\theta}_j - \theta_j \right ) +
 \left ( \theta_j - \theta_i \right )+
 \left (\theta_i - \hat{\theta}_i \right ) >
 -\frac{\Delta \left ( \tau_{b_j} \right ) }{2} +
 \Delta \left (\tau_{b_j} \right )
 -\frac{\Delta \left ( \tau_{b_i} \right ) }{2} > 0,
\ee
where the first inequality is obtained by applying the left hand side of inequality
\eref{eq:Aibardef} to the first two and the last two terms, respectively, and
by applying the left hand side of
\eref{eq:grid_spacing} to the two middle terms.  The last inequality
is from the monotonicity of $\Delta \left ( \tau_{b} \right )$ in $b$.
Hence, if $\tau_{b_j} > \tau_{b_i}$, then $\bar{A}$ implies that
we must also have $\hat{\theta}_j > \hat{\theta}_i$ (and event $B$ must occur).
This means
that if the ML estimates of two letters separated by at least one grid spacing unit are
within the boxes defined in \eref{eq:Aibardef}, then these ML estimates are still ordered
in the same order as the original letters.  Hence, the only case where ML estimates of two
different letters may not be in the original order of the letters is when
$\tau_{b_j} = \tau_{b_i}$ for $j > i$.  For $j < k$, this implies also that
$\theta_j = \theta_i$, and thus if $\avec_i \left (\pvece \right ) = \hat{\theta}_j$ but also
\eref{eq:Aibardef} holds for $\theta_j$, then,
\be
 \left | \avec_i \left ( \pvece \right ) - \theta_i \right | =
 \left | \hat{\theta}_j - \theta_i \right | =
 \left | \hat{\theta}_j - \theta_j \right | < \frac{\Delta\left ( \tau_{b_j} \right )}{2}
 = \frac{\Delta\left ( \tau_{b_i} \right )}{2}.
\ee
Therefore, for all $i\leq k$, except for at most $i=k$ and one value $i=l<k$, for which
$\tau_{b_l} = \tau_{b_k}$, if $\bar{A}$ occurs,
also $\overline{\Psi \left ( A_i \right )}$ occurs.  This is because except for permutations
with $\hat{\theta}_k$, the only permutations violating
the order of $\pvec$ in the resulting $\pvece$ can occur between letters with equal probabilities
in $\pvec$.  From the last inequality, such permutations still result in occurrence of
$\overline{\Psi \left ( A_i \right )}$.

The only case in which $\theta_j > \theta_i$ does not necessarily imply
$\hat{\theta}_j > \hat{\theta}_i$ is when $j = k$ and $\tau_{b_i} = \tau_{b_k}$.
Let us now consider this case when $\hat{\theta}_k$ is not the maximal component
of $\pvece$.  (If $\hat{\theta}_k$ is the maximal component of $\pvece$,
the order of the estimates in $\pvece$ is not violated beyond permutations of
equal components in $\pvec$, and
we are back in the previous cases, for which the lemma
has already been proved.)
Let $\hat{\theta}_i$ be the maximum component of $\pvece$.
Then, $\avec_k \left ( \pvece \right ) = \hat{\theta}_i$.  Also,
there exists $l$, for which
$\tau_{b_l} = \tau_{b_k}$, such that $\avec_l \left ( \pvece \right ) = \hat{\theta}_k$.
We show that if either $\Psi \left (A_l \right )$
or $\Psi \left (A_k \right )$ occur together with $B$,
then either $A_i$ or $A_k$ must occur as well.
%Hence, if neither $A_i$ nor $A_k$ occur, then
%neither $\Psi \left (A_l \right )$ nor $\Psi \left (A_k \right )$ can occur.

First, let $\Psi \left ( A_k \right )$ occur, i.e.,
\be
 \left | \hat{\theta}_i - \theta_k \right | \geq \frac{\Delta\left ( \tau_{b_k} \right )}{2}.
\ee
If $\hat{\theta}_i > \theta_k$,
\be
 \hat{\theta}_i - \theta_i = \left ( \hat{\theta}_i - \theta_k \right )+
 \left (\theta_k - \theta_i \right ) \geq
 \frac{\Delta\left ( \tau_{b_k} \right )}{2},
\ee
where the inequality is by definition of this case and by the ordering of $\pvec$.
The last inequality means that $A_i$ occurs.
If $\hat{\theta}_i < \theta_k$,
\be
 \theta_k - \hat{\theta}_k = \left (\theta_k - \hat{\theta}_i \right ) +
 \left ( \hat{\theta}_i - \hat{\theta}_k \right ) \geq
 \frac{\Delta\left ( \tau_{b_k} \right )}{2},
\ee
where the inequality is, again, by definition of the case, and by the assumption that
$\hat{\theta}_i$ is the maximum component of the ML estimate of $\pvec$.  This inequality
implies that $A_k$ occurs.
Now, let $\Psi \left (A_l \right )$ occur for $l$ defined above.  Then,
\be
 \left | \avec_l \left ( \pvece \right ) - \theta_l \right | =
 \left | \hat{\theta}_k - \theta_l \right | \geq \frac{\Delta\left ( \tau_{b_l} \right )}{2}
 = \frac{\Delta\left ( \tau_{b_k} \right )}{2}
 = \frac{\Delta\left ( \tau_{b_i} \right )}{2},
\ee
where the equalities are since the occurrence of
$B$ implies $\tau_{b_l} = \tau_{b_k} = \tau_{b_i}$.
If $\hat{\theta}_k < \theta_l$,
\be
 \theta_k - \hat{\theta}_k = \left ( \theta_k - \theta_l \right )+
 \left (\theta_l - \hat{\theta}_k \right ) \geq
 \frac{\Delta\left ( \tau_{b_l} \right )}{2}
 = \frac{\Delta\left ( \tau_{b_k} \right )}{2},
\ee
where the inequality is obtained similarly to the previous cases.  The last
equality is from the occurrence of $B$.  This implies
that $A_k$ occurs.
If $\hat{\theta}_k > \theta_l$,
in a similar manner,
\be
 \hat{\theta}_i - \theta_i = \left ( \hat{\theta}_i - \hat{\theta}_k \right ) +
 \left ( \hat{\theta}_k - \theta_i \right ) = \left ( \hat{\theta}_i - \hat{\theta}_k \right ) +
 \left ( \hat{\theta}_k - \theta_l \right )
 \geq \frac{\Delta\left ( \tau_{b_l} \right )}{2} =
 \frac{\Delta\left ( \tau_{b_i} \right )}{2},
\ee
where the second and last equalities are because of the occurrence of $B$.  This
implies $A_i$ occurs, and concludes the proof of Lemma~\ref{lemma_patterns_iid}.
{\hfill $\Box$}

The proof of Lemma~\ref{lemma_patterns_iid} considered three different cases
relating between two components $\theta_i$ and $\theta_j$; $j > i$, of $\pvec$.
Figure~\ref{fig:grid_regions} shows the projection of these three cases
onto a two dimensional subspace that contains only components $i$ and $j$.
The dots represent grid points.
A rectangular box surrounding a dot contains all the ML estimator points that are
in event $\bar{A}_i \cap \bar{A}_j$ if $(\theta_i, \theta_j)$ is on the dot.
The first two cases are in part $(a)$ of the figure, and the last in part $(b)$.
In the first case, the complete box
is contained in $\Psi \left (\avk \right )$.
This is the case in which $\theta_j > \theta_i; j < k$.  The occurrence of $\bar{A}$
implies event $B$, which means that the ML estimates in this case will remain
in the original ordering, i.e., estimating the components of $\avec \left ( \pvece \right )$
out of $\Psi \left (X^n \right )$ will give the same estimates as those obtained by
estimating $\pvece$ out of $X^n$.
Note that
$P_{\theta} \left \{ \Psi \left ( A_i \right ) \cup \Psi \left (A_j \right ) \right \} \leq
P_{\theta} \left (A_i \cup A_j \right )$, where a possible decrease is
because some un-typical sequences that have ML estimates
$\pvece \not \in \Psi \left ( \avk \right )$ will be projected into the
same box around $\pvec$ by estimating out of $\Psi \left (X^n \right )$
and will (insignificantly) increase the probability
of $\overline{\Psi \left (A_i \right ) \cup \Psi \left (A_j \right )}$ from that of
$\overline{A_i \cup A_j}$.
\bef
 \centerline{\includegraphics[bbllx=30pt,bblly=375pt,bburx=550pt,
  bbury=770pt,height=8cm,
    clip=]{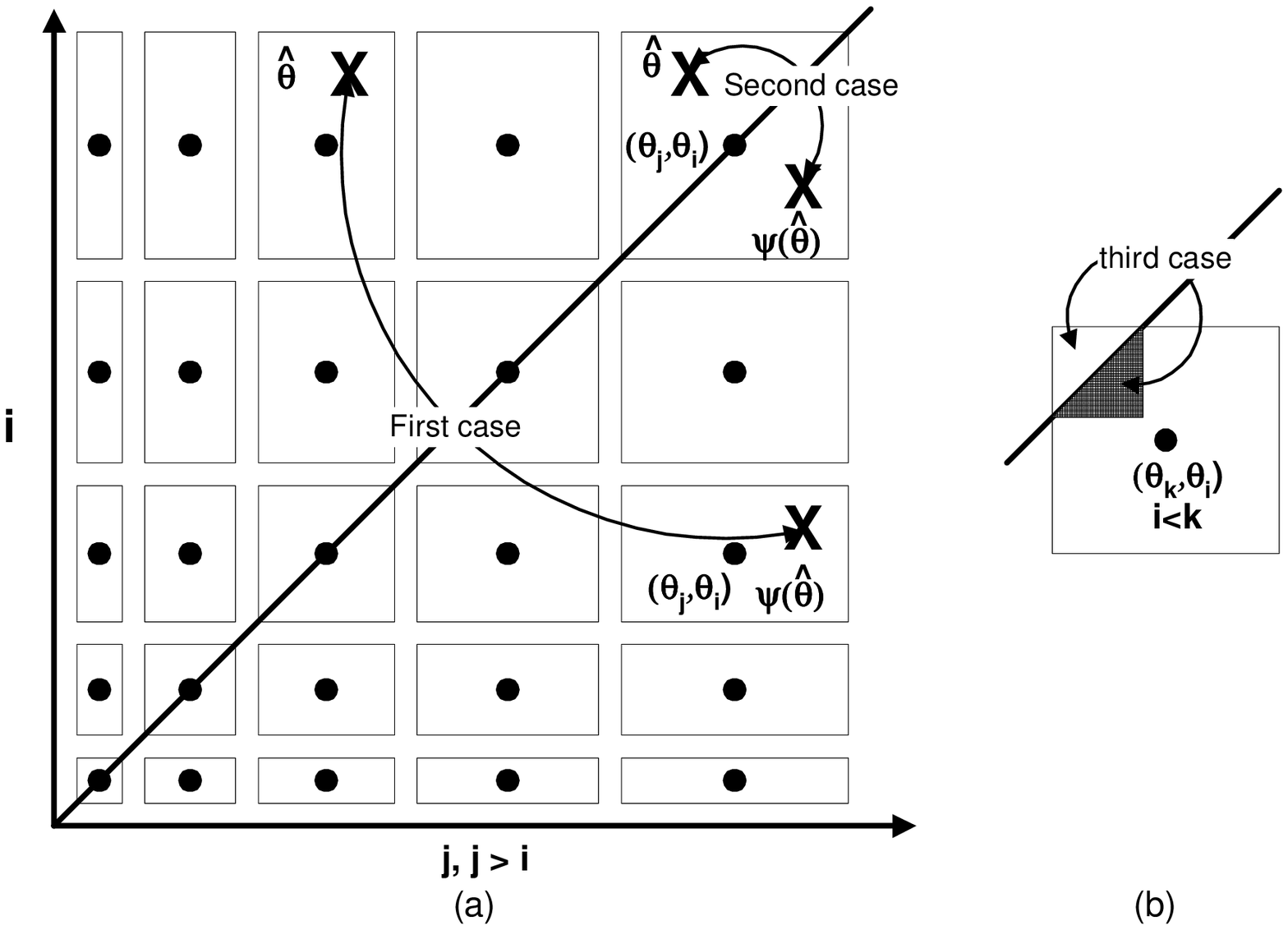}}
 \caption{Decision regions in a two dimensional projection of the pattern
 grid $\Psi \left ( \Agrid \right )$}
 \label{fig:grid_regions}
\enf

In the second and third cases, the box around $\pvec$ contains
a region that is in $\avk$ but outside $\Psi \left ( \avk \right )$, i.e.,
there exist sequences $x^n$ that can be generated by $\pvec$ and result in
an ML i.i.d.\ estimator $\pvece$ of $\pvec$ that is still within the box defined
above, but is not properly
ordered, and thus $\avec \left ( \pvece \right ) \neq \pvece$.
As shown in the proof of Lemma~\ref{lemma_patterns_iid}, this can only
occur when $\theta_i = \theta_j$, as in the second case in Figure~\ref{fig:grid_regions},
or when $j = k$, and $\tau_{b_i} = \tau_{b_k}$, as shown in the third case
of Figure~\ref{fig:grid_regions}.  As shown in the proof of Lemma~\ref{lemma_patterns_iid},
both cases still result in $\bar{A} \rightarrow \overline{\Psi \left ( A \right )}$ when
estimation is done according to $\Psi \left ( x^n \right )$.
From Figure~\ref{fig:grid_regions}, we see that this is the case, because
the re-ordering of $\pvece$ to generate $\avec \left ( \pvece \right )$ means
projection of components of $\pvece$ over the diagonal lines as shown for both cases in
the figure.

To conclude the proof of Theorem~\ref{theorem_maximin_patterns}, we need to
consider the estimator $\pvece^{\psi}_{\Omega} \in \Psi \left ( \Agrid \right )$,
which estimates
$\pvec = \avec \left ( \pvec \right )$ by the point in
$\Psi \left ( \Agrid \right )$ nearest to
$\avec \left ( \pvece \right )$.
Based on Lemmas~\ref{lemma_event_A_prob} and~\ref{lemma_patterns_iid}, we show that
the error probability for this estimator, which is solely based on the pattern
$\Psi \left (X^n \right )$ of $X^n$, vanishes with $n$.
An error occurs if $\pvece^{\psi}_{\Omega} \neq \pvec$.
If this happens, event $\Psi \left ( A \right )$ must happen, because
the distance between two adjacent grid points is not smaller than
$2 \Delta \left ( \tau_{b_i} \right )/2 = \Delta \left ( \theta_i \right )$.
(Note that now we only need to
estimate the first $k-1$ components of $\pvec$, since the last component $\theta_k$
is then determined by the others).
Also, for the second region of the bound,
no error is possible in the first $k - k_m$ small parameters
because they need not be estimated since they are equal for all points on the
grid $\Psi \left ( \Agrid \right )$, and the probability that any of
these letters occurs vanishes.
Hence,
\be
 P_{\theta} \left ( \pvece^{\psi}_{\Omega} \neq \pvec \right ) \leq
 P_{\theta} \left [ \Psi \left ( A \right ) \right ] \leq
 P_{\theta} \left ( A \right ) \rightarrow 0.
\ee
This concludes the proof of Theorem~\ref{theorem_maximin_patterns}.
{\hfill $\Box$}

\section{A Lower Bound for Most Sources}
\label{sec:lb_most}

The analysis in Section~\ref{sec:maximin} cannot be used to lower bound the
average pattern redundancy
for most sources.  This is because of the non-uniform grid.
The strong version of the redundancy-capacity theorem requires
the sources in each set of $M$ sources to be uniformly distributed
for the result in \eref{eq:red_cap} to hold.
However, randomly choosing a non-uniform
grid, generating a uniform distribution of the sources in the grid, results in an overall
non-uniform distribution of the sources in $\Psi \left ( \avk \right )$, because
sources in the dense areas are more likely to be chosen.
The redundancy-capacity theorem can still be used, but the bound that is obtained
will be a bound on the class, assuming the sources are distributed with a non-uniform
prior in the class $\Psi \left ( \avk \right )$.
Such a bound is not a bound for most sources in the
class in Rissanen's sense.

To derive a lower
bound on the redundancy for most sources in the class
$\Psi \left ( \avk \right )$, a different
approach from that in Section~\ref{sec:maximin}
must, therefore, be used.  Instead of a non-uniform grid,
we show that sources in the centers of disjoint
spheres with radius $r = n^{-0.5 ( 1 - \varepsilon )}$ in the
$k-1$ dimensional \emph{pattern\/}
space are distinguishable, and count the number of spheres that
can be packed in the space $\Psi \left ( \avk \right )$ (see
\cite{conway} for information about the sphere packing problem).  This sphere lattice can
be shifted to cover the whole class for different choices
of $M$ points.  Hence,
the conditions of the strong version of
the redundancy-capacity theorem are then satisfied, and the normalized logarithm
of the bound on the number of spheres becomes the lower bound on the redundancy for
most sources.  (This approach resembles Rissanen's pioneering work \cite{rissa} for
sources with a finite number of parameters.  However, here the asymptotics change due
to the consideration of patterns and large alphabets.)

Since we
no longer take advantage of the fact that sources that vary only in small parameters
are still distinguishable, the size of the grid that is constructed reduces
w.r.t.\ that of the minimax bound.  This leads to a smaller lower bound on
the redundancy for most sources, hinting that
it may be possible to compress most sources in the class better than the worst
sources.  This is reasonable because many sources, with large $k$ in particular, may
generate very compressible pattern sequences, that may decrease the overall average
redundancy.
On the other hand, however,
this redundancy reduction may also be due to looseness in the
bounding techniques.
The orders of the bounds obtained remain the same
as those of the minimax bound, but for large alphabets,
the coefficients become smaller.
For small alphabets, the decrease in the bound is reflected in a
smaller second order term.
We proceed with Theorem~\ref{theorem_most_patterns}, that lower bounds
the redundancy of patterns generated by most sources in the class
$\avk$ and conclude this section with its proof.

\begin{theorem}
\label{theorem_most_patterns}
 Fix an arbitrarily small $\varepsilon > 0$, and let $n \rightarrow \infty$.  Then,
 the $n$th-order average universal coding redundancy for
 coding patterns induced by i.i.d.\ sources with alphabet size $k$ is lower
 bounded by
 \be
  \label{eq:most_sources_pattern_bound}
  R_n \left [ L, \avec \left ( \pvec \right ) \right ]\geq
   \left \{ \begin{array}{ll}
    \frac{k-1}{2n} \log \frac{n^{1-\varepsilon}}{k^3} -
    \frac{k-1}{2n} \log \frac{8 \pi}{e^3} -
    O \left ( \frac{\log k}{n} \right ), & \mbox{for } k \leq
    \frac{1}{2} \cdot \left ( \frac{n^{1-\varepsilon}}{\pi} \right )^{1/3} \\
    \frac{1.5 \log e}{2 \pi^{1/3}} \cdot n^{-(2+\varepsilon)/3} -
    O \left ( \frac{\log n}{n} \right ), &
    \mbox{for } k > \frac{1}{2} \cdot
    \left ( \frac{n^{1-\varepsilon}}{\pi} \right )^{1/3}
   \end{array} \right .
 \ee
 for every code $L(\cdot)$ and
 almost every i.i.d.\ source $\pvec \in \avk$, except for a set of
 sources $A_{\varepsilon} \left ( n \right )$ whose volume goes to $0$
 as $n \rightarrow \infty$.
\end{theorem}

Theorem~\ref{theorem_most_patterns} shows similar
behavior of the redundancy for most sources to that shown
by Theorem~\ref{theorem_maximin_patterns} for the minimax redundancy.
For small $k$, each probability parameter, again, costs
$0.5 \log (n/k^3)$ extra code bits.  For large $k$'s (including $k>n$),
we obtain a redundancy bound of $O \left (n^{-2/3} \right )$,
identical for all large values of $k$.
The lower bound of Theorem~\ref{theorem_most_patterns} naturally is the strongest
sense bound and applies also to the minimax average and individual redundancies.
It is therefore smaller than the other two sets of bounds.
While the first order term in the first region of \eref{eq:most_sources_pattern_bound}
is equal to that of \eref{eq:minimax_pattern_bound}, the second order
term here is negative and decreases the redundancy for most sources linearly with $k$,
whereas the second order
term of the first region in \eref{eq:minimax_pattern_bound} is positive
and increases the minimax redundancy linearly with $k$.
In the second region of the bound in \eref{eq:most_sources_pattern_bound},
the coefficient of the redundancy which approximately equals
$0.74$ decreases w.r.t.\ that of the minimax redundancy
in \eref{eq:minimax_pattern_bound}, which approximately equals $2.52$.

The proof of Theorem~\ref{theorem_most_patterns} lower bounds the volume of the
space $\Psi \left ( \avk \right )$, and then uses sphere packing density results
\cite{conway} to lower bound the number of spheres that can be packed in this
volume.  Then, it is shown that sources at centers of disjoint spheres with radius
$r = n^{-0.5 ( 1 - \varepsilon )}$ are distinguishable also in the pattern
space, i.e., by observing $\Psi \left (X^n \right )$.
There are two methods that bound the volume of the space
$\Psi \left ( \avk \right )$.  The first takes the volume of $\avk$, which by condition
\eref{eq:prob_cond} must be $1/(k-1)!$, and divides it by $k!$ to extract all permutations
of the same sources, resulting in a volume of $1/[(k-1)! k!]$.  The other method
directly computes the volume of $\Psi \left ( \avk \right )$ from the
conditions defining an ordered vector $\avec \left ( \pvec \right )$.
Both methods obtain the same bound on the volume of $\Psi \left ( \avk \right )$.
We will, therefore, demonstrate only the second one.
Since the second method is tight, it hints to the fact that, unlike the reduction
of the grid $\Agrid$ in Section~\ref{sec:maximin} by a factor of $k!$ to form
the grid $\Psi \left (\Agrid \right )$, the reduction of the volume
of $\avk$ by a factor of $k!$ to bound the volume of $\Psi \left ( \avk \right )$
is tight.  This is because of the difference in considering a grid and a continuous
space.  In the continuous space $\avk$, sources with several \emph{exactly\/} identical
components make a negligible portion of the space (as the probability of any single
point is zero), whereas such sources are not negligible when we construct a grid
as in Section~\ref{sec:maximin}.

Although the bounding of the volume of $\Psi \left (\avk \right )$ is tight,
we still encounter a similar phenomenon to that in Section~\ref{sec:maximin}, where
there exists a constant $c$, such that for every $k > c n^{(1-\varepsilon)/3}$,
the bound becomes negative.  This is due to another step in the bounding.
In this analysis, we bound the number of spheres packed in $\Psi \left ( \avk \right )$
dividing the volume of $\Psi \left ( \avk \right )$ by a volume of a single sphere
and factoring a packing density factor.  However, as $k$ increases, most spheres
contained in $\Psi \left ( \avk \right )$ have only portions in the space, whereas
big portions of those spheres are outside the space.  Therefore, division by the
complete volume of a sphere results in loose bounding of the number of sources
that are still distinguishable in the space.  We solve this problem in a manner that
resembles the solution in Section~\ref{sec:maximin}.
Let $k_m$ be the value of $k$ for which the bound is maximal.  Then,
for $k> k_m$,
instead of considering the whole space $\Psi \left ( \avk \right )$ and
bounding the number of spheres in it, we bound the number
of spheres in a slice of this space,
in which there are only $k_m$ sufficiently large probability parameters, and
all the other $k -k_m$ probability parameters sum to an insignificantly small
total probability.  This idea is best pictured if one considers
packing spheres in a triangular based pyramid.  The number of circles that
can be packed on its basis is larger than the number of circles that can
be packed in any horizontal two dimension cut above the basis.  If the spheres
are very large, we may not be able to pack any complete two dimensional cuts of these
spheres above the basis. Since we are not interested in complete spheres in all
dimensions, it is sufficient to consider the number of dimensions that
will give the maximum number of sphere portions that are packed in the space.
This number is a lower bound on the total number of sphere portions that can
be packed in the space.
Using only $k_m$ dimensions in the sphere packing analysis,
we obtain the second region of
the bound.  Note that when we shift the sphere lattice to obtain
a covering of the whole space, some center points that represent
sources in the set will no longer be in the space, reducing $M$.  However,
the lower bound on $M$ obtained from the $k_m$ dimensional cut will not
be affected, when at the same time the shifting allows the space covering
condition of the strong version of the
redundancy-capacity theorem to be satisfied.

As in Section~\ref{sec:maximin}, we also need to show that distinguishability
in the i.i.d.\ space carries over to the pattern space.  This is, in fact,
easier than in the minimax case.  All we need to show is that
a point $\pvece$ in $\avk$ outside $\Psi \left ( \avk \right )$ but still in a sphere
that is centered inside $\Psi \left (\avk \right )$ projects onto
a point $\avec \left ( \pvece \right )$ that is still in the same sphere.  The point
$\avec \left ( \pvece \right )$ is the one that will be obtained directly from
$\Psi \left (X^n \right )$.
Therefore, if the ML i.i.d.\ estimator $\pvece$ of $\pvec$ based on $X^n$ is outside
$\Psi \left ( \avk \right )$ but still distinguishable in the i.i.d.\ space,
its projection $\avec \left ( \pvece \right )$ into $\Psi \left ( \avk \right )$, obtained
from $\Psi \left ( X^n \right )$,
is still in the same sphere.
This is shown by geometric considerations
demonstrated as a series of exchanges that rearrange
the components of $\pvece$ into $\avec \left ( \pvece \right )$ by exchanging
a pair in each step.
We conclude this section with the proof of Theorem~\ref{theorem_most_patterns}.

\noindent
{\bf Proof of Theorem~\ref{theorem_most_patterns}:}
We begin with bounding the volume of the $k-1$ dimensional space
$\Psi \left ( \avk \right )$.  Only ordered vectors $\pvec$
for which $\theta_1 \leq \theta_2 \leq \cdots \leq \theta_{k-1}$ are contained
in $\Psi \left ( \avk \right )$.  This can be used to set constraints
on a $k-1$ dimensional integral that bounds the volume
of $\Psi \left ( \avk \right )$.  By condition \eref{eq:prob_cond},
\be
 \label{eq:prob_cond_patterns}
 1 \geq \sum_{i=1}^{k-1} \theta_i \geq \left ( k - 1 \right ) \theta_1
 ~\Rightarrow~ \theta_1 \leq \frac{1}{k-1}.
\ee
Similarly (and more generally),
\be
 \label{eq:pattern_prob_constraints}
 1 - \sum_{j=1}^{i-1} \theta_j \geq \sum_{l = i}^{k-1} \theta_l  \geq
 (k - i) \theta_i
 ~\Rightarrow~ \theta_i \leq
 \frac{1 - \sum_{j=1}^{i-1} \theta_j}{(k - i )}.
\ee
Now, \eref{eq:pattern_prob_constraints} gives upper limits on every component
of $\pvec$.  The ordering condition of $\pvec$ that is necessary for $\pvec$ to
be in $\Psi \left ( \avk \right )$ gives lower limits on each component of
$\pvec$.  Ordering is maintained by the above
conditions except for the $k$th component $\theta_k$.  Therefore,
the volume obtained by a $k-1$ dimensional integral over $1$ within all these
limits needs to be reduced by a factor of $k$ to only take the $k$ dimensional
permutations for which $\theta_k$ is not smaller than all other components of
$\pvec$.  Including all the constraints, $V \left [ \Psi \left ( \avk \right ) \right ]$
is computed in the following equations:
\bea
 \nonumber
 V \left [ \Psi \left ( \avk \right ) \right ]
 &=&
 \frac{1}{k} \cdot
 \int_0^{\frac{1}{k-1}} d \theta_1
 \int_{\theta_1}^{\frac{1}{k-2}\left ( 1 - \theta_1 \right )} d \theta_2
 \int_{\theta_2}^{\frac{1}{k-3}\left ( 1 - \theta_1 - \theta_2 \right )} d \theta_3
 \cdots
 \int_{\theta_{k-2}}^{\frac{1}{1}\left ( 1 - \theta_1 - \theta_2 -
 \cdots - \theta_{k-2} \right )} d \theta_{k-1} \\
 \nonumber
 &=& \cdots~~ =
 \frac{1}{k} \cdot
 \int_0^{\frac{1}{k-1}} d \theta_1
 \left \{ \frac{1}{ \left [ \left ( k -2 \right )! \right ]^2}
 \left [ 1 - \left ( k - 1 \right ) \theta_1 \right ]^{k-2} \right \} \\
 \label{eq:pattern_space_volume}
 &=&
 \frac{1}{k} \cdot
 \left [ -\frac{1}{\left [\left ( k - 1 \right )! \right ]^2}
 \left [ 1 - \left ( k - 1 \right ) \theta_1 \right ]^{k-1} \right ]_0^{\frac{1}{k-1}}
 ~=~
 \frac{k}{\left [ k! \right ]^2}
 ~=~
 \frac{1}{\left ( k - 1 \right )! \cdot k!}
\eea
Now, consider packing of $k-1$ dimensional spheres with radius
$r = n^{-0.5\left(1 - \varepsilon \right )}$ in $\Psi \left ( \avk \right )$
so that no spheres share the
same point in the space.
The ratio between the volume $V \left [ \Psi \left ( \avk \right ) \right ]$
of $\Psi \left ( \avk \right )$
and the volume
of one sphere $V_{k-1}(r)$ is
\be
 \label{eq:sphere_pack1}
 \rho \dfn \frac{V\left [ \Psi \left ( \avk \right ) \right ]}
 {V_{k-1}(r)} = \frac{1}{(k-1)! \cdot k! \cdot V_{k-1}(r)} =
 \left \{
 \begin{array}{ll}
  \frac{\left [ \left ( k - 1 \right )/2 \right ]! \cdot
  n^{\frac{1}{2} \left ( 1 - \varepsilon \right ) \left (k - 1 \right )}}
  {\pi^{(k-1)/2} \cdot \left ( k - 1 \right )! \cdot k!}; & k~\mbox{odd} \\
  \frac{n^{\frac{1}{2} \left ( 1 - \varepsilon \right ) \left (k - 1 \right )}}
  {\left [ \left (k - 2 \right )/2 \right ]! \cdot 2^{k-1} \cdot
  \pi^{(k-2)/2} \cdot k!}; & k~\mbox{even},
 \end{array}
 \right .
\ee
where we substituted the volume of
$\Psi \left ( \avk \right )$ from
\eref{eq:pattern_space_volume}.
However, the number of spheres that can be packed in
$\Psi \left ( \avk \right )$ is
bounded by
\be
 \label{eq:sphere_bound}
 M \geq \Delta \rho \geq \frac{1}{(k-1)! \cdot k!
 \cdot V_{k-1} \left (r\right ) \cdot 2^{k-1}},
\ee
where the factor $\Delta = 2^{-(k-1)}$ is a lower bound on the sphere
packing density, i.e., the fraction of the space that is actually occupied by spheres
(see \cite{conway}).
Now, let us
choose a grid that contains the sources $\pvec$
at the centers of all the $M$ spheres
packed in $\Psi \left ( \avk \right )$.  We can lower bound the number
of sources in one such grid by using \eref{eq:sphere_pack1}-\eref{eq:sphere_bound}.
Taking the logarithm of the bound in \eref{eq:sphere_bound} and using
Stirling's formula to bound factorials, we obtain the bound
\be
 \label{eq:most_logM_bound}
 \log M \geq
 \left ( 1 - \varepsilon \right )
 \frac{k-1}{2} \log n -
 \frac{k-1}{2} \log k^3 -
 \frac{k-1}{2} \log \frac{8 \pi}{e^3} -
 \frac{3}{2} \log k +
 \frac{1}{2} \log \frac{e^3}{4\pi} -
 O \left ( \frac{1}{k} \right ).
\ee
As long as the lower bound on $M$ is large,
we can (cyclicly) shift the whole grid to allow different choices of
grids in $\Psi \left ( \avk \right )$ to cover the whole space, and satisfy
the conditions of the strong version of the redundancy-capacity theorem.
All random shifts of the original grid will
form a covering of $\Psi \left (\avk \right )$,
and can be designed so that
uniform distribution is preserved for choosing a point
$\pvec \in \Psi \left ( \avk \right )$ over the whole
class and also within every set of $M$ points that is chosen.
Hence, in this case we can use
the normalized logarithm of the number $M$ of points on this random grid
as a lower bound on the redundancy for most sources if all sources within any
shift of the grid
are distinguishable by the
observed random sequence.
This yields the first region of the bound in
\eref{eq:most_sources_pattern_bound}.
However, observing \eref{eq:most_logM_bound}, as in
the minimax case, the bound becomes negative and useless for large $k$'s.
As in Section~\ref{sec:maximin}, we solve this problem by
fixing the bound at its maximum value as a function of $k$.  Assume
this value is attained
at $k = k_m$.  Then, for every $k > k_m$, we will obtain the same bound,
resulting in the second region in \eref{eq:most_sources_pattern_bound}.
By straightforward differentiation it can be shown that
the bound in \eref{eq:most_logM_bound} attains its maximum value
for $k_m = 0.5 \left ( n^{1-\varepsilon} / \pi \right )^{1/3}$.  Substituting
this value of $k_m$ in \eref{eq:most_logM_bound}, normalizing by $n$,
we obtain the bound of the second region of \eref{eq:most_sources_pattern_bound}.

When $k_m$ is used to obtain the bound for a larger $k$, we still
shift the complete grid to create a covering of the space
$\Psi \left ( \avk \right )$ in which each source is contained in one
grid.  Unlike the minimax case, here we cannot simply discard
points in the grid with $k > k_m$ nonzero parameters.  These must be included
in the grid, and distinguishability between them and other points must be proven.
However, we
can lower bound the number of sources in the grid by the
number of spheres in $k_m$ dimensional cut of $\Psi \left ( \avk \right )$
for which all the other (first) $k-k_m$ parameters are very small, and insignificant.
This analysis is valid also if $k > n$, and thus the bound in the
second region is general, and applies also to such large alphabets.

Finally, to satisfy the covering of the whole space, we need to
show that every source in $\avk$ is included in a grid.  Demonstrating that
only for the ordered permutation is not sufficient.  This can be
done by taking different grids for each permutation vector, i.e.,
each ordered source $\avec \left (\pvec \right )$ will appear in $k!$
different grids through its permutations.  (Since the probability of a single
point is zero in a continuous space, sources for which identical components
exist do not pose a problem.)

To conclude the proof of Theorem~\ref{theorem_most_patterns}, we need to
show distinguishability of the grids defined above in the pattern space.
We show that this is a direct result
of distinguishability of the respective grids in the i.i.d.\ space.
First, we state a lemma showing distinguishability in the i.i.d.\ space, i.e.,
by observing $X^n$,
and then we prove another lemma that implies
that distinguishability in the i.i.d.\ space causes distinguishability in the
pattern space on the reduced pattern grid, obtained by observing only $\Psi \left (X^n \right )$.

\begin{lemma}
\label{lemma_distinct2}
Consider one choice of a random grid in the i.i.d.\ space
$\avk$ as defined above.
Let $\pvec \in \avk$ be a point on this grid,
and let the random sequence $X^n$ be generated by the conditional
probability $P_{\theta} (X^n)$ (given $\pvec$).
Then, the probability that the ML estimator of $\pvec$
from the observed $X^n$ is outside
the sphere of radius $1/\sqrt{n}^{1-\varepsilon}$ centered in $\pvec$ vanishes with $n$,
\be
\label{eq:lemma_distinct2}
 \lim_{n \rightarrow \infty}
 P_{\theta} \left \{ \left \| \pvece - \pvec \right \|
 > \frac{1}{\sqrt{n}^{1-\varepsilon}} \right \} = 0,
\ee
for every alphabet size $k$.
\end{lemma}
The proof of Lemma~\ref{lemma_distinct2} is presented in
\ref{ap:lemma_distinct2_proof}.
The next lemma shows that the distance between two points, one in
$\Psi \left ( \avk \right )$ and the other in $\avk$, can only decrease
if the latter is projected into $\Psi \left ( \avk \right )$.  This lemma
is necessary, because the ordered ML estimator $\avec \left ( \pvece \right )$
obtained directly from $\Psi \left (X^n \right )$ simply performs this projection
over the i.i.d.\ ML estimator $\pvece$.  Hence, this lemma implies that the ordered
estimator must be closer to the point estimated, which is in the pattern space.
\begin{lemma}
\label{lemma_pattern_space_distance}
 Let $\pvec$ and $\pvec'$ be two points in $\avk$, such that
 $\pvec \in \Psi \left ( \avk \right )$.  Then,
\be
 \label{eq:Euclid_pattern}
 \left \| \pvec - \avec \left ( \pvec' \right ) \right \| \leq
 \left \| \pvec - \pvec' \right \|.
\ee
\end{lemma}
\noindent
{\bf Proof:}
Vector $\avec \left ( \pvec' \right )$, which is ordered in non-decreasing order, can
be obtained from $\pvec'$ by a series of exchanges between two components $i_l$ and $j_l$;
$i_l < j_l$, where each exchange must decrease the (index) distances of both components from
their location in $\avec \left ( \pvec' \right )$.  Namely,
let $\pvec'^{(l)}$ denote the vector obtained after the $l$-th exchange.  Then,
$\theta'^{(l-1)}_{i_l} > \theta'^{(l-1)}_{j_l}$, and also $i_l \geq \iota$ and $j_l \leq \rho$,
where $\avec_{\iota} \left ( \pvec' \right ) = \theta'^{(l-1)}_{j_l}$
and $\avec_{\rho} \left ( \pvec' \right ) = \theta'^{(l-1)}_{i_l}$, i.e., the final
destination of each of the components in the ordered vector is in the same direction
as the exchange.
For simplicity,
we omit the index $l$ from $i_l$ and $j_l$ when it can be inferred from the context.
We show that each exchange can only decrease the Euclidean distance
to $\pvec$.  For notation simplicity, let $\varphi_i \dfn \theta'^{(l)}_i = \theta'^{(l-1)}_j$
and $\varphi_j \dfn \theta'^{(l)}_j = \theta'^{(l-1)}_i$.  Thus $\varphi_j > \varphi_i$.
The difference between the square of the Euclidean distance from $\pvec$ before and after
the exchange satisfies
\bea
 \nonumber
 \left \| \pvec - \pvec'^{(l-1)} \right \|^2 -
 \left \| \pvec - \pvec'^{(l)} \right \|^2
 &=&
 \left ( \theta_i - \theta'^{(l-1)}_i \right )^2 +
 \left ( \theta_j - \theta'^{(l-1)}_j \right )^2 -
 \left ( \theta_i - \theta'^{(l)}_i \right )^2 -
 \left ( \theta_j - \theta'^{(l)}_j \right )^2 \\
 \nonumber
 &=&
 \left (\theta_i -\varphi_j \right )^2 +
 \left (\theta_j -\varphi_i \right )^2 -
 \left (\theta_i -\varphi_i \right )^2 -
 \left (\theta_j -\varphi_j \right )^2 \\
 \label{eq:iid_pat_distinct_most}
 &=&
 2 \left ( \varphi_j - \varphi_i \right )
 \left ( \theta_j - \theta_i \right ) \geq 0,
\eea
where the last inequality is obtained since $\varphi_j > \varphi_i$ and
$\theta_j \geq \theta_i$ since $\pvec \in \Psi \left ( \avk \right )$.
Figure~\ref{fig:projection} shows a two dimensional projection of components
$i$ and $j$ of all vectors for one exchange as described above.  It
demonstrates the decrease in distance to $\pvec$ resulting from the exchange.
\bef
 \centerline{\includegraphics[bbllx=150pt,bblly=460pt,bburx=545pt,
  bbury=720pt,height=6cm,
    clip=]{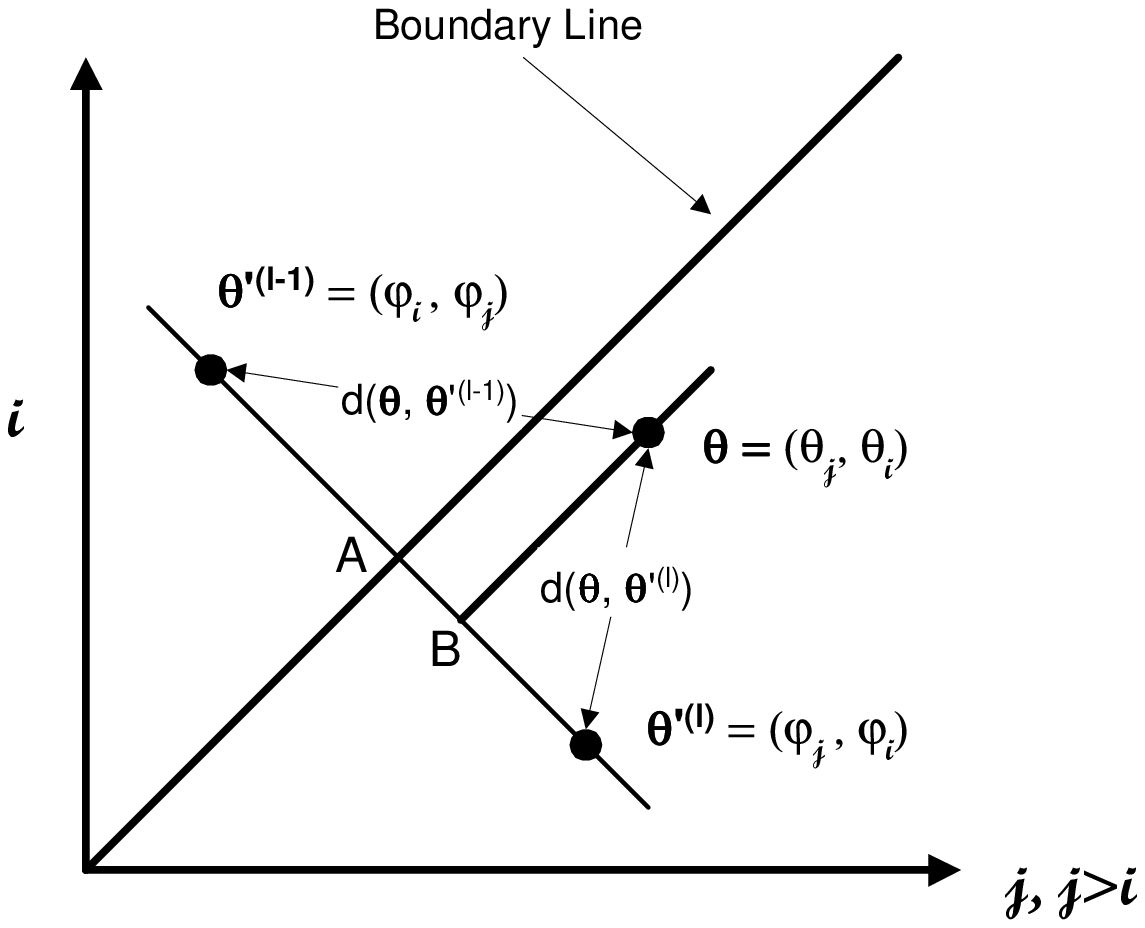}}
 \caption{One exchange step in projection of a source $\pvec' \in \avk$ onto the pattern
 space $\Psi \left ( \avk \right )$}
 \label{fig:projection}
\enf

Now, using \eref{eq:iid_pat_distinct_most},
\bea
 \nonumber
 \left \| \pvec - \pvec' \right \|^2 -
 \left \| \pvec - \avec \left ( \pvec' \right ) \right \|^2 &=&
 \sum_l \left \{
 \left \| \pvec - \pvec'^{(l-1)} \right \|^2 -
 \left \| \pvec - \pvec'^{(l)} \right \|^2 \right \} \\
 &=&
 2 \sum_l \left ( \theta'^{(l)}_{j_l} - \theta'^{(l)}_{i_l} \right )
 \left ( \theta_{j_l} - \theta_{i_l} \right ) \geq 0.
\eea
Since all components of the sum are non-negative, the sum is also non-negative.
This concludes the proof of Lemma~\ref{lemma_pattern_space_distance}.
{\hfill $\Box$}

From Lemma~\ref{lemma_pattern_space_distance}, if
$\left \| \pvec - \pvece \right \| \leq 1/\sqrt{n}^{1-\varepsilon}$, then also
$\left \| \pvec - \avec \left ( \pvece \right ) \right \| \leq 1/\sqrt{n}^{1-\varepsilon}$.
Similarly to the proof of Theorem~\ref{theorem_maximin_patterns}, now let
$\pvece^{\psi}_{\Omega}$ be the point in the random pattern
grid, denoted by $\Psi \left (\Agrid \right )$,
nearest to $\avec \left ( \pvece \right )$.  Then,
using Lemmas~\ref{lemma_distinct2} and~\ref{lemma_pattern_space_distance},
the probability
that a sequence generated by $\pvec$ will appear by $\Psi \left (X^n \right )$
to have been generated by
another source in the same grid is upper bounded,
as $n \rightarrow \infty$, by
\be
 P_{\theta} \left ( \pvece^{\psi}_{\Omega} \neq \pvec \right ) \leq
 P_{\theta} \left \{
 \left \| \avec \left ( \pvece \right ) - \pvec \right \|
 > \frac{1}{\sqrt{n}^{1-\varepsilon}} \right \} \leq
 P_{\theta} \left \{
 \left \| \pvece - \pvec \right \|
 > \frac{1}{\sqrt{n}^{1-\varepsilon}} \right \} \rightarrow 0.
\ee
The first bound is since not all points in $\Psi \left ( \Agrid \right )$ are contained
in spheres.
Hence, distinguishability is attained.
This concludes the proof of Theorem~\ref{theorem_most_patterns}.
{\hfill $\Box$}

\section{Upper Bounds}
\label{sec:upper_bounds}

We now show how to design codes that attain low redundancy for coding
patterns induced by i.i.d.\ sequences.
We propose a code with good performance for
smaller alphabets sizes, namely, $k \leq \sqrt{n}^{1-\varepsilon}$,
for an arbitrarily small $\varepsilon > 0$, and
combine it with the method in
\cite{orlitsky4} to asymptotically achieve the better compression of the two
for a specific pattern.
The new code uses Rissanen's \cite{rissa} two-part grid based coding approach
combined with a non-uniform grid that resembles that in Section~\ref{sec:maximin}.
For a given sequence with $\hat{k}$ distinct symbols,
we find the best $\hat{k}$-dimensional pattern probability vector
$\widehat{\avec \left ( \pvec \right )}$, which is the vector that gives
the $\hat{k}$th-order ML probability for the \emph{pattern\/} of the sequence.
Note that $\widehat{\avec \left ( \pvec \right )}$ may be different from
$\pvece$ and $\avec \left ( \pvece \right )$.
(Furthermore, the actual ML estimate of a pattern may contain more letters
than those actually observed.  However, in analyzing this code, we constrain
the analysis to the average case, in which our reference is the $k$-dimensional
pattern probability, and to
the class $\tilde{\avecspace}_k$ in which it is unlikely that
$\hat{k} < k$.)
Then, $\widehat{\avec \left ( \pvec \right )}$
is quantized to a grid.  The quantized components are first coded,
and then, the sequence is assigned a probability according to these
quantized
probability parameters.
In \cite{orlitsky4}, the number of
all different \emph{types\/} of patterns of length $n$ is shown to equal
the number of unordered partitioning of the integer $n$.  Given the type, the
pattern ML probability vector $\widehat{\avec_u \left ( \pvec \right )}$
can be computed, as well as its ML probability, which is used
to then encode the sequence using a number of bits that equals
its negative logarithm.  Hence, the redundancy is the logarithm of
the number of types, as shown in the upper bound of \eref{eq:orlitsky_bounds}.
The combined code can compute both description lengths, and then choose between them,
and use the one that requires fewer bits.  One bit is needed to relay to the decoder which
of the codes is used.
We summarize the performance of
the code combined of both codes in the next theorem.

\begin{theorem}
 \label{theorem_upper_bounds}
 Fix an arbitrarily small $\varepsilon > 0$, and let $n \rightarrow \infty$.
 Then, there exist codes with length function
 $L^* \left ( \cdot \right )$ that achieve redundancy
 \be
  \label{eq:upper_bounds_patterns}
  R_n \left [ L^*, \avec \left ( \pvec \right ) \right ] \leq
  \left \{
   \begin{array}{ll}
    \left ( 1 + \varepsilon \right ) \frac{k-1}{2n} \log \frac{n^{1+\varepsilon}}{k^2}, &
    \mbox{for } k \leq \sqrt{n}^{1 - \varepsilon} ~\mbox{and}~
    \pvec \in \tilde{\avecspace}_k \\
     \frac{\pi \sqrt{2/3} \log e}{\sqrt{n}} + O\left (\frac{1}{n}\right ), &
 %   \left ( 1 + \varepsilon \right ) \frac{2}{\sqrt{\ln 2}}
 %   \frac{\sqrt{\log n}}{\sqrt{n}}, &
    \mbox{for } k \geq \sqrt{n}^{1 - \varepsilon}
    ~\mbox{or}~
    \pvec \not \in \tilde{\avecspace}_k
   \end{array}
  \right .
 \ee
 for patterns induced by any i.i.d.\ source $\pvec \in \avk$,
 with alphabet of size $k$.
\end{theorem}

The first region of Theorem~\ref{theorem_upper_bounds} applies to the class
$\tilde{\avecspace}_k$, i.e., it is assumed that the probability that less than
$k$ letters will be observed in $X^n$ is $o(k/n)$.  If the probabilities of
all letters are greater than $1/n^{1-\varepsilon}$, this condition is satisfied.
Note that the proposed code should also achieve good performance even if less than
$k$ letters are likely to be observed in $X^n$.  However, further research
still needs to guarantee that the penalty does not increase in this case, and
is still bounded as in the first region of \eref{eq:upper_bounds_patterns}.
The bound of the first region of \eref{eq:upper_bounds_patterns} also applies
to the individual pattern redundancy under the assumption that the
underlying alphabet contains no symbols other than those observed.
A weaker upper bound, which is to first order twice the bound of the first region of
\eref{eq:upper_bounds_patterns} was subsequently derived in \cite{orlitsky4o} for coding
individual patterns with $k$ occurring indices
as long as $k = o \left ( n^{1/3} \right )$.
While the bound in \cite{orlitsky4o}
is larger (thus weaker) and applies only to smaller $k$'s, it is stronger
in the sense that it applies to a wider class containing all sequences
in which $k$ symbols occur, without restricting the pattern generating alphabet
to contain only symbols observed in $X^n$.
The bound in the second region of Theorem~\ref{theorem_upper_bounds} applies to
the class $\avk$.
%Unlike the bound of the first region of Theorem~\ref{theorem_upper_bounds},
%the bound in the second region applies to the class $\avk$.  Its individual sequence
%version applies to any sequence with $k$ occurring letters, even if the ML
%pattern estimate has more than $k$ letters.

The upper bounds in \eref{eq:upper_bounds_patterns} show that we can design universal codes
for patterns that require at most $0.5 \log (n/k^2)$ bits for each unknown
probability parameter, as long as $k$ is small enough, essentially of
$O \left ( \sqrt{n} \right )$ or less.  If $k$ is larger, we observe a similar
phenomenon as that of the lower bounds, in which we achieve the same
redundancy for every large $k$, which is of $O \left ( n^{-1/2} \right )$
bits per symbol overall.  This performance is better than that attainable
in standard i.i.d.\ compression.  In particular, in the first region
we gain $0.5 \log k$ bits for each parameter, and the gain increases
with $k$ in the second region.
In Section~\ref{sec:discuss}, we discuss a different method that can be used
to bound the redundancy in the second region.  The ideas considered can be used
(as in subsequent work \cite{shamir_ciss04}) to obtain stronger bounds in this
region.

As indicated earlier, we observe gaps between the
upper bounds and the lower bounds considered in the previous sections.
In the first region, the lower bound is
smaller by $0.5 \log k$ bits for each parameter, whereas in the
second region (as in the results in
\cite{orlitsky1}-\cite{orlitsky4o}), the lower bound is of $O \left ( n^{-2/3} \right )$
overall instead of $O \left ( n^{-1/2} \right )$.  Naturally, the
second region for the lower bounds starts with smaller $k$.
Gaps between the upper and the lower bounds are still an open problem
and will be discussed in Section~\ref{sec:discuss} in somewhat more detail.
This section is concluded with the proof of Theorem~\ref{theorem_upper_bounds}.

\noindent
{\bf Proof of Theorem~\ref{theorem_upper_bounds}:}
To prove Theorem~\ref{theorem_upper_bounds}, we demonstrate and analyze the
code that achieves
the redundancy bound for the first region of
\eref{eq:upper_bounds_patterns}. As mentioned earlier,
a given pattern is encoded by this code as well as the code in \cite{orlitsky4},
and the one with the smaller description
length is then chosen.  One bit is used to convey which code has been used (resulting in
the additional $O(1/n)$ term of the second region).  The rest of the proof is thus focused
on the first region and bounding the performance of the new code.  The proof for the
second region is concluded using \cite{orlitsky4}.

Using the code for the first region, we first need
$\left ( 1 + \varepsilon \right ) \log \hat{k}$ bits to encode the number
of occurring letters
$\hat{k}$ with Elias's coding for the integers \cite{elias}.
Let $\pvec \in \tilde{\avecspace}_k$
and $k \leq \sqrt{n}^{1-\varepsilon}$.
Let $\widehat{\avec \left ( \pvec \right )} =
\left ( \psi_1, \psi_2, \ldots, \psi_{\hat{k}} \right )$ be the $\hat{k}$-dimensional
probability vector
that maximizes the probability of $\Psi \left ( X^n \right )$
in \eref{eq:pattern_probability} for $X^n$.
Let $\pvece$ be the i.i.d.\ ML estimator of $\pvec$ from $X^n$.
%Without loss of generality, assume that $\pvece$ is ordered such that
%$\hat{\theta}_1 \leq \hat{\theta}_2 \leq \cdots \leq \hat{\theta}_k$.
%(If this is not the case, we can simply consider the
%permutation of $\pvece$ that satisfies this condition instead.)
Let
$\tgrid \dfn \left ( \tau_1, \tau_2, \ldots, \tau_b, \ldots, \tau_B \right )$
be a grid of $B$ points whose $b$th component is defined in a similar manner to
\eref{eq:grid_point}, where $-\varepsilon$ is replaced by
$\varepsilon$, i.e.,
\be
 \label{eq:grid_point_upper}
 \tau_b \dfn
 \sum_{j=1}^{b} \frac{2 (j - \frac{1}{2})}{n^{1+\varepsilon}} =
 \frac{b^2}{n^{1+\varepsilon}}.
\ee
Thus, there are
\be
 \label{eq:upper_grid_size}
 B = \sqrt{n}^{1+\varepsilon}
\ee
points in $\tgrid$.  Let $\vphivec \dfn \left (
\varphi_1, \varphi_2, \ldots, \varphi_{k-1}, \varphi_k \right )$
be a quantized version of
$\widehat{\avec \left ( \pvec \right )}$, for which
each of the first $k-1$ components
$\varphi_i$
takes one of the two nearest grid points surrounding $\psi_i$, i.e.,
if $\psi_i \in \left [ \tau_b, \tau_{b+1} \right ]$,
$\varphi_i$ equals either $\tau_b$ or $\tau_{b+1}$.  The point
that is chosen for $\varphi_i$ between the two grid points is the
one that minimizes the absolute value of the cumulative difference
between the first $k-1$ components of $\widehat{\avec \left ( \pvec \right )}$
and those of $\vphivec$ such that the non-decreasing order of the components
of $\vphivec$ is retained. This ensures that the last largest component
$\varphi_k$ of $\vphivec$ is within the defined grid spacing around
$\psi_k$, even if it does not take a value in $\tgrid$.

The code first codes the first $k-1$ components of $\vphivec$, and then
computes $P_{\varphi} \left [ \Psi \left ( X^n \right ) \right ]$, and
uses (up to integer length constraints)
$-\log P_{\varphi} \left [ \Psi \left ( X^n \right ) \right ]$ bits
to code the pattern.
The average code length for $\pvec \in \tilde{\avecspace}_k$
and $k \leq \sqrt{n}^{1-\varepsilon}$ is thus bounded (up to integer length constraints) by
\be
 E_{\theta} L^{*} \left [ \Psi \left (X^n \right ) \right ] \leq
 1 + \left ( 1 + \varepsilon \right ) \log k +
 E_{\theta} \left \{ L^*_R \left [ \widehat{\avec \left ( \pvec \right )} \right ] \right \} -
 E_{\theta} \left \{ \log P_{\varphi} \left [\Psi \left (X^n \right ) \right ] \right \},
\ee
where $L^*_R \left [ \widehat{\avec \left ( \pvec \right )} \right ]$ is the cost of representing
the quantized version $\vphivec$ of $\widehat{\avec \left ( \pvec \right )}$.
The first term of $1$ is the cost of one bit distinguishing between the two codes.  The second
term is a bound on the cost of representing $\hat{k} < k$.  The last term is the cost
of coding the pattern using the quantized ML estimates in $\vphivec$.  The inequality
is also since some patterns may be represented shorter by the code from \cite{orlitsky4}.
Denoting an upper bound on the representation cost of an up to $k$-dimensional vector $\vphivec$
by $\bar{L}^*_{R, k}$, the average redundancy for $\pvec \in \tilde{\avecspace}_k$
and $k \leq \sqrt{n}^{1-\varepsilon}$ is, therefore, upper bounded by
\bea
\nonumber
 n R_n \left [ L^*, \avec \left ( \pvec \right ) \right ] &\leq&
 1 + \left ( 1 + \varepsilon \right ) \log k +
 E_{\theta} \left \{ L^*_R \left [ \widehat{\avec \left ( \pvec \right )} \right ] \right \} +
 E_{\theta} \left \{ \log\frac{P_{\theta }
 \left [\Psi \left (X^n \right ) \right ]}
 {P_{\varphi} \left [\Psi \left (X^n \right ) \right ]} \right \} \\
 \nonumber
 &\leq&
 1 + \left ( 1 + \varepsilon \right ) \log k +
 \bar{L}^*_{R, k} +
 P_{\theta} \left (\hat{k} < k \right ) n \log k +
 E_{\theta} \left \{ \left . \log\frac{P_{\widehat{\psi \left (\theta \right )}}
 \left [\Psi \left (X^n \right ) \right ]}
 {P_{\varphi} \left [\Psi \left (X^n \right ) \right ]}  \right | \hat{k} = k \right \} \\
 &=&
 \label{eq:ub_proof3_1}
 \bar{L}^*_{R, k} +
 E_{\theta} \left \{ \left . \log\frac{P_{\widehat{\psi \left (\theta \right )}}
 \left [\Psi \left (X^n \right ) \right ]}
 {P_{\varphi} \left [\Psi \left (X^n \right ) \right ]}  \right | \hat{k} = k \right \} +
 o \left (k \log \frac{n}{k^2} \right ).
\eea
The second inequality is since at most $\log k$ bits are required to code every index, and
also because the pattern probability w.r.t.\ the $k$-dimensional ML estimate is not smaller
than the probability w.r.t.\ the actual parameter $\pvec$.  The next
equality is because of the assumption that $P_{\theta} \left (\hat{k} < k \right ) = o(k/n)$,
and since $o \left ( \log k \right ) = o \left (\log (n/k^2) \right )$ by
definition of the region.

To complete the bound in the first region, we now need to bound the remaining first
two terms of \eref{eq:ub_proof3_1}.  These two costs are
the cost of coding $\vphivec$, and the cost
of using the quantized version $\vphivec$ of the $k$-dimensional pattern ML
probability estimator
$\widehat{\avec \left ( \pvec \right )}$ instead
of using the actual $k$-dimensional pattern ML probability estimator.
For the remainder of the proof, we can now assume that $\hat{k} = k$ because
for the first term, we will obtain a bound that increases with $\hat{k}$, and
for the second term, we compute the expectation conditioned on this event.
We next bound the two costs and
show that the second is negligible w.r.t.\ the first in the first
region of the bound.  This together with \eref{eq:ub_proof3_1}
results in the upper bound for this region.

Instead of coding the first $k-1$ components $\varphi_i$ of
$\vphivec$, we can code their indices in $\tgrid$.  Let
$b \left ( \varphi_i \right )$ be the index in $\tgrid$ of
the grid point that equals $\varphi_i$.  Since the
vector $\vphivec$ is ordered, we can use a differential code which uses
\[
 \left ( 1 + \varepsilon' \right ) \log
 \left [ b\left( \varphi_i \right ) - b \left ( \varphi_{i-1} \right ) + c\right ]
\]
bits, where $c$ is a constant,
to represent the integer displacement to the index of $\varphi_i$ from
that of $\varphi_{i-1}$ with Elias's code, where
$\varepsilon' > 0$ is arbitrarily small,
$b \left ( \varphi_0 \right ) \dfn 0$,
and extra $c$ bits are added to apply for zero or small displacements.
Hence, altogether, we will need
\bea
 \nonumber
 L^*_R \left [\widehat{\avec \left ( \pvec \right )} \right ] &=&
 \sum_{i=1}^{k-1} \left ( 1 + \varepsilon' \right )
 \log \left [  b\left( \varphi_i \right ) - b \left ( \varphi_{i-1} \right ) + c\right ]
 \\ \nonumber
 &\leq&
 \left ( 1 + \varepsilon' \right )
 \left ( k - 1 \right ) \log \frac{B+ck}{k-1} \\
 \label{eq:varphi_cost}
 &\leq& \left ( 1 + \varepsilon_1 \right )
 \frac{k-1}{2} \log \frac{n^{1+\varepsilon}}{k^2}
 \dfn \bar{L}^*_{R,k}
\eea
bits to represent $\vphivec$, where the first inequality is obtained
by Jensen's inequality, and the second follows directly
from \eref{eq:upper_grid_size} and the assumption that
$k = o \left (\sqrt{n} \right )$ by absorbing low-order terms in $\varepsilon_1$.
Note that the last inequality in
\eref{eq:varphi_cost} holds only for $k = o \left (\sqrt{n} \right )$.
The bound of \eref{eq:varphi_cost} is used to bound the cost
of representing $\vphivec$ in \eref{eq:ub_proof3_1}.  (We note that
in Section~\ref{sec:discuss}, we
will demonstrate a method that
yields representation cost for $\vphivec$ which is fixed
at $O \left ( n^{(1+\varepsilon)/3} \right )$ bits.  For
$k \geq n^{1/3}$, this cost is better than that in \eref{eq:varphi_cost}.  However,
the cost of quantizing the pattern ML estimator, which is shown next,
will overwhelm this cost for large $k$.)
%resulting in just a slightly lower order bound
%from that of the first region of \eref{eq:upper_bounds_patterns}.

We now bound the second term of \eref{eq:ub_proof3_1}.
The probability of $\Psi \left ( x^n \right )$ can be expressed as
in \eref{eq:pattern_probability} by summing over all sequences that
have the same pattern with a fixed parameter vector.
On the other hand, we can express the same
probability by fixing the actual sequence and summing over all permutations of
the parameter vector
\be
 \label{eq:pattern_prob1}
 P_{\theta} \left [ \Psi \left ( x^n \right ) \right ] =
 \sum_{\sigvec} P_{\theta(\sigma)} \left (x^n \right ).
\ee
Now, to bound the cost of quantizing the pattern ML
estimator reflected in the second term of \eref{eq:ub_proof3_1},
we consider the logarithm of the ratio between
$P_{\widehat{\psi\left ( \theta \right )}}
\left [ \Psi \left ( X^n \right ) \right ]$ and
$P_{\varphi} \left [ \Psi \left ( X^n \right ) \right ]$.
We can express each of the two probabilities
using \eref{eq:pattern_prob1}.  Then, we discard permutations
of $\widehat{\psi\left ( \theta \right )}$ that give negligible
probability for $X^n$ (and their respective quantized
versions) from each of the sums in
the ratio.  Next, we bound the ratio between the probability of
$X^n$ given a non-negligible permutation of $\widehat{\psi\left ( \theta \right )}$
and that obtained by the quantized version of this permutation.  We obtain
the same bound for all these permutations.  This bound can, in turn,
be used to bound the ratio between
$P_{\widehat{\psi\left ( \theta \right )}}
\left [ \Psi \left ( X^n \right ) \right ]$ and
$P_{\varphi} \left [ \Psi \left ( X^n \right ) \right ]$.
To obtain the bound for all permutations, we need to bound the absolute
differences between $\psi_i$ and $\varphi_i$, and between
$\hat{\theta}_i$ and $\varphi(\sigma_i)$, which is the $\sigma_i$th component
of the permutation of $\vphivec$ according to permutation
vector $\sigvec$.  The first difference is a direct
result of the definition of the components of $\tgrid$
in \eref{eq:grid_point_upper}.  The second difference
is the reason we need to omit negligible permutations
of $\widehat{\psi\left ( \theta \right )}$
from the analysis.
If we do not omit such permutations,
we will be unable to bound this difference.
Lemma~\ref{lemma_negligible_permutations}, which is presented next,
demonstrates that if the distance of components of a permutation
of $\widehat{\psi\left ( \theta \right )}$ from the respective
non-permuted components
of $\pvece$ is too large, then the contribution of
the conditional probability of this permutation
to the probability of the pattern of $X^n$ in \eref{eq:pattern_prob1}
will be negligible.  A corollary to the
lemma
(which is shown in \ref{ap:lemma_q_proof} as part of the proof
of Lemma~\ref{lemma_quantization_ratio})
will give us a bound on the absolute difference
between components of a non-negligible permuted version of
$\widehat{\psi\left ( \theta \right )}$ and
those of $\pvece$, which, in turn, will lead
to a bound on the desired difference
between $\hat{\theta}_i$ and $\varphi(\sigma_i)$.

We begin by showing that there are permutations of the
pattern ML estimator that contribute negligibly to the pattern
probability.
The following lemma can be used to demonstrate that.  The lemma
is stated more generally.
\begin{lemma}
 \label{lemma_negligible_permutations}
 Let $n \rightarrow \infty$.
 Let $\pvece$ be the standard i.i.d.\ ML estimator with $k$ non-zero
 components of the
 probability vector that governs $X^n$.  Let
 $\phivec \dfn \left ( \phi_1, \phi_2, \ldots, \phi_k \right )$
 be another $k$-dimensional probability vector.  Define
 \be
  \label{eq:delta_lemma_def}
  \delta_i \dfn \hat{\theta}_i - \phi_i, ~~i = 1, 2, \ldots, k.
 \ee
 Assume that there exists a set $J$ of at least $j$ indices
 $i \in J$, $1 \leq i \leq k$,
 such that
 \be
  \label{eq:neg_lemma_bound}
  \left | \delta_i \right | \geq
  \left \{
  \begin{array}{ll}
   \frac{k}{j} \cdot \frac{\sqrt{\hat{\theta}_i}}{\sqrt{n}^{1 - \varepsilon/4}}; &
   \mbox{if}~ \phi_i > 2\hat{\theta}_i, \\
   \sqrt{\frac{k}{j}} \cdot \frac{\sqrt{\hat{\theta}_i}}{\sqrt{n}^{1 - \varepsilon/4}}; &
   \mbox{if}~ \phi_i \leq 2\hat{\theta}_i.
  \end{array}
  \right .
 \ee
 Then, as $n\rightarrow \infty$,
 \be
  \frac{k! P_{\phi} \left (X^n \right )}
  {P_{\hat{\theta}} \left (X^n \right )} \rightarrow 0.
 \ee
\end{lemma}
Lemma~\ref{lemma_negligible_permutations} shows that if there are too many
components of a vector $\phivec$ that are far from those of
$\pvece$, then even if we multiply
the probability of $X^n$ given $\phivec$ by $k!$ it still remains negligible w.r.t.\
the ML probability of $X^n$.  The lemma shows that this is true
for large distance with few components, as well as smaller distance with more
components.
Lemma~\ref{lemma_negligible_permutations}
is proved in \ref{ap:lemma_neglig_proof}.

For the sake of simple notation, let
$\avec \dfn \widehat{\avec \left ( \pvec \right )}$
denote the pattern ML probability parameter vector from this point
on to the end of the proof of the redundancy of the code for the first region.  (This is a
slight abuse of notation, but is much less tedious.)
Now, $\avec \left ( \sigvec \right )$ and
$\vphivec \left ( \sigvec \right )$ are the permutations
of $\avec$ and $\vphivec$, respectively, obtained by permutation vector
$\sigvec$.  Define the set $\eventA$ as the set of all permutation vectors
$\sigvec$, for which $\phivec \dfn \avec (\sigvec)$ satisfies
the condition in Lemma~\ref{lemma_negligible_permutations}
w.r.t.\ $\pvece$.
Note that Lemma~\ref{lemma_negligible_permutations} also implies
that $\avec$ cannot satisfy its conditions w.r.t.\ $\pvece$.
Then,
given that for every $\avec(\sigvec) \not \in \eventA$, we obtain
$P_{\psi \left (\sigma \right )} \left ( X^n \right ) /
P_{\varphi \left (\sigma \right )} \left ( X^n \right )
\leq \alpha$, for some expression $\alpha$,
the normalized contribution of the quantization of
$\avec$ to the redundancy for every $x^n$ with
$\hat{k}= k \leq \sqrt{n}^{1-\varepsilon}$ observed
symbols can be upper bounded by
\bea
 \nonumber
 \frac{1}{n} \log
 \frac{P_{\widehat{\psi(\theta)}} \left [ \Psi \left ( x^n \right ) \right ]}
 {P_{\varphi} \left [ \Psi \left ( x^n \right ) \right ]}
 &=&
 \frac{1}{n} \log
 \frac{\sum_{\sigvec} P_{\psi(\sigma)} \left ( x^n \right ) }
 {\sum_{\sigvec} P_{\varphi(\sigma)} \left ( x^n \right )} \\
 \nonumber
 &=&
 \frac{1}{n} \log
 \frac{\sum_{\sigvec:\avec(\sigvec) \not \in \eventA}
 P_{\psi(\sigma)} \left ( x^n \right )
 +
\sum_{\sigvec:\avec(\sigvec) \in \eventA}
 P_{\psi(\sigma)} \left ( x^n \right )}
 {\sum_{\sigvec:\avec(\sigvec) \not \in \eventA}
 P_{\varphi(\sigma)} \left ( x^n \right )
 +
\sum_{\sigvec:\avec(\sigvec) \in \eventA}
 P_{\varphi(\sigma)} \left ( x^n \right )} \\
 \nonumber
 &\leq&
 \frac{1}{n} \log
 \frac{ \left ( 1 + \varepsilon_2 \right )
 \sum_{\sigvec:\avec(\sigvec) \not \in \eventA}
 P_{\psi(\sigma)} \left ( x^n \right )}
 {
 \sum_{\sigvec:\avec(\sigvec) \not \in \eventA}
 P_{\varphi(\sigma)} \left ( x^n \right )
 } \\
 &\leq&
 \label{eq:pattern_quantize1}
 \frac{1}{n} \log
 \frac{ \left ( 1 + \varepsilon_2 \right )
 \sum_{\sigvec:\avec(\sigvec) \not \in \eventA}
 \alpha P_{\varphi(\sigma)} \left ( x^n \right )}
 {
 \sum_{\sigvec:\avec(\sigvec) \not \in \eventA}
 P_{\varphi(\sigma)} \left ( x^n \right )
 }
 ~\leq~
 \frac{\varepsilon_2 \log e}{n} + \frac{\log \alpha}{n}.
\eea
The first inequality is obtained from
Lemma~\ref{lemma_negligible_permutations}, using a fixed arbitrarily
small $\varepsilon_2 > 0$, and also by decreasing the denominator.
The last inequality is obtained since $\ln \left (1 + x \right ) \leq x$ for
every $x > -1$.  To complete the bound, we need to find $\alpha$.
This is done in the following lemma.
\begin{lemma}
 \label{lemma_quantization_ratio}
 Let $\avec$ be the $k$-dimensional pattern ML estimator obtained from $X^n$ for
 $k \leq \sqrt{n}^{1-\varepsilon}$, let
 $\vphivec$ be its quantized version, and
 let $\sigvec$ be a permutation vector such that
 $\avec(\sigvec) \not \in \eventA$.  Then,
 \be
  \label{eq:quantization_log_ratio}
  \log \frac{P_{\psi(\sigma)} \left ( X^n \right )}
  {P_{\varphi(\sigma)} \left ( X^n \right )} \leq
  \frac{c k \ln k}{n^{\varepsilon/4}},
 \ee
 where $c$ is a constant.
\end{lemma}
The proof of Lemma~\ref{lemma_quantization_ratio} is
in \ref{ap:lemma_q_proof}.  We can plug \eref{eq:quantization_log_ratio}
in \eref{eq:pattern_quantize1} for a particular $x^n$ to show that
\be
 \label{eq:quantize_cost2}
 \frac{1}{n} \log
 \frac{P_{\widehat{\psi(\theta)}} \left [ \Psi \left ( x^n \right ) \right ]}
 {P_{\varphi} \left [ \Psi \left ( x^n \right ) \right ]}
 \leq
 \frac{\varepsilon_2 \log e}{n} + \frac{c k \ln k}{n^{1+\varepsilon/4}} =
 o \left (\frac{k}{n} \right ),
\ee
and hence the quantization cost is negligible w.r.t.\ the
cost of representing $\vphivec$ in \eref{eq:varphi_cost}.
Plugging the bounds of \eref{eq:varphi_cost} and \eref{eq:quantize_cost2}
in \eref{eq:ub_proof3_1}, absorbing all low order terms in the leading $\varepsilon$,
normalizing by $n$, we obtain the upper bound of the first
region of \eref{eq:upper_bounds_patterns}, thus concluding the proof
of Theorem~\ref{theorem_upper_bounds}.
{\hfill $\Box$}

\section{Low Complexity Sequential Schemes and Pattern Entropy}
\label{sec:seq}

We now present two sub-optimal low-complexity
sequential algorithms for compressing patterns.
We are interested
in analyzing the performance for various alphabet sizes, and in the
total description length of a pattern,
which can be obtained by adding the modified redundancy we obtain here
to the i.i.d.\
entropy as in \eref{eq:pattern_description_length}.  The results in this section provide
bounds on the universal description length for coding patterns.  A very interesting
corollary is that for sufficiently large alphabets, the universal description length of
patterns is smaller than the i.i.d.\ entropy.  This points out to an interesting phenomenon
where the pattern entropy must decrease from the i.i.d.\ one for sufficiently
large alphabets.  Subsequently to the work reported in this paper, pattern entropy
and entropy rate
have been extensively studied, first in \cite{gil111}, and later
in \cite{gemelos04}-\cite{gemelos05isit}, \cite{orlitsky04itw}-\cite{orlitsky05isit},
\cite{gil11}, \cite{shamir_allerton04}-\cite{shamir_itw05}.

\subsection{Known Alphabet Size and A Mixture Code}

Let us first assume that although the alphabet $\Sigma$ itself is unknown,
its size $k$ is known.  For coding i.i.d.\ sequences, Krichevsky and Trofimov
\cite{krichevsky} demonstrated that the \emph{minimum description length\/}
(MDL) for i.i.d.\ sequences
\cite{rissa}, \cite{gil9} can be sequentially achieved using sequential
probability assignment, which when combined with arithmetic coding
\cite{arit1} results in an optimal sequential code.
In particular, they defined the probability
$Q_{KT} \left ( x^n \right )$
which is sequentially assigned to the sequence $x^n$ as
\be
 \label{eq:kt1}
 Q_{KT} \left ( x^n \right ) =
 \prod_{i=1}^n Q_{KT} \left ( x_i ~|~ x^{i-1} \right ),
\ee
where $Q_{KT} \left (x_i ~|~ x^{i-1} \right )$ is a conditional probability
assigned to the $i$th symbol $x_i$, given the subsequence of all the
preceding symbols
$x^{i-1}$. It is defined as
\be
\label{eq:kt}
 Q_{KT} \left ( x_i ~|~ x^{i-1} \right ) \dfn
  \frac{n^{i-1} \left ( x_i \right ) + 1/2}
  {i - 1 + k/2},
\ee
where $n^{i-1} \left ( x_i \right )$ is the number of occurrences of
the symbol $x_i$ in the subsequence $x^{i-1}$.

We can adopt this approach for coding patterns if we know that $k$ symbols
occur in the sequence.  If a letter (or index) has already occurred, we can still
update the probability as in \eref{eq:kt}.  However, once a new symbol
occurs, i.e., $x_i$ is not contained in the subsequence $x^{i-1}$,
$\Psi \left ( x_i \right )$ will be determined as the
next available index, regardless of the actual value of $x_i$.
This means that the event that $\Psi \left ( x_i \right )$ will take
a new value not in $\Psi \left (x^{i-1} \right )$ should be assigned
the sum of the probabilities of all letters
$u \in \Sigma$ that have not yet occurred.  Hence, similarly to
\eref{eq:kt1}, $\Psi \left (x^n \right )$ will
be assigned probability
\be
 \label{eq:kt1_index}
 Q_k \left [ \Psi \left ( x^n \right ) \right ] =
 \prod_{i=1}^n Q_k \left [ \Psi \left ( x_i \right ) ~|~ x^{i-1} \right ],
\ee
where
\be
\label{eq:kt_index}
 Q_k \left [ \Psi \left ( x_i \right ) ~|~ x^{i-1} \right ] =
 \left \{
 \begin{array}{ll}
  \frac{n^{i-1} \left ( x_i \right ) + 1/2}
  {i - 1 + k/2}, & \mbox{if~} n^{i-1} \left ( x_i \right ) > 0, \\
  \frac{\left ( k - C_{i-1} \right ) /2 }{i - 1 + k/2}, &
  \mbox{otherwise},
 \end{array}
 \right .
\ee
where $C_{i-1}$ is the number of distinct letters that occurred in the
subsequence $x^{i-1}$.
Theorem~\ref{theorem_seq1} summarizes the
performance of the
probability assignment in \eref{eq:kt1_index}-\eref{eq:kt_index}.

\begin{theorem}
 \label{theorem_seq1}
 Let $n \rightarrow \infty$.  Then,
 the individual modified redundancy of the
 probability assignment in \eref{eq:kt1_index}-\eref{eq:kt_index} is
 upper bounded by
 \be
  \label{eq:red_up_bound1}
  \tilde{R}_n \left [ Q_k, \Psi \left ( x^n \right ) \right ] \leq
  \frac{k}{2n} \log \frac{n}{k^3} +
  \left ( \frac{19}{12} \log e \right ) \frac{k}{n} -
  \frac{1}{2n} \log n + O \left ( \frac{k^2}{n^2} \right ).
 \ee
 for every pattern
 $\Psi \left (x^n \right )$ of a sequence $x^n$ with $k$ distinct
 indices and
 for every $k \leq n$.
\end{theorem}
The proof of Theorem~\ref{theorem_seq1} is purely technical relying on
Stirling's approximation and is
presented in \ref{ap:proof_theorem_seq1}.
From the proof, we can see that the last term is
$k^2 (\log e)/\left ( 4n^2 \right )$, which is always negligible w.r.t.\
the sum of all other terms.
We can also notice from the proof that if
almost all letters in $x^n$ (except $o(k)$) occur more than a fixed
number of occurrences, \eref{eq:red_up_bound1} reduces to
\be
 \label{eq:red_up_bound11}
 \tilde{R}_n \left [ Q_k, \Psi \left ( x^n \right ) \right ] \leq
 \frac{k}{2n} \log \frac{n}{k^3} +
 \left ( 1.5 \log e \right ) \frac{k}{n} -
 \frac{1}{2n} \log n + O \left ( \frac{k^2}{n^2} \right ),
\ee
for every $k$ (i.e., the second term is slightly smaller).
However, if there are $O(k)$ letters that occur only one time,
we must include the term of at most
$k (\log e)/12$, obtained from the upper bound of
Stirling's approximation.  Worst sequence case bounds
on the individual true redundancy can be easily obtained from
Theorem~\ref{theorem_seq1}.
If the alphabet size is limited to $k$,
the pattern probability of the worst sequence
will be at most $k!$ times its i.i.d.\ ML probability.  Hence, the redundancy will
increase by $\log (k!)$, yielding the same redundancy as that of the i.i.d.\
case of $0.5 (k - 1) \log (n/k)$ bits, which diminishes for $k = o(n)$.

The expression in \eref{eq:red_up_bound11}
attains a maximum for $k = n^{1/3}$ (neglecting the last two
terms).  The maximum of
\eref{eq:red_up_bound11} meets the performance of the minimax code
in \eref{eq:minimax_modified_red}.
In fact, a minimax code that does not distinguish between
different $k$'s adopts the
worst case performance of $k = n^{1/3}$ for every value of $k$.
For $k > e \cdot n^{1/3}$, $\tilde{R}_n \left [ Q_k, \Psi \left ( x^n \right ) \right ]$
in \eref{eq:red_up_bound11}
becomes negative.
(This is also true for \eref{eq:red_up_bound1} for $k > e^{19/18} \cdot n^{1/3}$.)
This means that the number of bits required to code the pattern
is smaller than the negative logarithm of the ML i.i.d.\ probability of
$x^n$, and that the pattern
entropy is much smaller than that of i.i.d.\ sequences
for large $k$'s.  This result cannot be observed from
the lower and upper bounds of the previous sections because
they refer to the
true redundancy w.r.t.\ the pattern entropy.  Further study of pattern entropy
\cite{gil11} extensively
characterized the behavior of the pattern entropy for different alphabet
sizes and arrangements of the letter probabilities in $\pvec$.

The main drawback of the code above is that it requires knowledge of $k$.
A ``semi-sequential'' two pass code that identifies $k$ during the first pass
can be used to achieve almost similar performance with additional
$(1 + \varepsilon ) \log k$ bits to inform the decoder of $k$.
Elias's coding of the integers \cite{elias} can be used to first
encode $k$, and then the scheme of \eref{eq:kt1_index}-\eref{eq:kt_index}
is used to code the pattern.
To avoid the use of a two pass
code, one can perform a mixture over all possible values $j$ of $k$.  This
can be done by assigning at every $i$; $1 \leq i \leq n$,
\be
 \label{eq:kt_mix}
 Q \left [ \Psi \left ( x^i \right ) \right ] \dfn
 \frac{1}{n-1} \sum_{j=2}^n
 \tilde{Q}_j \left [ \Psi \left ( x^i \right ) \right ],
\ee
where $\tilde{Q}_j \left [\Psi \left ( x^n \right ) \right ]$ is defined by
\be
\label{eq:kt_mix1}
 \tilde{Q}_j \left [ \Psi \left ( x_i \right ) ~|~ x^{i-1} \right ] \dfn
 \left \{
 \begin{array}{ll}
  \frac{n^{i-1} \left ( x_i \right ) + 1/2}
  {i - 1 + j/2}, & \mbox{if}~j > C_{i-1}~\mbox{and}~
  n^{i-1} \left ( x_i \right ) > 0, \\
  \frac{\left ( j - C_{i-1} \right ) /2 }
  { i - 1 + j/2}, &
  \mbox{if~} j > C_{i-1}~\mbox{and}~
  n^{i-1} \left ( x_i \right ) = 0, \\
  \frac{1}{C_{i-1} + 1}, & \mbox{if}~ j \leq C_{i-1},
 \end{array}
 \right .
\ee
i.e., as long as the number of distinct occurring letters does not exceed $j-1$,
$\tilde{Q}_j \left [\Psi \left ( x^i \right ) \right ]$ is equal
to $Q_j \left [\Psi \left ( x^i \right ) \right ]$.  Otherwise,
$\tilde{Q}_j \left [\Psi \left ( x^i \right ) \right ]$ assigns equal probability to
all existing indices and also to the innovation index.  Then,
$Q \left [ \Psi \left ( x^i \right ) \right ]$ is averaged over
$\tilde{Q}_j \left [\Psi \left ( x^i \right ) \right ]$.
The assigned probability satisfies for the actual $k$,
\be
 Q \left [ \Psi \left ( x^n \right ) \right ]
 \geq \frac{1}{n}
 Q_{k+1} \left [ \Psi \left ( x^n \right ) \right ],
\ee
where we must consider index $k+1$ in case all $k$ symbols occur first earlier than at
time $n$.
This leads to modified redundancy of
\be
 \label{eq:red_up_bound2}
 \tilde{R}_n \left [ Q, \Psi \left ( x^n \right ) \right ] \leq
 \frac{k}{2n} \log \frac{n}{k^3} +
 \left ( \frac{19}{12} \log e \right ) \frac{k}{n} +
 \frac{1}{2n} \log \frac{n^2}{k^3} + O \left ( \frac{k^2}{n^2} \right ),
\ee
where the third term diverges from \eref{eq:red_up_bound1} because of the mixing
and the use of $k+1$ instead of $k$,
for this linear per-symbol complexity scheme.

\subsection{Unknown Alphabet Size}

The assignment described
requires extra manipulations or complexity for an unknown $k$.  In \cite{shtarkov1},
a more generalized form of \eref{eq:kt} was presented, in which
\be
\label{eq:gkt}
 Q_{GKT} \left ( x_i ~|~ x^{i-1} \right ) \dfn
 \left \{
 \begin{array}{ll}
  \frac{n^{i-1} \left ( x_i \right ) + \nu}
  {i - 1 + C_{i-1} \nu + \chi_{i-1} }, & \mbox{if~}
  n^{i-1} \left ( x_i \right ) > 0, \\
  \frac{\chi_{i-1}}{\left ( M - C_{i-1} \right )
  \left ( i - 1 + C_{i-1} \nu + \chi_{i-1} \right ) }, &
  \mbox{otherwise}
 \end{array}
 \right .,
\ee
where $\nu > 0$ is some constant, $\chi_{i - 1}$ is some function
of the subsequence $x^{i-1}$, and $M$ is a bound on the maximum number
of alphabet letters.
This extension of \eref{eq:kt} allows asymptotically
optimal performance for coding i.i.d.\
sequences.  This performance depends only on the
actual number $k$ of alphabet letters that occur, and not on the total alphabet
size $M$.

It turns out that with correct modification of \eref{eq:gkt}, one can sequentially
(with fixed per symbol complexity) asymptotically (with
$k \rightarrow \infty$) achieve the same performance of
\eref{eq:red_up_bound1}-\eref{eq:red_up_bound11} for patterns.  Let us consider
the code in which
\be
\label{eq:gkt_index}
 Q \left [ \Psi \left ( x_i \right ) ~|~ x^{i-1} \right ] \dfn
  \left \{
  \begin{array}{ll}
   \frac{n^{i-1} \left ( x_i \right ) + 1/2}
   {i - 1 + C_{i-1}/2 + \left ( C_{i-1} + 1 \right )^{1 - \varepsilon}/2}, &
   \mbox{if~} n^{i-1} \left ( x_i \right ) > 0, \\
   \frac{\left ( C_{i-1} + 1 \right )^{1 - \varepsilon}/2}
   {i - 1 + C_{i-1}/2 + \left ( C_{i-1} + 1 \right )^{1 - \varepsilon}/2}, &
   \mbox{otherwise},
  \end{array}
  \right .
\ee
where $\varepsilon > 0$ can be chosen arbitrarily small.
Theorem~\ref{theorem_seq2} summarizes the performance of this code.
\begin{theorem}
 \label{theorem_seq2}
 Let $n \rightarrow \infty$.  Then,
 the individual modified redundancy of the
 probability assignment in \eref{eq:gkt_index} is
 upper bounded by
\bea
 \nonumber
 \tilde{R}_n \left [ Q, \Psi \left ( x^n \right ) \right ] &\leq&
 \frac{k}{2n} \log \frac{n}{k^3} +
 \left ( \frac{19}{12} - \varepsilon \right ) \left ( \log e \right ) \frac{k}{n} -
 \frac{1}{2n} \log n + \\
 \label{eq:red_up_bound3}
 & &
 \frac{k^{1-\varepsilon}}{2n} \log \frac{2n}{k} +
 \frac{\varepsilon k \log k}{n}  +
% \frac{k^{1-\varepsilon}}{2n} +
% \frac{\varepsilon}{2n} \log k +
 O \left ( \frac{k^2}{n^2} \right ),
\eea
for every pattern
$\Psi \left (x^n \right )$ of a sequence $x^n$ with $k$ distinct
indices and for every $k \leq n$.
\end{theorem}
The proof of Theorem~\ref{theorem_seq2}, again, relies on
Stirling's approximation.  It is presented in
\ref{ap:proof_theorem_seq2}.
The bound in \eref{eq:red_up_bound3} is shown such that the first row
contains the terms (up to $\varepsilon$) identical to the upper bound
in \eref{eq:red_up_bound1}.  The second row contains the additional
terms that increase the bound due to the reduced complexity.
If $k \rightarrow \infty$ and $\varepsilon$ is arbitrarily small,
the bound in \eref{eq:red_up_bound3}
asymptotically meets
the modified redundancy upper bound of \eref{eq:red_up_bound1}, even
if $k$ goes to infinity at a slower rate than $n$.
However, for smaller $k$'s, the two terms in the bottom row increase the
redundancy, and if $k > n^{1/3}$, work against
the dominant negative first term.
In practice, $k$ may be too small, and the gap between
the redundancy of the first scheme in \eref{eq:red_up_bound1} and that
of the second scheme in \eref{eq:red_up_bound3} will be noticed.
A mixture of the assigned probability of the new scheme
and of those for known small $k$'s in
\eref{eq:kt1_index}-\eref{eq:kt_index} can be used to achieve the
performance of \eref{eq:red_up_bound1} for every $k$.

Under the assumption leading to \eref{eq:red_up_bound11},
the worst case modified redundancy of the first scheme
is obtained when $k = n^{1/3}$, where the extra number of
bits required beyond $-\log P_{ML} \left ( x^n \right )$ is linear in $k$.
The additional terms of the redundancy of the second scheme shift the maximum
redundancy to a larger value of $k$, yielding larger redundancy.
For example, if $k \rightarrow \infty$, a value of $\varepsilon = 0.1$ will
attain the worst $k$ (under the assumption leading
to \eref{eq:red_up_bound11})
at $k \approx n^{0.5/(1.5 - \varepsilon)} = n^{0.5/1.4}
\approx n^{0.357}$, which is larger
than $n^{1/3}$. If $n$ is not as large, the maximal redundancy will be
attained for finite $k$'s, and will increase w.r.t.\ $k$.
For example, if $n = 10^6$, and $\varepsilon = 0.1$,
the worst case $k$ is slightly above $k = 400$, which
is approximately $n^{0.44}$.
Figure~\ref{fig:sim} shows the un-normalized
modified redundancy bounds (in bits) of both schemes
(using a second order term of $1.5 (\log e)k/n$, and for the first
scheme with a known $k$),
as well as the
individual modified redundancies obtained using the two proposed schemes for
patterns of actual sequences $x^n$.
The results are shown for
$n = 10^6$, for alphabets of sizes $k = 2$ to $k = 1000$, and
for $\varepsilon = 0.1$ in the second scheme.
For the second scheme, the results are also shown for the worst possible sequence, i.e.,
the one that is used to obtain the bound of Theorem~\ref{theorem_seq2}, in which
all the $k$ letters occur in the first $k$ symbols of $x^n$.
The figure shows that the bound in \eref{eq:red_up_bound11} is tight.
As expected, the first scheme performs better than the second.
The bound of \eref{eq:red_up_bound3}
is loose because the proof of Theorem~\ref{theorem_seq2} makes a
loose assumption in order to use Stirling's bounds.  The algorithm is
thus much better than the bound in \eref{eq:red_up_bound3}.
Since the performance is for an individual sequence, the simulation curve for the second
scheme is rather noisy.  The reason is that the behavior varies depending on where
in the sequence first occurrences are.  Since each point is for a different individual
sequence, the locations of the first occurrences vary.
Figure~\ref{fig:sim} also
verifies the worst values of $k$ mentioned above.
\bef
 \centerline{\includegraphics[bbllx=35pt,bblly=190pt,
  bburx=600pt,bbury=620pt,
  height=8.4cm,clip=]{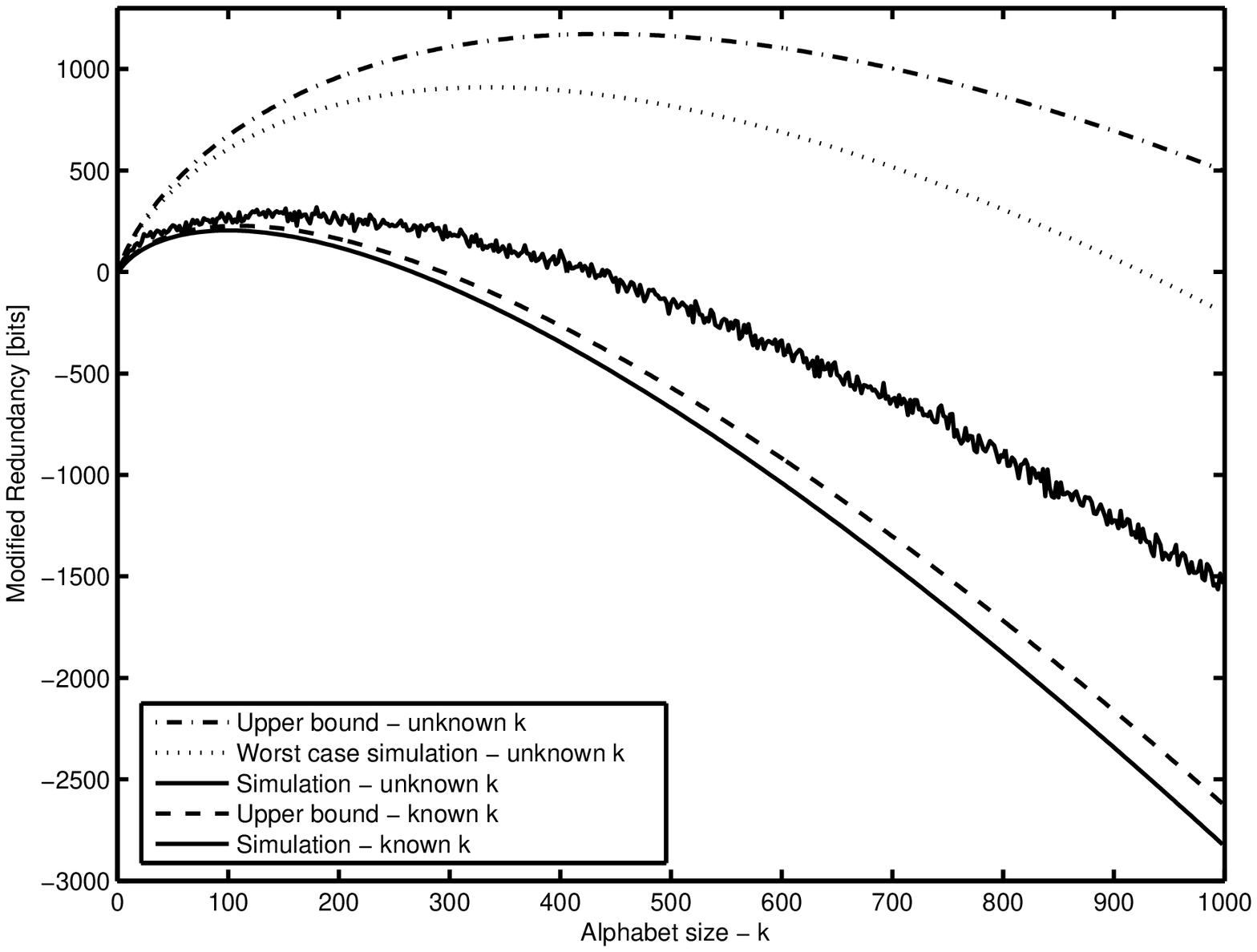}}
  \caption{Bounds and simulation results for the individual
  modified pattern redundancies of two sequential schemes with $n = 10^6$ and
  $\varepsilon = 0.1$}
  \label{fig:sim}
\enf

For a given sequence $x^n$,
the schemes described in this section assign probability based on
only a single permutation of $\pvece$.
However, a pattern probability can be expressed as a sum of all permutations
of its ML estimator.
Naturally, if the probabilities of
all $k!$ permutations
are included in the assigned probability,
better redundancy can be obtained.
Subsequently to the derivation of the schemes described here,
a class of computationally more demanding schemes
that accounts to many such permutations was obtained and described in
\cite{orlitsky2}-\cite{orlitsky4}.
%The algorithms proposed in this section can also be extended to belong to
%such a class of more computationally demanding schemes, whose true pattern
%redundancy is better in some cases.
Unlike those schemes, the methods proposed here can more easily be integrated into
efficient low-complexity implementations of adaptive arithmetic coding
(see, e.g., \cite{ryabko}).

\subsection{Pattern and I.I.D.\ Entropies}
\label{sec:entropy}

The description length in \eref{eq:red_up_bound1} can lead to an upper
bound on the pattern entropy in terms of the i.i.d.\ one.  In particular,
it follows from \eref{eq:red_up_bound1}, that if for an arbitrarily small
$\varepsilon > 0$, the probability that at least $k'$ distinct symbols will occur
in a sequence of length $n$ is at least $1 - \varepsilon$, then
\be
 \label{eq:entropy_bound}
 \frac{1}{n} H_{\theta} \left [ \Psi \left ( X^n \right ) \right ]
 \leq \left \{ \begin{array}{ll}
 H_{\theta} \left ( X \right ); & \mbox{if}~k' \leq e^{19/18} \cdot n^{1/3}, \\
 H_{\theta} \left ( X \right ) -
 \left ( 1 - \varepsilon \right )
 \frac{3}{2} \frac{k'}{n} \log \frac{k'}{e^{19/18} n^{1/3}} +
 O \left ( \frac{k'^2 + n\log n}{n^2} \right ); & \mbox{if}~ k' > e^{19/18} \cdot n^{1/3}.
 \end{array} \right .
\ee
The last equation can be shown by bounding the entropy by the average
description length of the following probability assignment code.
The code
assigns the pattern $\Psi \left (x^n \right )$ a codeword of length
$-\log P_{\theta} \left [ \Psi \left ( x^n \right ) \right ] \leq
-\log P_{\theta} \left ( x^n \right )$ bits if less than $k'$ indices occur
in $\Psi \left ( x^n \right )$.  Otherwise, it uses the code leading to
\eref{eq:red_up_bound1}.  One negligible bit is required to distinguish
between the two cases, and at most $O(\log n)$ bits are needed to
inform the decoder of the actual number of indices if greater than or equal
$k'$.
Hence, for a large $k'$, the pattern entropy is significantly smaller than the
i.i.d.\ one.
In fact,
from \eref{eq:entropy_bound} and the bounds derived in this paper,
we observe that not only does the pattern entropy decrease significantly
from the i.i.d.\ one, but also the true
pattern redundancy becomes negligible
compared to this decrease.
The extensive study of the pattern entropy has been the subject of several subsequent
works, first in \cite{gil111}, and later
in \cite{gemelos04}-\cite{gemelos05isit}, \cite{orlitsky04itw}-\cite{orlitsky05isit},
\cite{gil11}, \cite{shamir_allerton04}-\cite{shamir_itw05}.

\section{Discussion}
\label{sec:discuss}

This paper considered the average case of the problem of
universal pattern compression.
Both lower and upper bounds on the redundancies of codes for this
problem were obtained.  However, a gap still exists between the lower
bounds and the attainable upper bounds.  While we considered the
average problem, other work \cite{aberg}, \cite{orlitsky},
\cite{orlitsky1}-\cite{orlitsky4o} considered the individual sequence
case, using different techniques based on combinatorics.  Although the
aim and the techniques in those independent works were different, similar
qualitative results were obtained, and in particular the same gap
was shown to exist between the orders of magnitudes of the lower and the upper bounds.

Future work should try to bridge the two sets of bounds.  In
our work, it is clear that there is room for improvement and
tightening both lower and upper bounds.  The minimax lower bounds
derived in Section~\ref{sec:maximin} are not tight because we
decreased the grid size dividing by $k!$, which eliminated many
permutations of $\avec \left ( \pvec \right )$, more than once each.
In Section~\ref{sec:lb_most}, the assumption that complete spheres
are contained in the pattern space, and the division by their
complete volume to find the number of spheres packed in the space
resulted in a possibly loose bound for most sources.  It may be possible
to use techniques from combinatorics to tighten the bounds.  The question
is whether such techniques will improve the first order asymptotics or not.

On the other hand, the quantization approach of the upper bound in the
first region may be useful also for the second region with larger $k$'s.
The derivation of the
upper bound of the second region does not quantize the estimators of the
probability parameters,
possibly leading to a loose bound.  One can show that quantization of the ML
parameters into the vector $\vphivec$ whose components are on the
grid points of $\tgrid$
defined in \eref{eq:grid_point_upper} can result in representation cost of
$O\left (n^{(1+\varepsilon)/3} \right )$ even for large $k$'s.  More precisely, let $\beta$ be some
partitioning index
in the grid $\tgrid$.  Consider representing the quantized
ML pattern probability parameters of $\vphivec$ as
follows:  For each of
the first $\beta$ grid points in $\tau$ use up to $(1+\varepsilon) \log k$ bits to represent
how many letters have probabilities quantized in $\vphivec$ to this grid point.  In
the remaining grid points (bounded by $B = \sqrt{n}^{1+\varepsilon}$) there
are at most $n^{1+\varepsilon}/\beta^2$ quantized probability parameters in the
components of $\vphivec$
(see, e.g., \eref{eq:grid_point_upper}-\eref{eq:upper_grid_size}).  For each of
these components of $\vphivec$, one can use up to $(1 +\varepsilon) \log B$ bits to represent the
index in $\tgrid$ of the point that equals this component.
This results in a total representation
cost upper bounded by
\be
 \label{eq:discussion_upper_cost}
 \left ( 1 + \delta \right )\beta \log k +
 \left ( 1 + \delta \right ) \frac{n^{1+\varepsilon}}{2\beta^2} \log n
\ee
for some fixed $\delta > \varepsilon$.  Differentiating w.r.t.\ $\beta$ to find the
value of the partition point $\beta$ of $\tgrid$ that yields the minimum of the expression
above yields a bound on the representation cost of
\be
 \label{eq:small_code1_cost}
 \left ( 1+\delta \right ) 1.5
 \cdot n^{\frac{1}{3} (1 + \varepsilon)} \cdot \left (\log n\right )^{1/3} \cdot
 \left ( \log k \right )^{2/3}
\ee
bits for coding $\vphivec$.  A more complex analysis takes the second cost above as the number
of choices of at most $n^{1+\varepsilon}/\beta^2$ elements out of at most
$\sqrt{n}^{1+\varepsilon}$ with repetitions allowed.  Asymptotically, it yields the
bound of \eref{eq:small_code1_cost} divided by a factor of $3^{1/3}$.  Considering the
bound only for $k \geq n^{1/3}$, this stronger bound can be bounded by
$\left ( 1+\delta \right ) 1.5 \cdot n^{\frac{1}{3} (1 + \varepsilon)} \cdot \log k$.

From \eref{eq:small_code1_cost} we know that the representation of a quantized
version of the pattern ML estimator whose components are quantized as proposed in
Section~\ref{sec:upper_bounds} can cost
$O \left ( n^{(1+\varepsilon)/3} \right )$ bits for every $k$, including
all large values of $k$ up to $k = n$.
However, the quantization cost, in this case, increases and becomes
of $O \left (k/n^{1+\varepsilon} \right )$ bits per symbol.
While subsequent work
in \cite{shamir_ciss04} has already
improved the upper bound by trading off between
the two costs, a gap to the lower bound still remains.
Therefore,
further research should explore this direction more in order to attempt to reduce the
upper bound that consists of these two components even more.
%Assume that we include several
%permutations of the pattern ML estimator
%in one set of permutations.  Then, it may be possible
%to attain tighter bounds, if the ratio between
%the sum of
%the probabilities of the pattern given by the permutations
%in the set and the sum of the quantized
%versions of these probabilities is bounded instead of
%bounding this ratio for one permutation at a time.

\section{Summary and Conclusions}
\label{sec:summary}

We studied the average universal coding problem of
patterns of sequences generated by i.i.d.\ sources.  Lower bounds
on the average minimax redundancy and the redundancy for most sources
were obtained, as well as upper bounds obtained for specific codes.
It was shown that for essentially small
alphabet sizes, the redundancy cost in coding
patterns is between $0.5 \log (n/k^3)$ and $0.5 \log (n / k^2)$ bits
per each unknown probability parameter in all average senses.
For essentially large alphabets, this cost is
between $O \left ( n^{1/3} \right )$ and $O \left ( n^{1/2} \right )$
bits overall.  These redundancies are better than those attained in standard i.i.d.\
sequence compression.  In particular, for large $k$'s where
universal compression with vanishing redundancy is impossible in the i.i.d.\
case, here it has vanishing redundancy.  The gain over i.i.d.\ compression
increases with $k$, since for large $k$'s, a fixed cost is maintained,
regardless of the value of $k$.  This gain
is reflected even more in the existence of universal
pattern codes whose pattern universal description length is smaller
than the i.i.d.\ non-universal MDL of the underlying sequence if the
alphabet is large enough.  This
implies a decrease of the pattern entropy w.r.t.\ the underlying i.i.d.\ one.
%In addition to the reduced redundancy,
%it was shown that the gain over i.i.d.\ compression is reflected even more in the
%pattern entropy.  In fact, as $k$ increases this gain is the significant one,
%where the coding redundancy becomes the third order term of the description
%length.
This overall gain, of course, does not come for free, and the
cost is embedded in coding the unknown alphabet characters before the
patterns are obtained.  Two low-complexity sub-optimal sequential
algorithms were presented and were used to demonstrate the gain in
coding patterns over the
i.i.d.\ case.
Future work should attempt to bridge the gap between
the upper and lower bounds.  Also, the results on the i.i.d.\
problem can serve as a basis for further research on pattern compression
for patterns induced by non-memoryless sources.

\appendix
\renewcommand{\thesection}{Appendix \Alph{section}}
\renewcommand{\theequation}{\thesection.\arabic{equation}}

\section{--~~ Proof of Lemma \ref{lemma_integer_vectors}}
\label{ap:lemma_integer_vectors_proof}
\renewcommand{\theequation}{A.\arabic{equation}}
\renewcommand{\theproposition}{A.\arabic{proposition}}
\renewcommand{\thelemma}{A.\arabic{lemma}}
\setcounter{equation}{0}
\setcounter{lemma}{0}
\setcounter{proposition}{0}

The number of vectors $\bvec$ with integer components
that satisfy \eref{eq:ball_condition} is lower bounded by the number of
cubes with edge $1$ that are \emph{fully\/} contained in the positive
quadrant of the $k-1$ dimensional
sphere centered at the origin with radius $\sqrt{n}^{1-\varepsilon}$.  (This is a lower
bound, because there are, in fact, more vectors with zero components than this number).  The
number of these cubes equals the total volume of these cubes.  However, since there can
exist cubes that are only partially contained in the sphere, the volume of the
sphere with radius $\sqrt{n}^{1-\varepsilon}$ cannot be used to bound the total volume
of these cubes.  Instead, we can subtract the longest diagonal of these cubes $\sqrt{k-1}$
from the radius of the original sphere, and use the new sphere with
radius $\sqrt{n}^{1-\varepsilon} - \sqrt{k-1}$ to bound the total volume of these
cubes, or even use a shorter radius $\sqrt{n}^{1-\varepsilon'} \leq
\sqrt{n}^{1-\varepsilon} - \sqrt{k-1}$ for this bound (this radius is shorter
by the assumption that $k \leq n^{1-2\varepsilon} \ll n^{1-\varepsilon}$).
The volume of the positive quadrant of this sphere is bounded in
\eref{eq:M_ball_bound}.  It is thus only left to show that all cubes that are only partially
contained (or are not contained)
in the sphere with radius $\sqrt{n}^{1-\varepsilon}$ are completely outside
the sphere with radius $\sqrt{n}^{1-\varepsilon} - \sqrt{k-1}$.  Hence, the volume of
this sphere is a lower bound on the total volume of cubes for which the farthest points from the
origin satisfy \eref{eq:ball_condition}, and thus a lower bound on the number
of integer components vectors $\bvec$ that satisfy this condition.

Let $\aavec \dfn \left (a_1, a_2, \ldots, a_{k-1} \right )$ be the farthest
point from the origin in an edge $1$
cube that is either partially in the sphere with radius
$\sqrt{n}^{1-\varepsilon}$ centered at the origin or not in the sphere.  By definition,
\be
\label{eq:sphere_proof1}
 \sum_{i=1}^{k-1} a_i^2 > n^{1-\varepsilon} ~\Rightarrow~
 \sqrt{\sum_{i=1}^{k-1} a_i^2} > \sqrt{n}^{1-\varepsilon} ~\Rightarrow~
 \frac{1}{\sqrt{n}^{1-\varepsilon}} > \frac{1}{\sqrt{\sum_{i=1}^{k-1} a_i^2 }}.
\ee
By Jensen's inequality on the function $x^2$,
\be
\label{eq:sphere_proof2}
 \sum_{i=1}^{k-1} a_i^2 \geq
 (k-1) \left ( \sum_{i=1}^{k-1} \frac{a_i}{k-1} \right )^2 =
 \frac{1}{k-1} \left ( \sum_{i=1}^{k-1} a_i \right )^2 ~\Rightarrow~
 \sum_{i=1}^{k-1} a_i \leq \sqrt{k-1} \sqrt{ \sum_{i=1}^{k-1} a_i^2}.
\ee
The nearest point to the origin of the cube considered is the point
$\left (a_1 - 1, a_2 -1, \ldots, a_{k-1} - 1 \right )$.  To prove that the cube
is completely outside the new sphere, we need to show that this nearest point to the
origin is outside this
sphere, i.e., that $\sum \left (a_i - 1 \right )^2 > \rho^2$ where
$\rho = \sqrt{n}^{1-\varepsilon} - \sqrt{k-1}$ is the radius
of the new sphere.  The difference between the two sides of this equation is
\bea
 \nonumber
 \lefteqn{
 \sum_{i=1}^{k-1} \left ( a_i - 1 \right )^2 -
 \left ( \sqrt{n}^{1-\varepsilon} - \sqrt{k-1} \right )^2 =
 \left ( \sum_{i=1}^{k-1} a_i^2 - n^{1-\varepsilon} \right ) -
 2 \left ( \sum_{i=1}^{k-1} a_i - \sqrt{k-1}\sqrt{n}^{1-\varepsilon} \right )} \\
 \nonumber
 &\geq&
 \sqrt{n}^{1-\varepsilon} \left ( \sqrt{\sum_{i=1}^{k-1} a_i^2} -\sqrt{n}^{1-\varepsilon} \right )
 - 2 \sqrt{k-1} \left ( \sqrt{\sum_{i=1}^{k-1} a_i^2} - \sqrt{n}^{1-\varepsilon} \right ) \\
 &=&
 \left ( \sqrt{n}^{1-\varepsilon} - 2\sqrt{k-1} \right )
 \left ( \sqrt{\sum_{i=1}^{k-1} a_i^2} - \sqrt{n}^{1-\varepsilon} \right ) > 0.
 \label{eq:sphere_proof3}
\eea
The first inequality is obtained by using
\eref{eq:sphere_proof1} for the first term and
\eref{eq:sphere_proof2} for the second.  The next inequality is
because the left term is positive as long as
$k < n^{1-\varepsilon}/4 + 1$, and the right term is positive by
\eref{eq:sphere_proof1}.  This proves that any point in a cube that is not completely
inside the sphere with radius $\sqrt{n}^{1-\varepsilon}$ must be outside the sphere we defined
with a smaller radius, and thus the volume of this sphere in the positive quadrant lower bounds
the number of nonnegative integer components vectors that satisfy \eref{eq:ball_condition}.
This concludes the proof of Lemma~\ref{lemma_integer_vectors}.
{\hfill $\Box$}

\section{--~~ Proof of Lemma \ref{lemma_event_A_prob}}
\label{ap:lemma_event_A_prob_proof}
\renewcommand{\theequation}{B.\arabic{equation}}
\renewcommand{\theproposition}{B.\arabic{proposition}}
\renewcommand{\thelemma}{B.\arabic{lemma}}
\setcounter{equation}{0}
\setcounter{lemma}{0}
\setcounter{proposition}{0}

Let the observed data sequence $X^n$ be generated with
distribution $P_{\theta} (x^n)$, where $\pvec \in \Agrid$. Let
$\pvece$ be the ML estimate of $\pvec$ from $X^n$.
To bound the probability of event $A$, we will use the union bound on events $A_i$.
Define
\be
 \label{eq:delta_i}
 \delta_i \dfn \hat{\theta}_i - \theta_i.
\ee
As defined in \eref{eq:Aidef}, event $A_i$ for $1 \leq i \leq k$ occurs if
$\left | \delta_i \right | \geq \Delta\left (\tau_{b_i} \right )/2$,
where $\Delta\left (\tau_{b_i} \right )$ is as defined in
\eref{eq:grid_spacing}.
Recall that for $i < k$, $\tau_{b_i} = \theta_i$.
For $i = k$, the constrained parameter $\theta_k$ may
not be on $\tgrid$ and we define $\tau_{b_k}$ as the nearest
point in $\tgrid$ to $\theta_k$ that is smaller than or equal to
$\theta_k$.  Note that in order to generate a bound that can be useful
for the distinguishability of patterns, we must bound $P_{\theta}(A)$, which
is greater than $P_{\theta} \left (\pvece_{\Omega} \neq \pvec \right )$. The latter
is sufficient for distinguishability in the i.i.d.\ case (see, e.g.,
\cite{gil9}).  In particular, we must include $A_k$ in the error event,
although we can use the assumption that
$\theta_k \geq \theta_i$, for all
$i; 1 \leq i \leq k-1$.  Hence, for all $i$, $\theta_i \geq
1/n^{1-\varepsilon}$ from the definition of the minimum grid point
in \eref{eq:grid_point}.

In the following lemma, we lower bound $\left |\delta_i \right |$
as a function of $\hat{\theta}_i$ given event $A_i$ occurred.
Following the lemma and its proof, we use this bound and the union
bound on the components of $\pvec$ to show that the overall
probability of $A$ vanishes.
\begin{lemma}
\label{lemma_delta} If event $A_i$ occurs,
then event
\be
 \label{eq:Bidef}
 B_i ~:~
 \left | \delta_i \right | = \left |
 \hat{\theta}_i - \theta_i
 \right | \geq
 \frac{\sqrt{\hat{\theta}_i}}{10 \sqrt{n}^{1 - \varepsilon}}
\ee
must occur as well.
\end{lemma}
Note that if $\hat{\theta}_i = 0$, Lemma~\ref{lemma_delta} holds
simply because the right hand side of \eref{eq:Bidef} is zero.

\noindent
{\bf Proof:}
First, extend the definition
of the function $\Delta \left ( \theta \right )$, defined in
\eref{eq:grid_spacing}, to every value of $\theta \geq 1/n^{1-\varepsilon}$,
\be
 \label{eq:ex_grid_spacing}
 \Delta \left ( \theta \right ) \dfn
 \frac{2 \left ( \sqrt{\theta} \sqrt{n}^{1-\varepsilon} - 0.5 \right )}
 {n^{1 - \varepsilon}} \geq
 \frac{\sqrt{\theta}}{\sqrt{n}^{1-\varepsilon}}.
\ee
The function $\Delta \left ( \theta \right )$ is
increasing in $\theta$.  Now, given $A_i$
occurred, consider two separate cases: (1)
$\hat{\theta}_i \leq \tau_{b_i + 2}$, and (2) $\hat{\theta}_i >
\tau_{b_i + 2}$.  For case (1),
\be
 \left | \delta_i \right | =
 \left | \hat{\theta}_i - \theta_i \right | \geq
 \frac{\Delta \left ( \tau_{b_i} \right )}{2} \geq
 \frac{\Delta \left ( \tau_{b_i + 2} \right )}{10} \geq
 \frac{\sqrt{\tau_{b_i+2}}}{10\sqrt{n}^{1-\varepsilon}}\geq
 \frac{\sqrt{\hat{\theta}_i}}{10 \sqrt{n}^{1 - \varepsilon}}.
\ee
To obtain the second inequality, we use the fact that
$\Delta \left (\tau_{b_i + 2} \right ) \leq 5 \Delta \left (
\tau_{b_i} \right )$, which can be shown by observing the case
$b_i = 1$.  Then, inequality \eref{eq:ex_grid_spacing} is used.
Finally, the definition of this case leads to the last inequality.
(Note that we need to consider
$\tau_{b_i + 2}$ instead of $\tau_{b_i + 1}$ only in case
$\theta_k$ is closer to $\tau_{b_k + 1}$ than to $\tau_{b_k}$,
resulting in $\hat{\theta}_k > \tau_{b_k + 1}$ still satisfying
the complement event to $A_k$.) For case (2), let $\hat{b}_i$ be
the index of the largest grid point still smaller than
$\hat{\theta}_i$, i.e., $\hat{\theta}_i > \tau_{\hat{b}_i}$. Then,
since there is more than one unit of grid spacing between
$\theta_i$ and $\hat{\theta}_i$,
\be
 \left | \delta_i \right | =
 \hat{\theta}_i - \theta_i >
 \Delta \left ( \tau_{\hat{b}_i} \right ) =
 \frac{2 \left (\hat{b}_i - \frac{1}{2} \right )}{n^{1-\varepsilon}} \geq
 \frac{\hat{b}_i + \frac{1}{2}}{n^{1-\varepsilon}} =
 \frac{\Delta\left (\tau_{\hat{b}_i+1} \right )}{2} \geq
 \frac{\Delta \left ( \hat{\theta}_i \right )}{2} \geq
 \frac{\sqrt{\hat{\theta}_i}}{2 \sqrt{n}^{1 - \varepsilon}}.
\ee
The second inequality is obtained since $\hat{b}_i \geq 2$.
This concludes the proof of Lemma~\ref{lemma_delta}. {\hfill
$\Box$}

Using Lemma~\ref{lemma_delta} and the union bound,
\be
\label{eq:minimax_UB}
 P_{\theta} \left ( A \right )
 \leq
 \sum_{i=1}^{k} P_{\theta} \left ( A_i \right )
 \leq
 \sum_{i=1}^{k} P_{\theta} \left ( B_i \right ),
\ee
and we need to bound
$P_{\theta} \left ( B_i \right )$.  Consider the Bernoulli
$n$-sequence $\Yvec_i$, whose $j$th symbol is defined by
\be
\label{eq:Y_transformation}
 Y_{ij} = \left \{ \begin{array}{ll}
 1,&\mbox{if}~X_j = i \\
 0,&\mbox{otherwise}
 \end{array} \right .,
\ee
where $X_j$ is the $j$th symbol of $X^n$, and we assume,
without loss of generality, that the $k$ alphabet letters are $1,
2, \ldots, k$.  Let $P_{\theta_i}\left ( \Yvec_i = y^n \right )$
be the probability that $\Yvec_i$ takes value $y^n \dfn \left
(y_1, y_2, \ldots, y_n \right )$, where the symbols $y_j$ can be
either $0$ or $1$.  Let $P_{\hat{\theta}_i}$ be the empirical
distribution of $y^n$ which was drawn by $P_{\theta_i}\left (
\Yvec_i \right )$, i.e., the Bernoulli probability mass function
induced by the ML estimator $\hat{\theta}_i$ of $\theta_i$ on the
random vector $\Yvec_i$.  For a given value of the random sequence
$X^n$, the sequence $\Yvec_i$, defined in
\eref{eq:Y_transformation}, will give the exact same ML estimator
$\hat{\theta}_i$ as the one obtained from $X^n$.  Therefore, by
typical sequences analysis (see \cite{cover}, \cite{cizar}),
\be
\label{eq:typical_B_i_bound}
 P_{\theta} \left ( B_i \right ) =
 P_{\theta_i} \left (B_i \right ) \leq
 n \cdot 2^{-n \cdot \min_{y^n \in B_i}
 D \left ( P_{\hat{\theta}_i} ~||~ P_{\theta_i} \right )},
\ee
where $D \left ( P_{\hat{\theta}_i} ~||~ P_{\theta_i} \right)$
is the \emph{divergence\/} (relative entropy) between the two
distributions, and the coefficient $n$ bounds the number of
possible different $n$-sequence types, for which event $B_i$
occurs.

We now need to lower bound $D \left ( P_{\hat{\theta}_i} ~||~
P_{\theta_i} \right )$ given event $B_i$ has occurred.  This is
done as follows: First, let us define the function
\be
 \label{eq:fdef}
 f\left (x \right ) \dfn \left \{
 \begin{array}{ll}
 \frac{x^2}{4}; & 0 \leq x \leq 1, \\
 \left ( 1 - \ln 2 \right ) x; & x > 1.
 \end{array}
 \right .
\ee
Using Taylor series expansions, it can be shown that
\bea
 \label{eq:log1plusx}
 -\log \left ( 1 + x \right ) &\geq& \left [ -x + f(x) \right ] \log e,~~\mbox{if}~
 x \geq 0 \\
 \label{eq:log1minusx}
 -\log \left ( 1 - x \right ) &\geq& \left ( x + \frac{x^2}{2} \right ) \log e,~~
 \mbox{if}~0\leq x < 1.
\eea
We will use these inequalities in the following derivations.
Given $B_i$ has occurred, for $\hat{\theta}_i > 0$,
\bea
 \nonumber
 D \left ( P_{\hat{\theta}_i} ~||~ P_{\theta_i} \right ) &=&
  \hat{\theta}_i \log \frac{\hat{\theta}_i}{\theta_i} +
  \left ( 1 - \hat{\theta}_i \right ) \log
  \frac{1 - \hat{\theta}_i}{1 - \theta_i} \\
 \nonumber
 &=& \hat{\theta}_i \log \frac{\hat{\theta}_i}{\hat{\theta}_i - \delta_i} +
 \left ( 1 - \hat{\theta}_i \right ) \log
 \frac{1 - \hat{\theta}_i}{1 - \hat{\theta}_i + \delta_i} \\
 \nonumber
 &=& -\hat{\theta}_i \log \left ( 1 - \frac{\delta_i}{\hat{\theta}_i} \right ) -
 \left ( 1 - \hat{\theta}_i \right ) \log \left ( 1 +
 \frac{\delta_i}{1 - \hat{\theta}_i} \right )  \\
 \nonumber
 &\geq&
 \left \{ \begin{array}{ll}
  \log e \cdot \left [
  \hat{\theta}_i \left ( \frac{\delta_i}{\hat{\theta}_i} +
  \frac{\delta_i^2}{2\hat{\theta}_i^2} \right ) +
  \left ( 1 - \hat{\theta}_i \right )
  \left [ -\frac{\delta_i}{1 - \hat{\theta}_i} +
  f \left (\frac{\delta_i}{1 - \hat{\theta}_i} \right ) \right ]
  \right ]; & \mbox{if}~\delta_i > 0 \\
  \log e \cdot \left [
  \hat{\theta}_i \left [ \frac{\delta_i}{\hat{\theta}_i} +
  f \left ( \frac{\left |\delta_i \right |}{\hat{\theta}_i} \right ) \right ]
  + \left ( 1 - \hat{\theta}_i \right )
  \left ( - \frac{\delta_i}{1 - \hat{\theta}_i} +
  \frac{\delta_i^2}{2 \left ( 1 - \hat{\theta}_i \right )^2} \right )
  \right ]; & \mbox{otherwise}
 \end{array} \right . \\
 \nonumber
 &=&
 \left \{ \begin{array}{ll}
 \log e \cdot \left [
 \frac{\delta_i^2}{2\hat{\theta}_i} +
 \left ( 1 - \hat{\theta}_i \right )
 f \left (\frac{\delta_i}{1 - \hat{\theta}_i} \right )
 \right ]; & \mbox{if}~\delta_i > 0 \\
 \log e \cdot \left [
 \hat{\theta}_i f \left ( \frac{\left |\delta_i \right |}{\hat{\theta}_i} \right ) +
 \frac{\delta_i^2}{2 \left ( 1 - \hat{\theta}_i \right )}
 \right ]; & \mbox{otherwise}
 \end{array} \right . \\
 &\geq&
 \left \{ \begin{array}{ll}
 \frac{\log e}{200 n^{1-\varepsilon}}; & \mbox{if}~ \delta_i > 0, \\
 \frac{\log e}{400 n^{1-\varepsilon}}; & \mbox{if}~ \delta_i < 0
 ~\mbox{and}~ 0 < \frac{\left |\delta_i \right |}{\hat{\theta}_i} < 1, \\
 \frac{\left (\log e \right ) \left ( 1 - \ln 2 \right )}{10n^{1-\varepsilon/2}}; &
 \mbox{if}~ \delta_i < 0
 ~\mbox{and}~ \frac{\left |\delta_i \right |}{\hat{\theta}_i} \geq 1.
 \end{array} \right .
\eea
The first inequality is obtained by applying
\eref{eq:log1plusx}-\eref{eq:log1minusx}. (Note that it is true
also in the limits of $\hat{\theta}_i \rightarrow 0$ and
$\hat{\theta}_i \rightarrow 1$.) Then, the first order terms
cancel each other out.  Finally, since all remaining terms are
positive, we only use the first term in each case with the
definition of event $B_i$ in \eref{eq:Bidef} to obtain the last
inequality.  In the third case, we also assume that a nonzero ML
estimator must satisfy $\hat{\theta}_i \geq 1/n$ to obtain the
bound. For $\hat{\theta}_i = 0$,
\be
 D \left ( P_{\hat{\theta}_i} ~||~ P_{\theta_i} \right ) =
 - \log \left ( 1 - \theta_i \right )
 \geq \theta_i \log e \geq \frac{\log e}{n^{1-\varepsilon}},
\ee
where the first inequality is obtained since $-\ln (1-x) \geq
x$ for $0 \leq x \leq 1$, and the second by the definition of the
minimum grid point in \eref{eq:grid_point}.

We can now plug the lower bounds on $D \left ( P_{\hat{\theta}_i}
~||~ P_{\theta_i} \right )$ in \eref{eq:typical_B_i_bound} to
bound $P_{\theta} \left ( B_i \right )$
\be
 P_{\theta} \left ( B_i \right ) \leq n \cdot
 2^{-n \cdot \frac{c}{n^{1-\varepsilon/2}}} =
 2^{\log n - cn^{\varepsilon/2}},
\ee
where $c$ is a constant that is the minimum over all the cases
described above. Finally, by the union bound in
\eref{eq:minimax_UB}, we obtain
\be
 P_{\theta} \left ( A \right )
 \leq k \cdot \max_i \left \{ P_{\theta} \left ( B_i \right ) \right \}\leq
 2^{(\log k) + (\log n) - cn^{\varepsilon/2}}
 \rightarrow 0.
\ee
This concludes the proof
of Lemma~\ref{lemma_event_A_prob}.
{\hfill $\Box$}

\section{--~~ Proof of Lemma \ref{lemma_distinct2}}
\label{ap:lemma_distinct2_proof}
\renewcommand{\theequation}{C.\arabic{equation}}
\renewcommand{\theproposition}{C.\arabic{proposition}}
\renewcommand{\thelemma}{C.\arabic{lemma}}
\setcounter{equation}{0}
\setcounter{lemma}{0}
\setcounter{proposition}{0}

Let $X^n$ be the observed random data sequence, which was generated by
point $\pvec$ on the uniform random grid.  Let $\pvece$ be
the ML estimator of $\pvec$ from $X^n$.
Let $\delta_i$ be defined as in \eref{eq:delta_i}.  Then, for the event
in \eref{eq:lemma_distinct2}, we have
\be
 \label{eq:random_error_event}
 \left \| \pvece - \pvec \right \| =
 \sqrt{\sum_{i=1}^{k-1} \delta_i^2} >
 \frac{1}{\sqrt{n}^{1-\varepsilon}}.
\ee
As in \ref{ap:lemma_event_A_prob_proof}, we will show that the event in
\eref{eq:random_error_event} is a union of events, and use the union
bound on these events to bound the error probability.  However, here, the
events are more complicated.  We start
with a lemma, that will be used to define the events.
\begin{lemma}
\label{lemma_j_events}
Let $n \rightarrow \infty$ and let
$\pvece$ and $\pvec$ satisfy \eref{eq:random_error_event}.  Then, there
exists $j$; $1 \leq j \leq k' \dfn \min \left \{2n^{1-\varepsilon/2}, k-1 \right \}$;
such that for at least $j$ components
$\theta_i$ of $\pvec$,
\be
 \label{eq:lemma_j_events}
 \left (\hat{\theta}_i - \theta_i \right )^2 \geq \frac{1}{j n^{1-\varepsilon/2}}.
\ee
\end{lemma}

\noindent
{\bf Proof:}
Let $\eventF$ contain all the indices $i$ for which
either $\theta_i \geq 1/n^{1-\varepsilon/2}$ or
$\hat{\theta}_i \geq 1/n^{1-\varepsilon/2}$.   The cardinality of
$\eventF$ is bounded by $\left | \eventF \right | \leq 2n^{1-\varepsilon/2}$.
We separate the components of $\pvec$ in $\eventF$ from those outside of it.
For $i \in \bar{\eventF}$, let $\theta_i \dfn \alpha_i/ n^{1 - \varepsilon/2}$
and $\hat{\theta}_i \dfn \hat{\alpha}_i /n^{1-\varepsilon/2}$, where
$0 < \alpha_i < 1$ and $0 < \hat{\alpha}_i < 1$.
The contribution of all $i \in \bar{\eventF}$ to the sum in
\eref{eq:random_error_event} is negligible, and thus the event defined in
\eref{eq:random_error_event} depends mostly on
$i \in \eventF$.  This step is necessary to show that distinguishability includes
also sources with more than $k_m$ letters for larger $k$.

Now, assume that no $j$ as defined above exists.
Then, for every $i$,
\be
 \label{eq:j_proof1}
 \left ( \hat{\theta}_i - \theta_i \right )^2 < \frac{1}{n^{1-\varepsilon/2}}.
\ee
Then, for every component $i$, but one,
\be
 \label{eq:j_proof2}
 \left ( \hat{\theta}_i - \theta_i \right )^2 < \frac{1}{2n^{1-\varepsilon/2}},
\ee
and for at most one component $\theta_i$ of $\pvec$,
\be
 \label{eq:j_proof3}
 \frac{1}{2n^{1-\varepsilon/2}} \leq
 \left ( \hat{\theta}_i - \theta_i \right )^2  <
 \frac{1}{n^{1-\varepsilon/2}}.
\ee
Next, there are at most two components $\theta_i$ of $\pvec$, for which
\be
 \label{eq:j_proof4}
 \left ( \hat{\theta}_i - \theta_i \right )^2  \geq
 \frac{1}{3n^{1-\varepsilon/2}},
\ee
but for at least one of them, \eref{eq:j_proof2}
must also be satisfied, and for both
\eref{eq:j_proof1} must be satisfied.  We can proceed this process up to
$j=\min \left \{k-1, 2n^{1-\varepsilon/2} \right \} \leq 2n^{1-\varepsilon/2}$.
Using this process, \eref{eq:j_proof1}-\eref{eq:j_proof4}, and
the following similar equations that can be obtained for larger $j$'s, we can
obtain the upper bound
\bea
 \nonumber
 \left \| \pvece - \pvec \right \|^2 &=&
 \sum_{i \in \bar{\eventF}} \left ( \hat{\theta}_i - \theta_i \right )^2 +
 \sum_{i \in \eventF} \left ( \hat{\theta}_i - \theta_i \right )^2 ~<~
% \\
% \nonumber
% &<&
 \sum_{i \in \bar{\eventF}} \theta_i^2 +
 \sum_{i \in \bar{\eventF}} \hat{\theta}_i^2 +
 \sum_{j=1}^{2n^{1-\varepsilon}} \frac{1}{j \cdot n^{1 - \varepsilon/2}} \\
 \nonumber
% &=&
% \frac{1}{n^{1-\varepsilon/2}} \sum_{j=1}^{k-1} \frac{1}{j}
 &\leq&
 \sum_{i \in \bar{\eventF}} \frac{\alpha_i^2}{n^{2-\varepsilon}} +
 \sum_{i \in \bar{\eventF}} \frac{\hat{\alpha}_i^2}{n^{2-\varepsilon}} +
 \frac{1}{n^{1-\varepsilon/2}}
 \left ( 1 + \int_1^{2n^{1-\varepsilon/2} +1} \frac{1}{x} dx \right ) \\
 \nonumber
 &<&
 \sum_{i \in \bar{\eventF}} \frac{\alpha_i}{n^{2-\varepsilon}} +
 \sum_{i \in \bar{\eventF}} \frac{\hat{\alpha}_i}{n^{2-\varepsilon}} +
 \frac{1}{n^{1-\varepsilon/2}} \ln \left [ e \left ( 2n^{1-\varepsilon/2} +1 \right ) \right ]\\
 \label{eq:j_proof5}
 &<&
 \frac{2}{n^{1-\varepsilon/2}} +
 \frac{n^{\varepsilon/4}}{n^{1 - \varepsilon/2}} ~<~
 \frac{1}{n^{1-\varepsilon}}.
\eea
The first inequality is by applying the above relations and by bounding the
square distance for small probabilities by the sum of their squares.
The third inequality is since $\alpha_i^2 < \alpha_i$ since $\alpha_i < 1$ for
$i \in \bar{\eventF}$, and
the same applies for $\hat{\alpha}_i$.  The next inequality is since
$\sum_{i\in \bar{\eventF}} \alpha_i \leq n^{1-\varepsilon/2}$ since
$\sum_{i\in \bar{\eventF}} \theta_i \leq 1$, and again, the same is applied
for $\hat{\alpha}_i$.  In addition, we apply $n \rightarrow \infty$ to bound the
last term for this inequality and for the last inequality.
Inequality \eref{eq:j_proof5}
contradicts \eref{eq:random_error_event}.  This concludes the proof
of Lemma~\ref{lemma_j_events}.
{\hfill $\Box$}

Before we use Lemma~\ref{lemma_j_events}, let us partition $\pvec$ into
the set $\pvec^-$ containing all letters with $\theta_i < 1/n^{2+\varepsilon}$ and
$\pvec^+$, containing all the remaining letters.  This is, again, necessary
for the case in which $k > k_m$.  Let event $\eventT$ contain all
$x^n$ for which any of the letters in $\pvec^-$ occurs more than once.
We can now use \eref{eq:lemma_j_events} to define event $A_j$ as all
sequences $x^n$ for which there are (at least) $j$ components $\hat{\theta}_i$ of
$\pvec^+$, for which \eref{eq:lemma_j_events} is satisfied.  Thus,
\be
\label{eq:most_UB}
 P_{\theta} \left \{
 \left \| \pvece - \pvec \right \| > \frac{1}{\sqrt{n}^{1-\varepsilon}}
 \right \}
 \leq
 P_{\theta} \left ( \eventT \right ) +
 \sum_{j=1}^{k''} P_{\theta} \left ( A_j \right ),
\ee
where $k'' = \min \left \{\left | \pvec^+ \right |, k' \right \} \leq
2n^{1-\varepsilon/2}$.  Inequality \eref{eq:most_UB} is because if \eref{eq:random_error_event}
is satisfied,
either $A_j$ occurs for some $j$, or there exist components in $\pvec^-$ for which
\eref{eq:lemma_j_events} is satisfied.  For such components, the occurrence of
\eref{eq:lemma_j_events} means that the letter occurred (significantly) more than
once.

First, let us bound the first term of \eref{eq:most_UB}.  The probability that letter
$i \in \pvec^-$ occurs in $X^n$ is given by
\be
 P_{\theta} \left ( i \in X^n \right ) =
 1 - \left ( 1 -\theta_i \right )^n \geq n \theta_i -
 \comb{n}{2} \theta_i^2.
\ee
The average re-occurrences (beyond the first occurrence)
of such a letter is then upper bounded by
\be
 E N_x \left ( i \right ) - P_{\theta} \left ( i \in X^n \right )  \leq
 \comb{n}{2} \theta_i^2,
\ee
where $E N_x \left ( i \right )$ is the expected number of occurrences of letter $i$.
Then, the average re-occurrence of any of the letters in $\pvec^-$ is bounded
by
\be
 \sum_{\theta_i \in \theta^-}
 \left \{E N_x \left ( i \right ) - P_{\theta} \left ( i \in X^n \right )
 \right \} \leq
 \comb{n}{2} \sum_{\theta_i \in \theta^-} \theta_i^2 \leq
 \frac{n^2}{2} \sum_{\theta_i \in \theta^-} \frac{\alpha_i^2}{n^{4+2\varepsilon}} \leq
 \frac{1}{2n^{\varepsilon}} \to 0,
\ee
where $\theta_i \dfn \alpha_i/n^{2+\varepsilon}$, $\alpha_i < 1$, and the
last inequality is obtained similarly to the derivation in
\eref{eq:j_proof5}, where $\sum_i \alpha_i^2 < \sum_i \alpha_i < n^{2+\varepsilon}$.
Using Markov inequality, the probability of $\eventT$ is bounded by the bound above,
i.e., $P_{\theta} \left ( \eventT \right ) \to 0$.

Event $A_j$ in \eref{eq:most_UB}
is the union of all events for which any $j$ components
of $\pvec^+$ satisfy \eref{eq:lemma_j_events}.  This applies to any
choice of $j$ components out of $\tilde{k} = \left | \pvec^+ \right | \leq
n^{2+\varepsilon}$.
Let $A_{jl}$ be the event in which
the $j$ components of the $l$th choice out of
\be
 L \leq \comb{\tilde{k}}{j}
\ee
choices of components
of $\pvec^+$ satisfy \eref{eq:lemma_j_events}.
Using the union bound, again,
\be
 \label{eq:most_UB2}
 P_{\theta} \left ( A_j \right ) \leq \sum_{l=1}^{L}
 P_{\theta} \left ( A_{jl} \right ) \leq
 \comb{\tilde{k}}{j} \cdot \max_{l} P_{\theta} \left (A_{jl} \right ) <
 \tilde{k}^{j} \cdot \max_{l} P_{\theta} \left (A_{jl} \right ).
\ee
To bound $P_{\theta} \left (A_{jl} \right )$ for given $j$ and $l$, let us
define a transformation of the alphabet of $\avk$ to the alphabet
$\Sigma_{jl}$ with cardinality $j+1$.  The $j$ letters denoted by
$u_1, u_2, \ldots, u_j$, whose ML estimates
satisfy \eref{eq:lemma_j_events} will be numbered from $1$ to
$j$, and all other letters will be transformed into the letter
$j+1 \in \Sigma_{jl}$.  Let $\Yvec_l$ be an $n$-dimensional transformation of
$X^n$ that takes a value $y^n$, such that, in a
similar manner to \eref{eq:Y_transformation},
\be
\label{eq:Y_transformation2}
 Y_{lm} = \left \{ \begin{array}{ll}
 i, &\mbox{if}~X_m = u_i, \\
 j+1, &\mbox{otherwise}
 \end{array} \right ..
\ee
Let $\varphi_i = \theta_{u_i}$ for $1 \leq i \leq j$, be the probability
of $Y_{lm}$ taking the value $i$, where $\varphi_{j+1}$ is the sum of all
the remaining probability components of $\pvec$.
Let $\hat{\varphi}_i$ be the ML estimate of $\varphi_i$ from
the vector $\Yvec_l$.  Let $\vphivec$ be the $j$ dimensional vector
that defines the i.i.d.\ distribution of vector $\Yvec_l$.
Since the probability of $A_{jl}$ depends only on the $j$ parameters that satisfy
\eref{eq:lemma_j_events}, we can now use the new parameter vector $\vphivec$, which
is a permutation of these $j$ parameters with all other components of
$\pvec$ condensed into one probability parameter, to bound this probability.
By typical sets analysis,
\be
 \label{eq:Ajl_prob}
 P_{\theta} \left ( A_{jl} \right ) =
 P_{\varphi} \left ( A_{jl} \right ) \leq
 \left (n + 1 \right )^j
 2^{-n \min_{y^n \in A_{jl}} D \left ( P_{\hat{\varphi}} ~||~ P_{\varphi} \right )},
\ee
where the polynomial coefficient is a bound on the number of types.
To bound the expression in \eref{eq:Ajl_prob}, we can lower bound
the divergence in its exponent.  Let $U_1$ be the set of components of
$\vphivec$ for which $\hat{\varphi}_i \geq \varphi_i$, and $U_2$ the set
for which $\hat{\varphi}_i < \varphi_i$.
Also define $\delta_i$ now w.r.t.\
$\vphivec$ and $\hat{\vphivec}$.  Then,
\bea
 \nonumber
 D \left ( P_{\hat{\varphi}} ~||~ P_{\varphi} \right )
 &=&
 \sum_{\varphi_i \in U_1} \hat{\varphi_i} \log \frac{\hat{\varphi_i}}{\varphi_i} +
 \sum_{\varphi_i \in U_2} \hat{\varphi_i} \log \frac{\hat{\varphi_i}}{\varphi_i} \\
 \nonumber
 &=&
 - \sum_{\varphi_i \in U_1} \hat{\varphi}_i \log
 \left ( 1 - \frac{\delta_i}{\hat{\varphi}_i} \right )
 - \sum_{\varphi_i \in U_2} \hat{\varphi}_i \log
 \left ( 1 - \frac{\delta_i}{\hat{\varphi}_i} \right ) \\
 \nonumber
 &\geq&
 \log e \cdot \left \{
 \sum_{\varphi_i \in U_1} \hat{\varphi}_i
 \left [ \frac{\delta_i}{\hat{\varphi}_i} +
 \frac{\delta_i^2}{2 \hat{\varphi}_i^2} \right ] +
 \sum_{\varphi_i \in U_2} \hat{\varphi}_i
 \left [ -\frac{-\delta_i}{\hat{\varphi}_i} +
 f \left (\frac{-\delta_i}{\hat{\varphi}_i} \right ) \right ]
 \right \} \\
 \label{eq:diver_deriv}
 &=&
 \log e \cdot \left \{
 \sum_{\varphi_i \in U_1}
 \frac{\delta_i^2}{2 \hat{\varphi}_i} +
 \sum_{\varphi_i \in U_2}
 \hat{\varphi}_i f \left (\frac{-\delta_i}{\hat{\varphi}_i} \right )
 \right \}.
\eea
The inequality is obtained from \eref{eq:log1plusx}
and \eref{eq:log1minusx} and the definition of the function $f(\cdot)$ in
\eref{eq:fdef}.  The last equality is since all the first order terms
cancel each other.
Now, define the set $U'_1$ as the union of all components $\varphi_i$,
$1 \leq i \leq j$, in $U_1$ and these components in
$U_2$ for which $\left |\delta_i \right | \leq \hat{\varphi}_i$, and $U'_2$
as the set of all the remaining components in $U_2$.  (Note that
we extract $\varphi_{j+1}$ from both sets.)  Assume that there are
$\alpha j$, $0\leq \alpha \leq 1$, components in $U'_2$.
Then, by definition of $f \left ( \cdot \right )$ in \eref{eq:fdef}, and
since all $j$ components in both sets satisfy
\eref{eq:lemma_j_events}, we obtain from \eref{eq:diver_deriv},
\bea
 \nonumber
 D \left ( P_{\hat{\varphi}} ~||~ P_{\varphi} \right )
 &\geq&
 \sum_{\varphi_i \in U'_1}
 \frac{\log e}{4 j n^{1 - \varepsilon/2} \hat{\varphi}_i} +
 \sum_{\varphi_i \in U'_2}
 \frac{\log (e/2)}{\sqrt{j} \sqrt{n}^{1- \varepsilon/2}} \\
 \nonumber
 &\geq&
 \frac{\left ( \log e \right ) \left ( 1 - \alpha \right ) j}
 {4 j n^{1 - \varepsilon/2}}
  \sum_{\varphi_i \in U'_1} \frac{1}{\left ( 1 - \alpha \right ) j} \cdot
  \frac{1}{ \hat{\varphi}_i} +
  \frac{\alpha \sqrt{j} \log (e/2)}{ \sqrt{n}^{1- \varepsilon/2}} \\
 \label{eq:diver_deriv1}
 &\geq&
 \frac{\left ( \log e \right ) \left ( 1 - \alpha \right )^2 j^2}
 {4 j n^{1 - \varepsilon/2}} +
 \frac{4 \alpha j^2 \log (e/2)}{\sqrt{2} \cdot 4 j n^{1 - \varepsilon/2}}
 ~\geq~
 \frac{c j}{n^{1 - \varepsilon/2}}.
\eea
The first term of the
third inequality is obtained by Jensen's inequality over the convex function
$1/x$, and since the sum on all $\hat{\varphi}_i \in U'_1$ is not larger than $1$.
The second term is obtained since $j \leq 2 n^{1- \varepsilon/2}$.
The last inequality is obtained since the expression is greater than
$0$ for every value of $\alpha$, and we can choose a proper
constant $c>0$, for which the inequality is satisfied.
(We note that the derivation above also applies in the
limit if there exist components
of $\vphivec$ whose ML estimates are $0$.)
Combining \eref{eq:most_UB} and the bound on
$P_{\theta} \left ( \eventT \right )$, \eref{eq:most_UB2}, \eref{eq:Ajl_prob},
and \eref{eq:diver_deriv1}, we conclude that
\bea
 \nonumber
 P_{\theta} \left \{
 \left \| \pvece - \pvec \right \| > \frac{1}{\sqrt{n}^{1-\varepsilon}}
 \right \}
 &\leq& \frac{1}{n^{\varepsilon}} +
 \sum_{j=1}^{k''}
 2^{-j \cdot \left [ c n^{\varepsilon/2} - \log (n + 1) - \log \tilde{k} \right ]} \\
 &\leq&
 \frac{1}{n^{\varepsilon}} +
 2^{-\left [ c n^{\varepsilon/2} - \log (n + 1) - (2+ \varepsilon)\log n -
 \log \left (2n^{1-\varepsilon/2}\right )
 \right ]}
 \rightarrow 0.
\eea
This concludes the proof of Lemma~\ref{lemma_distinct2}.
{\hfill $\Box$}

\section{--~~ Proof of Lemma \ref{lemma_negligible_permutations}}
\label{ap:lemma_neglig_proof}
\renewcommand{\theequation}{D.\arabic{equation}}
\renewcommand{\theproposition}{D.\arabic{proposition}}
\renewcommand{\thelemma}{D.\arabic{lemma}}
\setcounter{equation}{0}
\setcounter{lemma}{0}
\setcounter{proposition}{0}

Let us first bound the logarithm of the ratio between
the probability
given by the parameter vector $\phivec$ and
the ML probability of $X^n$.  Similarly to \eref{eq:log1plusx}-\eref{eq:log1minusx},
if $x < 1$, then
\be
 \label{eq:log_small}
 \log \left ( 1 - x \right ) \leq
 \left ( \log e \right ) \cdot \left [ - x - f'(x) \right ],
\ee
where
\be
 \label{eq:f_def}
 f'(x) \dfn \left \{
 \begin{array}{ll}
  \frac{x^2}{2}; & \mbox{if}~ x \geq 0, \\
  \frac{x^2}{4}; & \mbox{if}~ 0 > x \geq -1, \\
  - \left ( 1 - \ln 2 \right ) x; & \mbox{if}~ x < -1.
 \end{array}
 \right .
\ee
Using the above,
\bea
 \log \frac{P_{\phi} \left ( X^n \right )}
 {P_{\hat{\theta}} \left ( X^n \right )}
 &=&
 \log \prod_{i=1}^k \left ( \frac{\phi_i}{\hat{\theta}_i}
 \right )^{n \hat{\theta}_i}
 ~=~
 n \sum_{i=1}^k \hat{\theta}_i \log \left (
 1 - \frac{\delta_i}{\hat{\theta}_i} \right ) \\
 &\leq&
 \label{eq:lemma_n_deriv2}
 \left ( \log e \right ) n
 \sum_{i=1}^k \hat{\theta}_i
 \left [
  -\frac{\delta_i}{\hat{\theta}_i} -
  f' \left (\frac{\delta_i}{\hat{\theta}_i} \right )
 \right ]
 ~=~
 -\left ( \log e \right ) n
 \sum_{i=1}^k \hat{\theta}_i f' \left ( \frac{\delta_i}{\hat{\theta}_i}\right ) \\
 &=&
 -\left ( \log e \right ) n
 \sum_{i=1}^k \hat{\theta}_i \cdot
 \left \{
 \begin{array}{ll}
  \frac{\delta_i^2}{2 \hat{\theta}_i^2}; & \mbox{if}~\frac{\delta_i}{\hat{\theta}_i}
  \geq 0, \\
  \frac{\delta_i^2}{4 \hat{\theta}_i^2}; & \mbox{if}~0 \geq
  \frac{\delta_i}{\hat{\theta}_i}
  \geq -1, \\
  - \left ( 1 - \ln 2 \right ) \frac{\delta_i}{\hat{\theta}_i}; &
  \mbox{if}~ \frac{\delta_i}{\hat{\theta}_i} < -1
 \end{array}
 \right . \\
 &\leq&
 \label{eq:lemma_n_deriv4}
 -\left ( \log e \right ) n
 \sum_{i=1}^k \hat{\theta}_i \cdot
 \left \{
 \begin{array}{ll}
  \frac{\delta_i^2}{4 \hat{\theta}_i^2}; &
  \mbox{if}~ \phi_i \leq 2 \hat{\theta}_i, \\
  - \left ( 1 - \ln 2 \right ) \frac{\delta_i}{\hat{\theta}_i}; &
  \mbox{if}~ \phi_i > 2 \hat{\theta}_i
 \end{array}
 \right . \\
 &\leq&
 \label{eq:lemma_n_deriv5}
 -\left ( \log e \right ) n
 \sum_{i \in J}
 \left \{
 \begin{array}{ll}
  \frac{k}{4jn^{1-\varepsilon/4}};&
  \mbox{if}~ \phi_i \leq 2 \hat{\theta}_i, \\
  \left ( 1 - \ln 2 \right ) \frac{k}{j} \cdot
  \frac{\sqrt{\hat{\theta}_i}}{\sqrt{n}^{1-\varepsilon/4}}; &
  \mbox{if}~ \phi_i > 2 \hat{\theta}_i
 \end{array}
 \right . \\
 &\leq&
 \label{eq:lemma_n_deriv6}
 -\left ( \log e \right ) n
 \sum_{i \in J}
 \frac{k}{4jn^{1-\varepsilon/8}}
 ~\leq~
 -\frac{kn^{\varepsilon/8}}{4 (\ln 2)}.
\eea
The inequality in \eref{eq:lemma_n_deriv2} is obtained
from \eref{eq:log_small}, and the equality since the summation
on all $\delta_i$ must be zero.  The boundaries in
\eref{eq:lemma_n_deriv4} are obtained from the definition of
$\delta_i$ in \eref{eq:delta_lemma_def}.
To obtain \eref{eq:lemma_n_deriv5}, we bound all (negative) elements of
the sum for which $i \not \in J$ by zero, and all elements
$i \in J$ using \eref{eq:neg_lemma_bound}.  Then, to obtain
\eref{eq:lemma_n_deriv6}, we take the maximum over the different
regions, and also use the fact that by the definition of a $k$-dimensional
i.i.d.\ ML vector,
$\hat{\theta}_i \geq 1/n$.  The last inequality follows the fact that there
are at least $j$ elements in $J$.

Taking the bound of \eref{eq:lemma_n_deriv6}, we obtain
\be
 \frac{k! P_{\phi} \left ( X^n \right )}{P_{\hat{\theta}_i} \left ( X^n \right )}
 \leq k! \cdot \exp
 \left \{ - \frac{kn^{\varepsilon/8}}{4} \right \} \leq
 \exp \left \{ - k \left ( \frac{n^{\varepsilon/8}}{4} - \ln k \right ) \right \}
 \rightarrow 0.
\ee
This concludes the proof of Lemma~\ref{lemma_negligible_permutations}.
{\hfill $\Box$}

\section{--~~ Proof of Lemma \ref{lemma_quantization_ratio}}
\label{ap:lemma_q_proof}
\renewcommand{\theequation}{E.\arabic{equation}}
\renewcommand{\theproposition}{E.\arabic{proposition}}
\renewcommand{\thelemma}{E.\arabic{lemma}}
\setcounter{equation}{0}
\setcounter{lemma}{0}
\setcounter{proposition}{0}

To prove Lemma~\ref{lemma_quantization_ratio}, we express the logarithm of the
desired ratio as a function of the components of $\vphivec$ and of distances between
components of $\avec (\sigvec)$,
$\vphivec (\sigvec)$, and $\pvece$.  First, we bound
distances between corresponding components of the three vectors under
the assumption that $\avec \left (\sigvec \right ) \not \in \eventA$,
and use these bounds to bound the
logarithm of the ratio in \eref{eq:quantization_log_ratio}.
Let $\delta (a, b) \dfn a - b$ be the difference between $a$ and $b$.
Then, by definition of $\vphivec$ as the quantized form
of $\avec$, quantized onto points in $\tgrid$,
and by definition of $\tgrid$, we must have for every
$i$, $1 \leq i \leq k-1$,
\be
 \label{eq:delta_psi_phi}
 \left | \delta \left ( \psi_i, \varphi_i \right ) \right | \leq
 \Delta \left ( \tau_{b(\varphi_i) + 1}\right ) =
 \frac{2 \left [ b \left ( \varphi_i \right ) + \frac{1}{2} \right ]}
 {n^{1+\varepsilon}} \leq
 \frac{2.5 b \left ( \varphi_i \right )}{n^{1 + \varepsilon}} =
 \frac{2.5\sqrt{\varphi_i}}{\sqrt{n}^{1 + \varepsilon}},
\ee
where $\Delta \left (\cdot \right )$ is defined as in
\eref{eq:grid_spacing} but w.r.t.\ $\tgrid$ defined in
\eref{eq:grid_point_upper}.
The first inequality is obtained since
either $\psi \in \left [ \tau_{b(\varphi_i) - 1},  \tau_{b(\varphi_i)} = \varphi_i \right ]$
or $\psi \in \left [ \tau_{b(\varphi_i)},  \tau_{b(\varphi_i) + 1} \right ]$.
In either case, $\psi_i$ is at most
$\Delta \left ( \tau_{b(\varphi_i) + 1}\right )$ away from $\varphi_i$.
The last equality is obtained using a similar equation to
\eref{eq:grid_point2} where $-\varepsilon$ is replaced by $\varepsilon$ for
the proper grid.  The distance between the
last $k$th components of $\avec$ and $\vphivec$ can be bounded similarly by
\be
 \label{eq:delta_psi_phik}
 \left | \delta \left ( \psi_k, \varphi_k \right ) \right | \leq
 \frac{2.5 \sqrt{\varphi_{k-1}}}{\sqrt{n}^{1 + \varepsilon}} \leq
 \frac{2.5 \sqrt{\varphi_k}}{\sqrt{n}^{1 + \varepsilon}}.
\ee
This is because of the procedure used to quantize $\avec$ into $\vphivec$, that
ensures that the absolute
value of the cumulative difference between
the components of $\avec$ and those of $\vphivec$ is minimized, and is therefore
bounded by the maximal spacing around the largest free component.

From Lemma~\ref{lemma_negligible_permutations},
in order for $P_{\psi \left (\sigma \right )} \left ( X^n \right )$, the probability
of $X^n$ that is given by a permutation
$\avec(\sigvec)$, not to be negligible w.r.t.\
the ML probability of $X^n$,
$\avec(\sigvec)$
must have, for every $j$, no more than $j-1$ components for which
\eref{eq:neg_lemma_bound} is satisfied
(where $\delta_i$ is replaced by
$\delta \left [ \hat{\theta}_i, \psi(\sigma_i) \right ]$, and
$\phi_i$ by $\psi (\sigma_i)$).
This implies that
if a permutation $\avec(\sigvec)$ of
$\avec$ is not negligible,
it must have at least $k - j + 1$ components for every $j$,
$1 \leq j \leq k$, that satisfy
\be
 \label{eq:neg_lemma_ubound}
 \left | \delta \left [ \hat{\theta}_i, \psi(\sigma_i) \right ] \right | \leq
 \left \{
 \begin{array}{ll}
  \frac{k}{j} \cdot \frac{\sqrt{\hat{\theta}_i}}{\sqrt{n}^{1 - \varepsilon/4}}; &
  \mbox{if}~ \psi(\sigma_i) > 2\hat{\theta}_i, \\
  \sqrt{\frac{k}{j}} \cdot \frac{\sqrt{\hat{\theta}_i}}{\sqrt{n}^{1 - \varepsilon/4}}; &
  \mbox{if}~ \psi(\sigma_i) \leq 2\hat{\theta}_i.
 \end{array}
 \right .
\ee
Hence, in the worst case, there is one
distance component for which the tightest upper bound
is obtained from \eref{eq:neg_lemma_ubound} with $j = 1$,
one for $j = 2$, and so on, up to $j = k$, i.e.,
for each $j$, the inequality is satisfied for a distinct component $i$.  Conversely,
for the worst case, we can denote the distinct value of $j$ for each $i$
as a function of $i$ and of the
two vectors $\pvece$ and $\avec\left ( \sigvec \right )$, i.e., as
$j\left (\pvece, \avec\left ( \sigvec \right ),i \right )$.

We can now express $\delta \left [ \hat{\theta}_i, \varphi(\sigma_i) \right ]$ as
\bea
 \nonumber
 \delta \left [ \hat{\theta}_i, \varphi(\sigma_i) \right ] &=&
 \hat{\theta}_i - \varphi(\sigma_i) =
 \hat{\theta}_i - \psi(\sigma_i) + \psi(\sigma_i) - \varphi(\sigma_i) \\
 &=&
 \delta \left [ \hat{\theta}_i, \psi(\sigma_i) \right ] +
 \delta \left [ \psi(\sigma_i), \varphi(\sigma_i) \right ].
\eea
By the triangle inequality, \eref{eq:delta_psi_phi},
\eref{eq:delta_psi_phik}, and \eref{eq:neg_lemma_ubound},
if $\avec (\sigvec)\not \in \eventA$, for the $k -j + 1$ components of
$\psi \left ( \sigma_i \right )$ that satisfy \eref{eq:neg_lemma_ubound},
\bea
 \nonumber
 \left |
 \delta \left [ \hat{\theta}_i, \varphi(\sigma_i) \right ]
 \right |
 &\leq&
 \left |
 \delta \left [ \hat{\theta}_i, \psi(\sigma_i) \right ]
 \right | +
 \left |
 \delta \left [ \psi(\sigma_i), \varphi(\sigma_i) \right ]
 \right | \\
 \nonumber
 &\leq&
 2 \cdot \max \left \{
  \left |
 \delta \left [ \hat{\theta}_i, \psi(\sigma_i) \right ]
 \right |,
 \left |
 \delta \left [ \psi(\sigma_i), \varphi(\sigma_i) \right ]
 \right |
 \right \} \\
 \label{eq:delta_theta_phi}
 &\leq&
 \left \{ \begin{array}{ll}
% \frac{5\sqrt{\varphi(\sigma_i)}}
% {\sqrt{n}^{1-\varepsilon/4}} \cdot \frac{k}{j}; &
% \mbox{if}~ \varphi(\sigma_i) \geq \hat{\theta}_i, \\
 \frac{5\sqrt{2\varphi(\sigma_i)}}
 {\sqrt{n}^{1-\varepsilon/4}} \cdot \frac{k}{j}; &
 \mbox{if}~
% \varphi(\sigma_i) <
 \hat{\theta}_i \leq 2\varphi(\sigma_i),\\
 \frac{5\sqrt{\hat{\theta}_i}}
 {\sqrt{n}^{1-\varepsilon/4}} \cdot
 \sqrt{\frac{k}{j}}; &
 \mbox{if}~ \hat{\theta}_i > 2\varphi(\sigma_i).
 \end{array} \right .
\eea
The first region is obtained by combining the worse bound of
\eref{eq:neg_lemma_ubound} with that of \eref{eq:delta_psi_phi}.
The bound in the second region is true because
$2\hat{\theta}_i > 4 \varphi(\sigma_i) > \psi (\sigma_i)$ since
it can be shown by definition of $\tgrid$ that always
$\psi(\sigma_i) \leq 3 \varphi (\sigma_i)$.

The first region of the bound in \eref{eq:delta_theta_phi} is expressed as a
function of $\varphi \left ( \sigma_i \right )$.  However, the second region is
in terms of $\hat{\theta}_i$.  In order to obtain the bound
of \eref{eq:quantization_log_ratio}, we need to express both bounds in terms
of $\varphi \left ( \sigma_i \right )$.  Hence, we need to first bound
the second region of \eref{eq:delta_theta_phi} in terms of
$\varphi \left ( \sigma_i \right )$, or alternatively bound
$\sqrt{\hat{\theta}_i}$ in terms of $\sqrt{\varphi \left ( \sigma_i \right )}$.
To achieve that,
we observe that $\varphi(\sigma_i)$ is smaller
than half $\hat{\theta}_i$.  This means that
we represent the i.i.d.\ ML probability component $\hat{\theta}_i$ using
$\avec(\sigvec)$ by a probability that is roughly smaller than its half
(since $\psi(\sigma_i)$ is asymptotically much closer to
$\varphi(\sigma_i)$).
If $\hat{\theta}_i$ is large, this must yield a
negligible probability because
$\psi(\sigma_i)$ will be too far from $\hat{\theta}_i$.
Therefore, there must be an upper bound on $\hat{\theta}_i$ for
which the second region of \eref{eq:delta_theta_phi}
still applies while $\avec \left (\sigvec \right ) \not \in \eventA$.
By the bound in the second region of \eref{eq:delta_theta_phi}, we must
have
\be
 \frac{5\sqrt{\hat{\theta}_i}}
 {\sqrt{n}^{1-\varepsilon/4}} \cdot
 \sqrt{\frac{k}{j}} \geq
 \left |
 \delta \left [ \hat{\theta}_i, \varphi(\sigma_i) \right ]
 \right |
 =
 \hat{\theta}_i - \varphi(\sigma_i) \geq
 \frac{\hat{\theta}_i}{2}.
\ee
Hence, by rearranging terms of the last inequality,
\be
 \label{eq:theta_i_bound}
 \hat{\theta}_i \leq \frac{100k}{jn^{1-\varepsilon/4}}.
\ee
Now, we need the following lemma.
\begin{lemma}
\label{lemma_phi_i_bound}
Let $\hat{k} = k \leq \sqrt{n}^{1-\varepsilon}$, and let $\xi > 0$ be
arbitrarily small.  Then, for all $i$; $1 \leq i \leq k$,
\be
 \label{eq:phi_i_bound}
 \varphi_i \geq (1-\xi)/n.
\ee
\end{lemma}

\noindent
{\bf Proof:}
Let $\hat{\theta}_k = n_x (k)/n$ be the maximal component of $\pvece$, where $n_x(k)$
is the occurrence count of the respective letter.  Then, first, we must have
$\psi_k \geq (1 - \xi/2) \hat{\theta}_k$.  Otherwise,
$\widehat{\avec \left ( \pvec \right )} \in \eventA$, and cannot be the pattern
ML estimate, using Lemma~\ref{lemma_negligible_permutations}.  This is shown below.
Assume $\psi_k < (1 - \xi/2) \hat{\theta}_k$.  Then,
\be
 \delta \left ( \hat{\theta}_k, \psi_k \right ) >
 \frac{\xi\hat{\theta}_k}{2} \geq
 \frac{\xi \sqrt{\hat{\theta}_k}}{2 n^{(1-\varepsilon)/4}} \geq
 \sqrt{\frac{k}{1}} \cdot
 \frac{\sqrt{\hat{\theta}_k}}{\sqrt{n}^{1-\varepsilon/4}} \cdot
 \frac{\xi}{2} \sqrt{n}^{3\varepsilon/4} >
 \sqrt{\frac{k}{1}} \cdot
 \frac{\sqrt{\hat{\theta}_k}}{\sqrt{n}^{1-\varepsilon/4}}.
\ee
The second inequality is since $\hat{\theta}_k \geq 1/k \geq 1/\sqrt{n}^{1-\varepsilon}$.
The next inequality is again by the assumption that $k \leq \sqrt{n}^{1-\varepsilon}$.
The right hand side above shows that if $\psi_k < (1 - \xi/2) \hat{\theta}_k < 2 \hat{\theta}_k$,
the condition of Lemma~\ref{lemma_negligible_permutations} is satisfied w.r.t.\
$\widehat{\avec \left ( \pvec \right )}$, thus contradicting the fact that
$\widehat{\avec \left ( \pvec \right )}$ is the pattern ML probability vector.

Using the fact that $\psi_k \geq (1 - \xi/2) \hat{\theta}_k$, we now show
by differentiation that $P_{\psi} \left [\Psi \left (X^n \right ) \right ]$
attains its maximum w.r.t.\ $\psi_1$ for $\psi_1 \geq (1 - \xi/2)/n$.  First,
from \eref{eq:pattern_prob1},
\be
 \frac{d P_{\psi} \left [\Psi \left (X^n \right ) \right ]}{d \psi_1} =
 \sum_{\sigvec} \left [ \frac{n_x\left ( \sigma_1 \right )}{\psi_1} -
 \frac{n_x\left ( \sigma_k \right )}{\psi_k} \right ]
 P_{\psi\left ( \sigma \right )} \left (X^n \right ),
\ee
where $n_x\left ( \sigma_i \right )$ is the permuted entry of the occurrence vector at index $i$.
This derivative is a weighted sum of decreasing functions in $\psi_1$, each
attaining the value $0$ at
\be
 \psi_1 = \frac{n_x \left ( \sigma_1 \right )}{n_x \left ( \sigma_k \right )} \psi_k
 \geq \left ( 1 - \xi/2 \right ) \frac{n_x(k)}{n} \cdot
 \frac{n_x \left ( \sigma_1 \right )}{n_x \left ( \sigma_k \right )} \geq
 \frac{1 - \xi/2}{n},
\ee
where the first inequality is since $\psi_k \geq (1 - \xi/2) \hat{\theta}_k$, and
the second is because $n_x\left ( \sigma_1 \right ) \geq 1$ and
$n_x(k) \geq n_x \left ( \sigma_k \right )$.  Finally, by the quantization of $\psi_1$ to
$\varphi_1$ (using the definition of $\tgrid$) and the ordering in vector $\vphivec$,
we obtain $\varphi_i \geq \varphi_1 \geq (1 - \xi)/n$.
{\hfill $\Box$}

Using Lemma~\ref{lemma_phi_i_bound} (in particular, \eref{eq:phi_i_bound}),
we can now bound $\sqrt{\hat{\theta}_i}$ for the
second region of the bound in \eref{eq:delta_theta_phi}.  From
\eref{eq:theta_i_bound} and bounding, we have
\be
 \label{eq:sqrt_theta_varphi}
 \sqrt{\hat{\theta}_i} \leq
 10 \cdot \sqrt{\frac{k}{j}} \cdot \frac{1}
 {\sqrt{n}^{1 - \varepsilon/4}} \leq
 10 \cdot \sqrt{\frac{k}{(1-\xi)j}} \cdot n^{\varepsilon/8} \cdot
 \sqrt{\varphi(\sigma_i)}.
\ee

We can now bound the logarithm of the desired ratio.  This is done below:
\bea
 \log \frac{P_{\psi(\sigma)} \left ( X^n \right )}
 {P_{\varphi(\sigma)} \left ( X^n \right )}
 &=&
 \log \left \{
 \prod_{i=1}^k
 \left [
 \frac{\psi(\sigma_i)}{\varphi(\sigma_i)}
 \right ]^{n\hat{\theta}_i}
 \right \} \\
 &=&
 n \sum_{i=1}^k
 \hat{\theta}_i \log \frac{\psi(\sigma_i)}{\varphi(\sigma_i)} \\
 \label{eq:ratio_deriv3}
 &=&
 n \sum_{i=1}^k
 \hat{\theta}_i \log
 \left ( 1 + \frac{\delta \left [ \psi(\sigma_i), \varphi(\sigma_i) \right ]}
 {\varphi(\sigma_i)} \right ) \\
 \label{eq:ratio_deriv4}
 &\leq&
 n (\log e) \sum_{i=1}^k
 \hat{\theta}_i \cdot
 \frac{\delta \left [ \psi(\sigma_i), \varphi(\sigma_i) \right ]}{\varphi(\sigma_i)} \\
 \label{eq:ratio_deriv5}
 &=&
 n (\log e) \sum_{i=1}^k
 \varphi(\sigma_i) \cdot
 \left [ 1 +
 \frac{\delta \left [\hat{\theta}_i, \varphi(\sigma_i) \right ]}{\varphi(\sigma_i)}
 \right ]
 \cdot
 \frac{\delta \left [ \psi(\sigma_i), \varphi(\sigma_i) \right ]}{\varphi(\sigma_i)} \\
 \label{eq:ratio_deriv6}
 &=&
 n (\log e) \sum_{i=1}^k
 \frac{\delta \left [\hat{\theta}_i, \varphi(\sigma_i) \right ] \cdot
 \delta \left [ \psi(\sigma_i), \varphi(\sigma_i) \right ]}
 {\varphi(\sigma_i)} \\
 \label{eq:ratio_deriv7}
 &\leq&
 n (\log e) \sum_{i=1}^k
 \left \{
 \begin{array}{ll}
 \frac{12.5\sqrt{2}}{n^{1+3\varepsilon/8}} \cdot \frac{k}
 {j \left (\pvece, \avec (\sigvec), i \right )}; &
 \mbox{if}~\hat{\theta}_i \leq 2 \varphi(\sigma_i), \\
 \frac{12.5}{n^{1+3\varepsilon/8}} \cdot \sqrt{
 \frac{\hat{\theta}_i}{\varphi(\sigma_i)} \cdot
 \frac{k}
 {j \left (\pvece, \avec (\sigvec), i \right )}}; &
 \mbox{if}~\hat{\theta}_i > 2 \varphi(\sigma_i).
 \end{array}
 \right . \\
 \label{eq:ratio_deriv8}
 &\leq&
 \left ( \log e \right ) \cdot
 \sum_{j=1}^k \frac{125 k}{\sqrt{1-\xi}n^{\varepsilon/4} j}
 ~\leq~
 \frac{125}{\sqrt{1-\xi} \ln 2} \cdot
 \frac{k \ln (e(k+1))}{n^{\varepsilon/4}}.
\eea
Equalities \eref{eq:ratio_deriv3} and \eref{eq:ratio_deriv5} are
obtained by using
$\psi(\sigma_i) = \varphi(\sigma_i) +
\delta \left [ \psi(\sigma_i), \varphi(\sigma_i) \right ]$
and
$\hat{\theta}_i = \varphi(\sigma_i) +
\delta \left [ \hat{\theta}_i, \varphi(\sigma_i) \right ]$, respectively.
Inequality \eref{eq:ratio_deriv4} is true because
$\ln (1 + x ) \leq x$ for $x > -1$.
The sum on all displacements of one distribution w.r.t.\ the other
is zero, yielding \eref{eq:ratio_deriv6}.
Then, the bounds in \eref{eq:delta_psi_phi}, \eref{eq:delta_psi_phik},
and \eref{eq:delta_theta_phi} result in
\eref{eq:ratio_deriv7}, where we use the worst case defined following
\eref{eq:neg_lemma_ubound}.
Then, we rearrange the sum by $j$ instead of $i$ and use
\eref{eq:sqrt_theta_varphi} to obtain
\eref{eq:ratio_deriv8}.  The last inequality of
\eref{eq:ratio_deriv8} is obtained since
$\sum_{j=1}^k 1/j \leq \ln (e(k+1))$.  This concludes the proof of
Lemma~\ref{lemma_quantization_ratio}.
{\hfill $\Box$}

\section{--~~ Proof of Theorem \ref{theorem_seq1}}
\label{ap:proof_theorem_seq1}
\renewcommand{\theequation}{F.\arabic{equation}}
\renewcommand{\theproposition}{F.\arabic{proposition}}
\renewcommand{\thelemma}{F.\arabic{lemma}}
\setcounter{equation}{0}
\setcounter{lemma}{0}
\setcounter{proposition}{0}

The individual modified redundancy of the
code defined in \eref{eq:kt1_index}-\eref{eq:kt_index} is obtained by
\be
 \label{eq:modified_KT_red}
 n\tilde{R}_n \left [ Q_k, \Psi \left ( x^n \right ) \right ] =
 - \log Q_k \left [ \Psi \left ( x^n \right ) \right ] +
 \log P_{ML} \left (x^n \right ).
\ee
From \eref{eq:kt1_index}-\eref{eq:kt_index}, it can be observed that
\be
 Q_k \left [ \Psi \left ( x^n \right ) \right ] =
 k! \cdot Q_{KT} \left ( x^n \right ).
\ee
Therefore, the individual modified redundancy
of this code is $\log(k!)$ bits less
than the i.i.d.\ redundancy of the KT code, which is well known.  This yields
the bound of \eref{eq:red_up_bound1}.  However, for the sake of completeness, we show
the main steps of the derivation of the bound from
$Q_k \left [ \Psi \left ( x^n \right ) \right ]$ itself.

By definition of $Q_k \left [ \Psi \left ( x^n \right ) \right ]$ in
\eref{eq:kt1_index}-\eref{eq:kt_index},
\be
 \label{eq:KT_index_complete}
 - \log Q_k \left [ \Psi \left ( x^n \right ) \right ] =
 \left \{ \begin{array}{ll}
 -\log \left [\frac{\left ( \frac{k}{2} - 1 \right )! \cdot k!}
 {\left ( n + \frac{k}{2} - 1 \right )!} \cdot
 \prod_{j=1}^k
 \frac{\left [2n_x(j)\right ]!}
 {2^{2n_x(j)} \left [ n_x(j) \right ]!} \right ]; &
 \mbox{for even}~k, \\
 -\log \left [ \frac{(k-1)! \cdot \left ( n + \frac{k-1}{2} \right )! \cdot
 2^{2n+k-1} \cdot k!}
 {\left ( \frac{k-1}{2} \right )! \cdot \left ( 2n + k - 1 \right )! \cdot
 2^{k-1}}
 \cdot \prod_{j=1}^k
 \frac{\left [2n_x(j)\right ]!}
 {2^{2n_x(j)} \left [ n_x(j) \right ]!} \right ]; &
 \mbox{for odd}~k,
 \end{array} \right .
\ee
where $n_x(j)$ is the number of occurrences of index $j$ in the
pattern $\Psi \left ( x^n \right )$, and the $k!$ factor is the only different
additional factor to the expression above beyond that of the standard KT probability.
The terms to the left of the product on the right hand side of the equation (except
the $k!$ term) are
the result of multiplying the values of the denominator at all time points from $1$
to $n$.  The product on the right hand side with the $k!$
term are the result of multiplying the
numerators.  To complete the derivation of the
bound in \eref{eq:red_up_bound1}, we
plug \eref{eq:KT_index_complete} into
\eref{eq:modified_KT_red} to compute the redundancy, use Stirling's
approximation \eref{eq:stirling}
to upper and lower bound factorials, use the
relationship $\ln (1 + x ) \leq x$, and combine similar order terms.
The ML i.i.d.\ probability is reduced by the occurrence of the same factors
in $Q_k \left [ \Psi \left ( x^n \right ) \right ]$ resulting from the
product term on the right hand side of \eref{eq:KT_index_complete}.
This concludes the proof of Theorem~\ref{theorem_seq1}.
{\hfill $\Box$}

\section{--~~ Proof of Theorem \ref{theorem_seq2}}
\label{ap:proof_theorem_seq2}
\renewcommand{\theequation}{G.\arabic{equation}}
\renewcommand{\theproposition}{G.\arabic{proposition}}
\renewcommand{\thelemma}{G.\arabic{lemma}}
\setcounter{equation}{0}
\setcounter{lemma}{0}
\setcounter{proposition}{0}

To prove Theorem~\ref{theorem_seq2}, we need to make one key observation.
Let
\be
 \label{eq:m_def}
 m = \left \lceil \frac{k}{2} + \frac{(k+1)^{1 -\varepsilon}}{2} \right \rceil.
\ee
Then, we can use \eref{eq:m_def} to upper bound the product in the
denominator of $Q \left [ \Psi \left (x^n \right ) \right ]$ by
$(n + m - 1)!/(m-1)!$.  This bound bounds each term of the product over
the time $n$ by an expression that is larger than each such term, resulting in a
somewhat loose bound.  This is true even for the worst sequence in which
all the $k$ distinct letters occur in the first $k$ time units, for which the
denominator of $Q \left [ \Psi \left (x^n \right ) \right ]$ is maximal.  Using
this bound,
\be
 \label{eq:low_comp_Q}
 Q \left [ \Psi \left ( x^n \right ) \right ] \geq
 \frac{(m-1)! \cdot (k!)^{1-\varepsilon}}{(n+m-1)! \cdot 2^{2n}} \cdot
 \prod_{j=1}^k \frac{\left [ 2n_x(j) \right ]!}{\left [ n_x(j) \right ]!}.
\ee
The remaining steps use Stirling's bounds \eref{eq:stirling} to
bound factorial terms, and the bound
$\ln (1 + x) \leq x$, and then combine similar order terms, eventually substituting
\eref{eq:m_def} to express the bound as a function of $k$.  Finally,
we plug an upper bound on the negative logarithm of
$Q \left [ \Psi \left ( x^n \right ) \right ]$ in \eref{eq:modified_KT_red}
instead of $Q_k \left [ \Psi \left ( x^n \right ) \right ]$.
The components of the i.i.d.\ ML probability cancel out, and
\eref{eq:red_up_bound3} is attained.
This concludes the proof of Theorem~\ref{theorem_seq2}.
{\hfill $\Box$}

\section*{Acknowledgments}

The author gratefully acknowledges Alon Orlitsky,
Prasad Santhanam, and Junan Zhang.  The discussions the author
had with them, not only motivated this work, but helped significantly
in advancing it.  In particular, the author acknowledges their comments
that led to the correction of the
second region of \eref{eq:minimax_pattern_bound} and the improvements
of the proofs of Theorems~\ref{theorem_maximin_patterns}
and~\ref{theorem_most_patterns}.   The author acknowledges the associate editor
Serap Savari, for her help and patience with this paper, and the anonymous reviewers
for their very helpful comments, that helped improve the manuscript.
The author also acknowledges Frans
Willems, Tjalling Tjalkens, Serap Savari, and Lihua Song for very helpful
discussions, and Li Wang for performing the simulations for this
paper.

\end{document}